\documentclass[preprint, 3p]{elsarticle}

\usepackage{lineno,hyperref}
\modulolinenumbers[5]

\usepackage{dcolumn}
\usepackage{amsmath}
\usepackage{amssymb}
\usepackage{amsthm}
\usepackage{lmodern}
\usepackage{graphicx}
\usepackage{bm}
\usepackage[T1]{fontenc}
\usepackage{color}
\usepackage{url}
\usepackage[bf]{subfigure}
\usepackage{rotating}

\usepackage{scalefnt}
\usepackage{multirow}
\usepackage{hyperref}
\usepackage{threeparttable}
\usepackage{adjustbox}

\newcommand{\M}{\mathbf{M}}
\newcommand{\R}{\mathbf{R}}

\newcommand{\A}{\mathbf{A}}
\newcommand{\B}{\mathbf{B}}
\newcommand{\D}{\mathbf{D}}

\newcommand{\Lap}{\mathbf{L}}
\newcommand{\J}{\mathbf{J}}

\newcommand{\La}{\mathbf{L}}
\newcommand{\Hm}{\mathbf{H}}
\newcommand{\W}{\mathbf{W}}
\newcommand{\Pd}{\mathbf{P}}
\newcommand{\Gm}{\mathbf{G}}
\newcommand{\V}{\mathbf{V}}
\newcommand{\Cuop}{\mathbf{C}}

\DeclareMathOperator{\I}{\mathbf{I}}
\DeclareMathOperator{\Prob}{\mathbb{P}}

\newcommand{\E}[1]{\left\langle #1 \right\rangle}
\newcommand{\expm}[1]{\text{exp}\left( #1 \right)}
\newcommand{\bigO}[1]{O \left( #1 \right)}

\newcommand{\vertiii}[1]{{\left\vert\kern-0.25ex\left\vert\kern-0.25ex\left\vert #1 
    \right\vert\kern-0.25ex\right\vert\kern-0.25ex\right\vert}}

\graphicspath{{./figures/}}

\begin{document}

\title{Fundamentals of spreading processes in single and multilayer complex networks}

\author[icmc,isi]{Guilherme Ferraz de Arruda}
\ead{gui.f.arruda@gmail.com}

\author[icmc,warwick1,warwick2]{Francisco A. Rodrigues}
\ead{francisco@icmc.usp.br}

\author[bifi,zgz,isi]{Yamir Moreno}
\ead{yamir.moreno@gmail.com}

\address[icmc]{Departamento de Matem\'{a}tica Aplicada e Estat\'{i}stica, Instituto de Ci\^{e}ncias Matem\'{a}ticas e de Computa\c{c}\~{a}o, Universidade de S\~{a}o Paulo - Campus de S\~{a}o Carlos, Caixa Postal 668, 13560-970 S\~{a}o Carlos, SP, Brazil.}

\address[warwick1]{Mathematics Institute, University of Warwick, Gibbet Hill Road, Coventry CV4 7AL, UK.}
\address[warwick2]{Centre for Complexity Science, University of Warwick, Coventry CV4 7AL, UK.}

\address[bifi]{Institute for Biocomputation and Physics of Complex Systems (BIFI), University of Zaragoza, Zaragoza 50009, Spain}
\address[zgz]{Department of Theoretical Physics, University of Zaragoza, Zaragoza 50009, Spain}
\address[isi]{ISI Foundation, Turin 10126 , Italy}

\begin{abstract}
Spreading processes have been largely studied in the literature, both analytically and by means of large-scale numerical simulations. These processes mainly include the propagation of diseases, rumors and information on top of a given population. In the last two decades, with the advent of modern network science, we have witnessed significant advances in this field of research. Here we review the main theoretical and numerical methods developed for the study of spreading processes on complex networked systems. Specifically, we formally define epidemic processes on single and multilayer networks and discuss in detail the main methods used to perform numerical simulations. Throughout the review, we classify spreading processes (disease and rumor models) into two classes according to the nature of time: (i) continuous-time and (ii) cellular automata approach, where the second one can be further divided into synchronous and asynchronous updating schemes. Our revision includes the heterogeneous mean-field, the quenched-mean field, and the pair quenched mean field approaches, as well as their respective simulation techniques, emphasizing similarities and differences among the different techniques. The content presented here offers a whole suite of methods to study epidemic-like processes in complex networks, both for researchers without previous experience in the subject and for experts.
\end{abstract}

\maketitle
\tableofcontents

\section{Introduction}

Historically, the classical modeling of epidemic processes assumes homogeneously mixed populations, which implies that the probability of two individuals in the population being connected is the same. However, such an approach has proved to be unrealistic, as most real systems present heterogeneously mixed patterns, in which contacts among individuals frequently follow a scale-free organization~\cite{Newman03, Boccaletti06:PR, Costa07:AP}, have degree-degree correlations \cite{Newman02:PRL} and cycles \cite{watts:nature98}, and many other non-trivial patterns of connections~\cite{Boccaletti06:PR, Costa07:AP}. This limitation on the mathematical modeling is overcome by representing the  population as a network, where individuals are portrayed as nodes and their relationships are given by edges connecting them. Our knowledge of the structure and dynamics of networks has significantly increased since the end of the 90's~\cite{watts1998collective,Barabasi99}.

Modeling complex systems as networks \cite{Boccaletti06:PR, Costa07:AP, Barrat08:book, Newman010:book, Costa011:AP} has proved to be a successful approach to characterize the structure and dynamics of many real-world systems, from the biological domain to social and technological applications \cite{Costa011:AP}.  However, on single-layer networks, only one type of connection is accounted for. This modeling is also limited because most natural and artificial systems, such as the brain, our society or modern transportation networks \cite{Kivela2014, BoccalettiPR2014} are made up of different constituents and/or various types of interactions. For instance, in social networks, individuals can be connected according to different social ties, such as friendship or family relationships~(e.g. \cite{Verbrugge79}). In transportation networks, routes of a single airline can be represented as a network, whose vertices (destinations) can be mapped into networks of different companies \cite{Cardillo2013}. Gene co-expression networks consist of layers, each one representing a different signaling pathway or expression channel \cite{Li011:PLoS}. Mapping out these and similar systems as monoplex networks can miss relevant information, which is not captured if the single layers are analyzed separately -- neither when all the layers collapse together in a single aggregated graph. In most of these interconnected systems, the information travels not only among vertices of the same layer but also between pairs of layers. In order to overcome the aforementioned limitations of the single layer representation and to allow for a proper modeling of the previously presented examples, multilayer networks have been introduced in recent years~\cite{Kivela2014}. Systems are represented as a collection of layers, which are made of nodes and edges. In this context, edges can be of two types, intra or inter-layer, accounting for connections inside or between two different layers. Note that the original definition of networks is also included in the multilayer case, thus the latter generalize the previous one, but also including more complex and richer structures \cite{Kivela2014, BoccalettiPR2014}. 

On the other hand, the network organization has been considered in the modeling of many dynamical systems. For instance, rumor and epidemic spreading, synchronization of coupled oscillators, and diffusion processes have been studied by considering the structured populations and networks~\cite{BoccalettiPR2014, Kivela2014}. Interestingly, the introduction of heterogeneous structures has changed our understanding of the dynamical behavior of many systems. Take the case of disease spreading as an example. Pastor-Satorras and Vespignani \cite{Satorras01:PRL, Satorras2001} showed in 2001 that a disease outbreak takes place when the spreading rate, $\lambda$, is larger than the epidemic threshold that is related to the network structure~\cite{Satorras01:PRL, Satorras2001}, i.e., if $\lambda > \lambda_c^{HMF} = \E{k} / \E{k^2}$, where $\E{ k^m }$ is the $m$-th moment of the degree distribution. The authors introduced the heterogeneous mean-field (HMF) approximation in which all the nodes with a given degree are considered to be statistically equivalent, neglecting dynamic correlations. Therefore, in scale-free networks, which present degrees distributions that follow a power law $P(k)\sim k^{-\zeta}$ with $\zeta \leq 3$, are particularly prone to spreading of diseases, since the second moment of the degree distribution diverges and, therefore, $\lambda_c \rightarrow 0$ when $N\rightarrow\infty$. Contrasting with the HMF, in 2003 Wang proposed the QMF (quenched mean field) approach \cite{Wang03}, in which the individual node probability is evaluated, including the full network structure, by means of the adjacency matrix $\A$. This approach predicts the critical point as $\lambda_c^{QMF} = \frac{1}{\lambda_1(\A)}$, where $\lambda_1(\A)$ is the largest eigenvalue of $\A$. Complementing this analysis with the results of Chung et. al. \cite{ChungPNAS2003} on the spectral properties of $\A$, one gets $\lambda_c^{QMF} = \sqrt{k_{max}}^{-1}$ if $\zeta > \frac{5}{2}$  and $\lambda_c^{QMF} \approx \E{k} / \E{k^2}$ if $2 < \zeta < \frac{5}{2}$. Intriguingly, both the HMF and the QMF agree in their predictions for $2 < \zeta < \frac{5}{2}$, however, the first predicts a finite threshold for $\zeta > 3$, whilst the second predicts that the threshold vanishes when $k_{max}$ diverges, regardless of the exponent $\zeta$. Numerical results suggested that the QMF is qualitatively correct on power-law networks, therefore representing an improvement of the HMF approach \cite{Ferreira2012}. In order to explain the contradiction between both theories, in 2012 Goltsev et.al. \cite{Goltsev2012} proposed a spectral theory approach explaining the observed discrepancies among the critical point predictions as a localization phenomenon. Accordingly, if $\lambda_1(\A)$ corresponds to a localized eigenstate, the disease remains in a finite set of nodes, e.g., a subgraph containing a huge hub. Conversely, if $\lambda_1(\A)$ corresponds to a delocalized eigenstate, the disease affect a finite fraction of nodes. All these works reinforce the importance of network structure on epidemic modeling.

In addition to the aforementioned results for single-layer networks, also the theory of multilayer networks has been developed and used to study dynamical processes. For a theoretical analysis, a mathematical formulation following the tensorial \cite{DeDomenico2013} and matrix \cite{BoccalettiPR2014, Kivela2014} representation were proposed. For instance, in 2013, Gómez et. al. studied a diffusion process on multiplex networks \cite{GomezPRL2013}. Interestingly, they observed  super-diffusion, meaning that the time scale of the multiplex is smaller than that obtained when each layer is considered independently. Moreover, in the same year Cozzo et. al. studied disease spreading processes on top of multiplex networks \cite{Cozzo:2013}, extracting important results using a perturbative approach. Among their main results, they observed a shift to the left in the phase diagram of the non-dominant layer, as a consequence of the activity of the dominant one. More precisely, the non-dominant layer experiences an unexpected earlier outbreak, where the number of infected individuals on such a layer is larger than zero even before its expected critical point. This phenomenon appears as a consequence of the disease being spread through inter-layer edges. Besides, they also proposed an approximation for the leading eigenvalue of the adjacency matrix, which helped them to explain their results. Indeed, the previously mentioned results emphasize the different scales of multiplex systems, which can be one of three phases: (i) the decoupled phase, when the layers behave independently, (ii) the multiplex phase, when the system acts as a whole and (iii) the network of layers phase, when the inter-layer connections dominate the processes.

The importance of modeling epidemic processes realistically can be emphasized recalling the Influenza A pandemic, and specially, Dr Harvey Fineberg's presentation on May 2011 at the World Health Assembly: \emph{``Conclusion 3: The world is ill-prepared to respond to a severe influenza pandemic or to any similarly global, sustained and threatening public health emergency. Beyond the implementation of core public health capacities called for in the IHR (International Health Regulations), global preparedness can be advanced through research, strengthened health-care delivery systems, economic development in low- and middle-income countries and improved health status’’} \cite{H1N1}. Moreover, recently, the World Health Organization reported a new outbreak of Ebola in May 2017, two years after the first big outbreak \cite{Ebola}. Summarizing, epidemic and information spreading are important processes from a phenomenological point of view and are directly applicable to social problems and are especially interesting both in mathematics and physics, as many of the results could be extended to other different dynamical processes on top of networks. From this perspective, the goal of this review is to contribute, both socially and technically, to the fields of mathematical epidemiology and information spreading. 

Throughout this review, we study two types of networks: single and multilayer networks. Firstly we address the structural characterization of these systems by network measures. In Section \ref{sec:classical}, we formally define these networks in terms of their adjacency matrices. In general, there are many ways to represent them. Here we focus on the matrix representation because it allows us to naturally link the structure with its spectral properties, which are especially interesting when analyzing dynamical processes. Additionally, we present some key concepts of the structural characterization of networks, regarding two levels: the nodal characterization, i.e., centrality measures, and global measures. Moreover, we also briefly present network models, which are fundamental to analyze dynamical processes in networks, as they can be considered as null models. Furthermore, a special section is aimed at studying spectral properties of these systems, i.e., Section \ref{sec:spectral_characterization}, as they play a key role in dynamical processes, connecting the structure and dynamics. These sections allow the reader to become acquainted with the main fundamental points in the structural characterization of single and multilayer networks, paving the way for the next sections. 

In summary, this review can provide not only an up to date compendium of the central concepts related to disease and epidemic-like modeling on single-layer and multilayer networks, but also will present the necessary tools for other researchers (and notably for young researchers and scientists starting their career in the topic) to develop their methods and applications in a didactical manner. The methods described here offer a whole framework to study these processes, from the theoretical and computational points of view.

\section{Classical network theory} \label{sec:classical}

In this section, we provide a formal description of networks, representing their structure as graphs. We also present some basic network measurements and the main network models to generate random networks.

\subsection{Network representation} \label{sec:net_representation}

\begin{figure*}
\begin{center}
\includegraphics[width=0.96\textwidth]{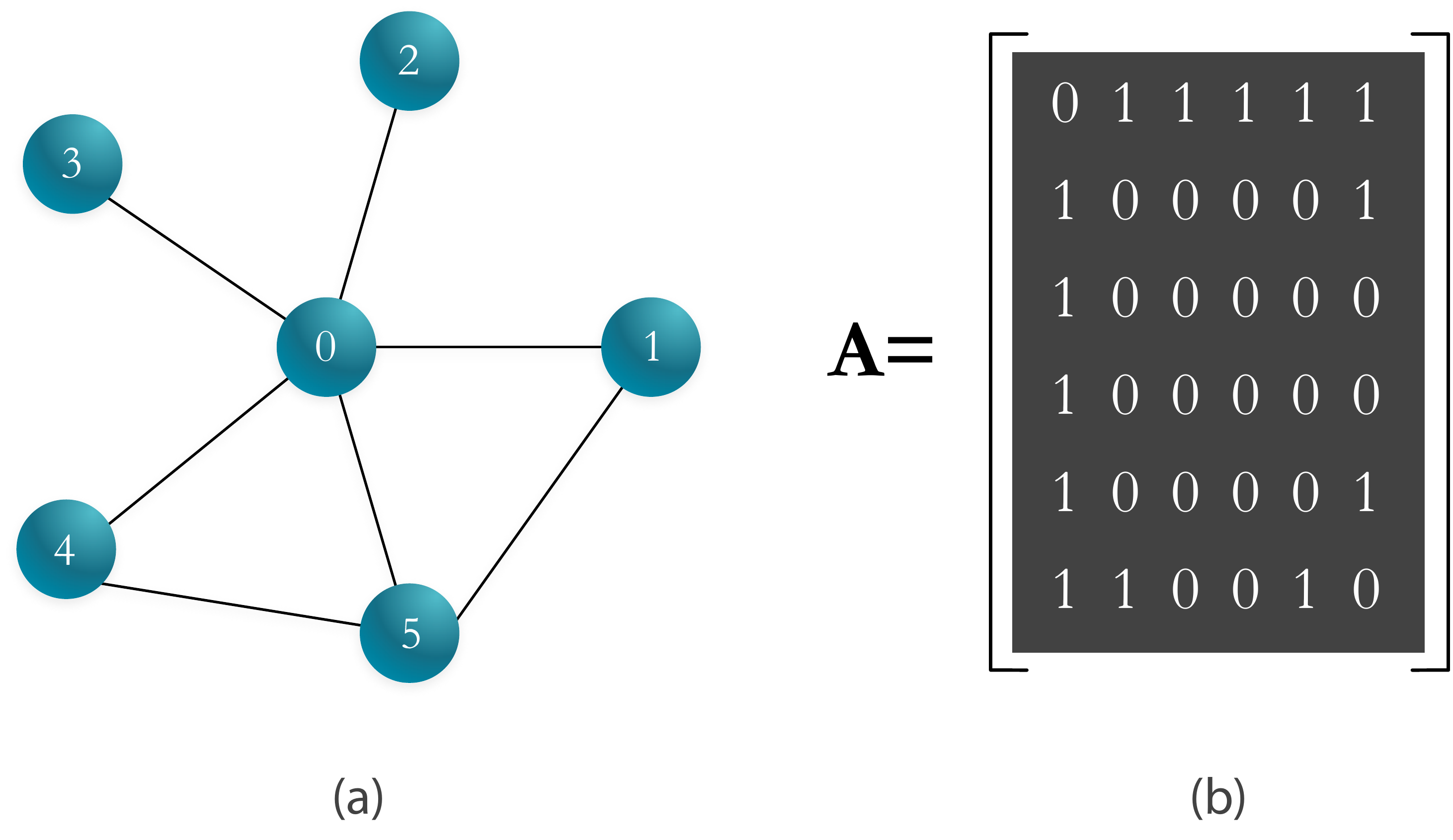}
\caption{The undirected network $G = ([N], [L])$, where $[N] = \{ 0, 1, 2, 3, 4, 5\}$ and $[L] = \{ l_{0 \rightarrow 1}, l_{0 \rightarrow 2}, l_{0 \rightarrow 3}, l_{0 \rightarrow 4}, l_{0 \rightarrow 5}, l_{5 \rightarrow 1}, l_{4 \rightarrow 5}, ... \}$, where the reciprocal edges were omitted. Its graphical representation is shown in (a), while its adjacency matrix is shown in (b).} \label{fig:net_ex}
\end{center}
\end{figure*}

The structure of a network is formally represented as a graph \cite{Boccaletti06:PR}. A graph consists of two sets, $G = ([N], [L])$, i.e., the set of nodes $[N]:=\{1,2,\ldots,N\}$ and the set of edges, which are ordered pairs of $[N]$, $l_{i \rightarrow j} := (i,j)$, $[L] := \{l_{i \rightarrow j}, \forall i,j \hspace{2mm} \text{are connected}\}$ and $|L| = M$. Observe that in undirected graphs, the pairs $l_{i \rightarrow j}$ are unordered implying that both $l_{i \rightarrow j} \in [L]$ and $l_{j \rightarrow i} \in [L]$. Often a graph is represented by its adjacency matrix, $\A$, where $\A_{ij} = 1$ if there is an edge from node $i$ to node $j$, i.e. $l_{i \rightarrow j} \in [L]$, and $\A_{ij} = 0$ otherwise. One might also associate a weight to each connection, making a so-called weighted network. This network is represented by $\W$, where the elements, $\W_{ij}$, of this matrix represent the weight between nodes $i$ and $j$. We remark that in the undirected case, $\A$ is symmetric, while in the directed case, this property may not hold. It is worth mentioning that other representations are also possible, such as edgelists, adjacency lists, among other matrices. However, here we focus on the adjacency matrix due to its mathematical elegance and potential. Figure \ref{fig:net_ex} exemplifies an undirected network, showing its adjacency matrix and its graphical representation.

Throughout this review, we restrict ourselves to some specific types of networks. We focus on finite undirected, unweighted, and connected networks. In other words, we consider a finite number of nodes, where all the edges present unitary weights and are undirected. Moreover, there is a path from any node $i$ to any other node $j$ on the network. In the case we can extend our results to other types of networks, we make it explicit, otherwise, those are our standard constraints.

\subsection{Characterization measures} \label{sec:Characterization}

The proper characterization of the network structure is an important task in the field of complex networks, allowing us to describe their topology and also to support studies on the analysis/evaluation of the relationship between structure, dynamics and function of networks. In a general statement, we might define the characterization task as the mapping of a system into a vector of measurements~\cite{Costa07:AP}. This representation is important to quantify different patterns of connections in networks and to verify how these patterns influence the evolution of dynamical process and the function of complex systems.

Many measures to characterize the network structure have been proposed in the literature \cite{Boccaletti06:PR, Costa07:AP}. They can be divided according to many criteria. However, here we divide them into centrality-based metrics and global measures. The first category, which is discussed in Section \ref{sec:cent}, enables us to compare different nodes within the same network. In other words, it quantifies the importance of a given node with respect to the other nodes. On the other hand, global measurements, which are discussed in Section \ref{sec:global}, enable us to compare different networks.

\subsubsection{Centrality measures} \label{sec:cent}

As mentioned before, some metrics can be considered to define the centrality of a node. Here, we provide the basic definitions of those used in the rest of the review. For more, please refer to \cite{Boccaletti06:PR, Costa07:AP}. The most basic definition of centrality is the node \emph{degree}, $k_i$, formally given in terms of its adjacency matrix as
\begin{equation}
 k_i = \sum_j \A_{ij}.
\end{equation}
In this case, the most central node has the largest number of connections. In directed networks, the in-degree of the node is defined by $k_i^{in} = \sum_{i=1}^N A_{ij}$ and the out-degree, $k_i^{out} = \sum_{j=1}^N A_{ij}$. Note that the degree is given by $k_i = k_i^{in} + k_i^{out}$. Regarding weighted networks, this concept is generalized in the so-called strength of the node $s_i = \sum_{j=1}^N W_{ij}$. Observe that the strength can also be generalized to directed networks. 

Alternatively, the centrality of a node can be defined in terms of the degree of its second neighbors, as strongly connected vertices can surround a central node. In this case, the \emph{average degree of the nearest neighbors} of $i$ is defined as
\begin{equation}
r_i = \frac{1}{k_i}\sum_{j \in \partial i}k_j,
\end{equation}
where $\partial i$ is the set of nodes connected to $i$. Interestingly, it can be observed that the average neighborhood degree is related to epidemic spreading in networks \cite{Gleeson2012, Barrat08:book}. 

Another property of interest are the shortest paths of the network. We can also define centrality based on the distance between pairs of nodes. From this perspective, the more central the node, the shorter its distance to all other nodes. The corresponding measure is the \emph{closeness centrality} defined as \cite{Newman010:book}
\begin{equation}
C_i = \frac{N}{\sum_{j=1, j \neq i}^N d_{ij}},
\end{equation}
where $d_{ij}$ is the length of the shortest distance between nodes $i$ and $j$. Alternatively, the effective load of a node (also defined in terms of shortest paths) can be considered as a centrality measure. This measurement is called \emph{betweenness centrality} and it quantifies the load as the number of times a node acts as a bridge along the shortest path between two other nodes \cite{Girvan02:PNAS}. Thus, for a node $i$,
\begin{equation}
B_i = \sum_{(a,b)} \frac{\sigma(a,i,b)}{\sigma(a,b)},
\label{betweenness}
\end{equation}
where $\sigma(a,i,b)$ is the number of shortest paths connecting nodes $a$ and $b$ that pass through node $i$ and $\sigma(a,b)$ is the total number of shortest paths between $a$ and $b$. The sum is over all pairs $(a,b)$ of distinct nodes. In this case, a central node should be crossed by many paths and shows the highest values of $B_i$.

\begin{figure*}
\includegraphics[width=0.96\textwidth]{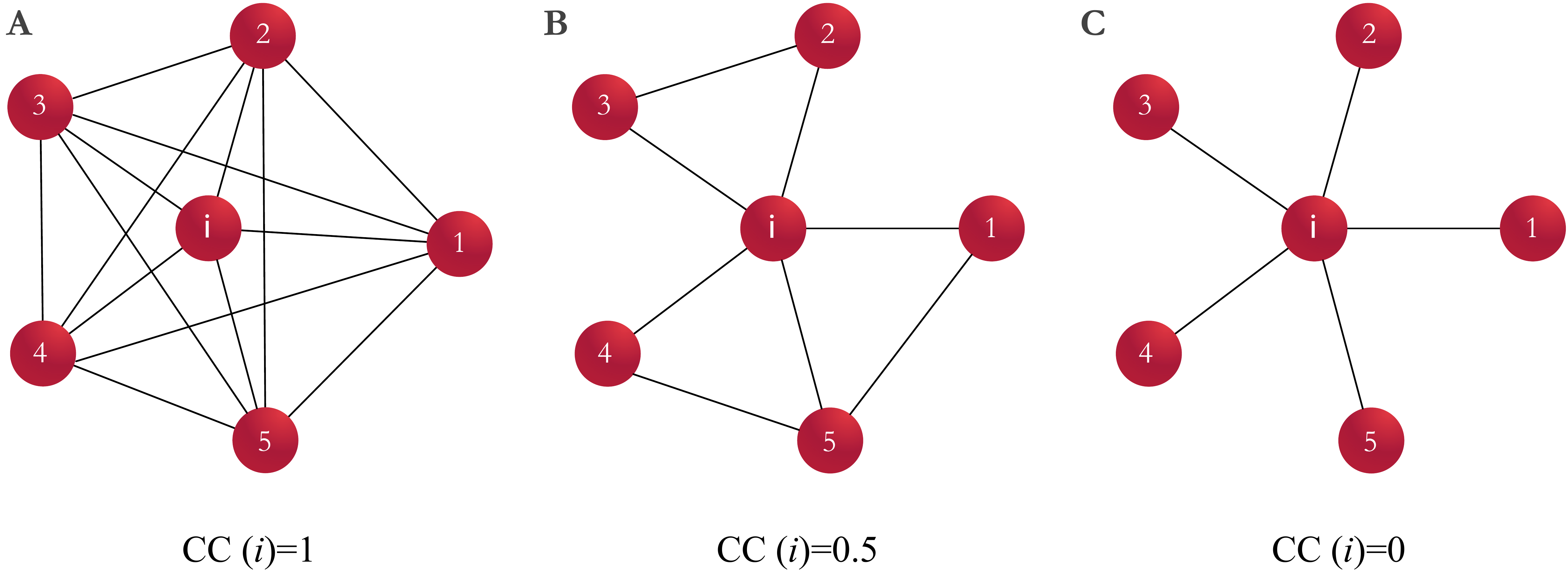}
\caption{An example of clustering coefficient. Three network configurations that result in different values of the clustering coefficient.}\label{fig:cc} 
\end{figure*}

Another important quantity of interest is the number of triangles in a given network. We call the \emph{clustering coefficient} the measure that quantifies the occurrence of triangles in the networks. It is defined as \cite{Boccaletti06:PR}
\begin{equation}
cc(i) = \frac{3N_{\triangle}(i)}{N_3(i)},
\label{eq:cci}
\end{equation}
where $N_{\triangle}(i)$ is the number of triangles involving the node $i$ and $N_3(i)$ is the number of triples centered around $i$. As argued in \cite{Newman010:book}, $cc(i)$ can be understood as a centrality measure in the sense that if two nodes are connected only via node $i$, this node can control the information flow. Thus, the clustering coefficient can be thought of as a local version of the betweenness centrality. Note that $cc(i)$ takes smaller values for more central nodes, contrary to the other centrality measures. Figure \ref{fig:cc} illustrates the calculus of the clustering coefficient.

The \emph{$k$-shell} decomposition removes nodes in a network to obtain a subgraph (called \emph{$k$-core}) whose vertices have at least $k$ interconnections. The centrality of each node is defined by assigning an integer index to each node $i$, $k_c(i)$, where $k_c(i) = k$ if $i$ belongs to $k$-core, but does not belong to $(k + 1)$-core \cite{Seidman83}. Nodes with low values of $k_c$ are located at the periphery of the network. The most central nodes should have the highest values of coreness, whereas high-degree nodes located in the network periphery should display small values of coreness \cite{Kitsak10:NP}. Therefore, only hubs at the main core of the networks present the highest values of $k_c$. Importantly, this measure has been shown to be correlated with epidemic spreading in networks~\cite{Kitsak10:NP}.

Considering the eigenstructure of the adjacency matrix we can define the so-called \emph{eigenvector centrality}. This measure considers that the centrality of each node is the sum of the centrality values of the nodes to which it is connected. Thus, the eigenvector centrality is defined as the components of the leading eigenvector, i.e., the eigenvector associated to the largest eigenvalue of the adjacency matrix $\A$. In formal terms,
\begin{equation}
x_i = \lambda_1^{-1} \sum_j \A_{ij} x_j,
\end{equation}
alternatively, in the matrix form we have the eigenvalue problem, $\A \textbf{x} = \lambda_1 \textbf{x}$, where $\textbf{x}$ is the right leading eigenvector \cite{Newman010:book} and $\lambda_1$ is the largest eigenvalue. Observe that, in contrast to the previous metrics, which considers only local information, the eigenvector centrality considers the whole network structure. Besides, this measure is closely connected to spreading processes in networks, as they relate to the spectra of the adjacency matrix.

The number of visits that a given node receives when an agent travels through the network without a preferential route can also be taken into account to quantify the node centrality. In this case, a possible measure is the \emph{Google PageRank} \cite{Brin98theanatomy}. Formally, PageRank is calculated as 
\begin{equation}
\pi^T = \pi^T \Gm,
\end{equation}
where $\Gm$ is the Google matrix, i.e.,
\begin{equation}
\Gm = \kappa \left(\Pd + \frac{ae^T}{N} \right) + \frac{(1-\kappa)}{N} u u^T,
\end{equation}
and $a$ is the binary vector called dangling node vector ($a_i$ is equal to one if $i$ is a dangling node and 0 otherwise), $u$  is a vector with unitary elements and $\Pd_{ij} = \frac{\A_{ij}}{k_j}$ is the transition probability matrix of the respective network (for more on this matrix see Section \ref{sec:Pd}). The original version of the algorithm considers $\kappa = 0.85$ \cite{Brin98theanatomy}. The PageRank of a node $i$, $\pi_i$, is given by the $i$-th entry of the dominant eigenvector $\pi$ of $\Gm$, given that $\sum_i \pi_i = 1$. $\pi_i$ can be understood as the probability of arriving at node $i$ after numerous steps following a random walk navigation of the network.

\subsubsection{A visual example of centrality measures}\label{Sec:cent_comp}

\begin{figure*}
\includegraphics[width=\textwidth]{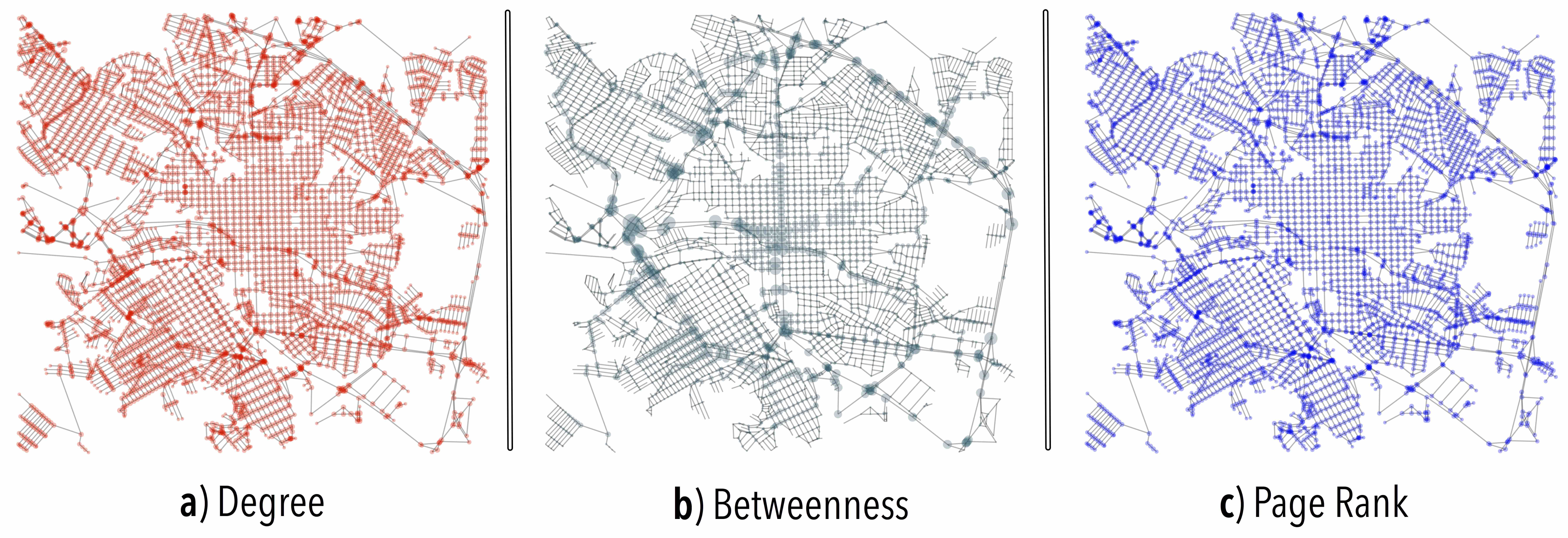}
\caption{Comparison of three different centrality measures: degree in (a), betweenness centrality in (b) and Page Rank in (c). The size of the nodes are proportional to the centrality of the node. The network used is the S\~{a}o Carlos road map, extracted using the Python library OSMNX (Version: 0.5.1 -- \href{https://github.com/gboeing/osmnx}{https://github.com/gboeing/osmnx}).}\label{fig:cent_comp} 
\end{figure*}

In order to illustrate the concept of centrality in networks, in Figure \ref{fig:cent_comp} we show different centrality measures. The network used here is the S\~{a}o Carlos street network (in Brazil). Regarding the centralities, we chose three metrics, degree, betweenness centrality and PageRank. Note that the first is based on a local property, the second on shortest paths and the last one on navigation properties. As commented previously, there is no single measurement that summarizes the importance of a single node. Thus, each metric defines the centrality in different ways. Therefore, the most central nodes in terms of a given measure might not be the same ones defined by other measures.

S\~{a}o Carlos is a city in S\~{a}o Paulo state, Brazil and its estimated population in 2017 was about 246 thousand inhabitants\footnote{Data From  \href{http://cidades.ibge.gov.br/xtras/perfil.php?lang=&codmun=354890&search=sao-paulo|sao-carlos}{IBGE/2017}.}. Observe that the city center presents a regular pattern, where most of the nodes have four connections, while in the periphery we have different patterns even with long roads connecting distant parts of the city. This information is well captured by the betweenness centrality, while neglected by local measures such as the degree centrality. Thus, due to the nature of the analyzed network, we observe that the most suitable measure here is the betweenness centrality, which captures the main avenue of S\~{a}o Carlos (and its crossings) as the most important roads. Furthermore, we also notice a strong correlation between degree and the Page Rank centrality in this specific analysis.

\subsubsection{Generalized random walk accessibility}\label{Sec:Acc}

\begin{figure} [t]\centering
\includegraphics[width=0.65\columnwidth]{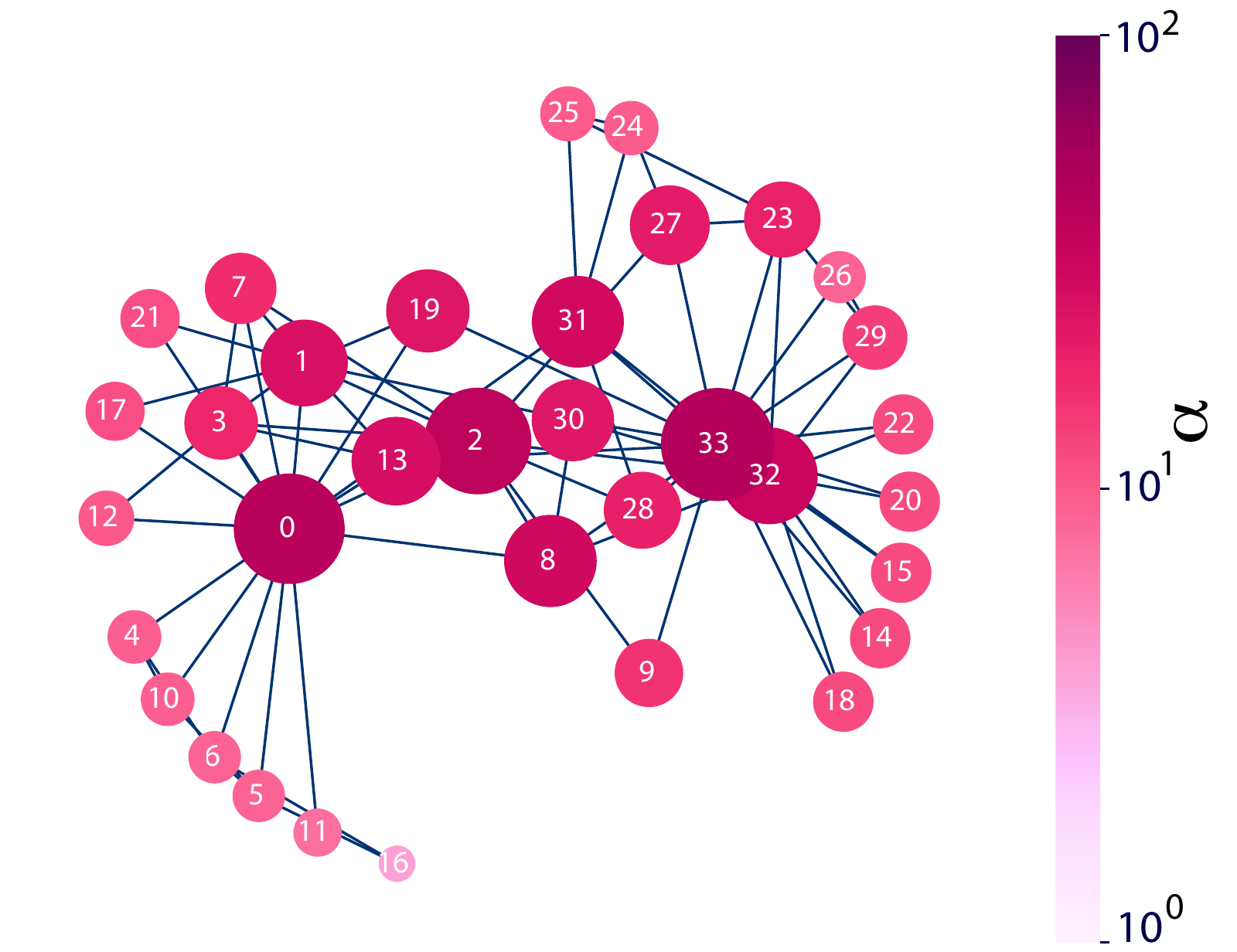}
\caption{Illustration of the concept of accessibility (values calculated from Eq.\ref{Eq:acc}) in the Zachary Karate-club network \cite{Zachary77}. Node size is also proportional to its accessibility values. Nodes at the center of the network present the highest accessibility.}
\label{fig:ex_acc}
\end{figure}

In this section, we present another way in which random walks can be applied to study centrality. More specifically, we use the accessibility \cite{Travenccolo2008:PLA, Travencolo2009:NJP}, which is related to the diversity of access of individual nodes through random walks \cite{Travenccolo2008:PLA}. This measure has been considered to identify the border of complex networks \cite{Travencolo2009:NJP}. Let $\Pd^{(h)}(i,j)$ be the probability of reaching node $j$ by performing random walks of length $h$ departing from $i$ \footnote{Note that $\Pd^{(h)}$ is the $h$-th power of the transition matrix $\Pd_{ij} = \frac{\A_{ij}}{k_j}$.}. The accessibility of node $i$ for a given distance $h$, is given by the exponential of the Shannon entropy \cite{Travenccolo2008:PLA}, i.e.,
\begin{equation}
\alpha_h(i) = \exp\left({-\sum_{j}P^{(h)}(i,j)\log P^{(h)}(i,j)}\right),
\label{Eq:acch}
\end{equation}
where $1 \leq \alpha_h(i) \leq N$. The maximum value corresponds to the case in which all nodes are reached with the same probability $1/N$. Note that this metric was defined in a multilevel fashion, depending on the parameter $h$ that defines the scale of the dynamics \cite{Travenccolo2008:PLA, Travencolo2009:NJP}. In addition, although we are constrained to random walks, virtually any other type of dynamics yielding transition probabilities between adjacent nodes can be considered in the accessibility, which makes this measurement adaptable to the dynamics of each problem being studied.

In order to generalize the accessibility, here we introduce a new version of this metric, which is based on the matrix exponential operation \cite{Bhatia1997:book}. This matrix enables the calculation of the probability of transition considering walks of all lengths between any pair of nodes. Thus, if $\Pd$ is the transition matrix, the exponential of $\Pd$ is defined as \cite{deArruda2014}
\begin{equation}\label{Eq:W}
\textbf{W} = \sum_{k=0}^{\infty} \frac{1}{k!} {\Pd}^k = e^{\Pd}.
\end{equation}
The matrix $\textbf{W}$ is based on a modified random walk, which penalizes longer paths. To construct this stochastic process we consider an usual random walk $(X_n)_{n\geq 0}$, where $X_n$ represents the node visited by the agent at time $n$. We take a collection of independent and identically distributed uniform random variables in the interval $(0,1)$, i.e. $\{U_1, U_2,\ldots\}$, which represents a kind of ``fitness'' associated to each step of the walk. Moreover, we assume independence between the collection of uniform random variables and the random walk. This modified random walk, which we call accessibility random walk (ARW), considers walks through the network such that all associated fitnesses along a trajectory are in ascending order. We say that node $j$ is visited by the ARW, at time $n$, if $X_n =j$ and $U_1 < U_2 < U_3 < \cdots < U_n$. We denote by $(\tilde{X}_n)_{n\geq 0}$ the new process and note that $\{\tilde{X}_n = j\}$ implies $\{X_n = j\}$, but the opposite is not necessarily true. A quantity of interest is the number of visits that a given node $j$ receives when an agent travels through the network according to the ARW. This quantity can be written as $\sum_{n=1}^{\infty}I_{\left\{\tilde{X}_n =j\right\}}$, where $I_{A}$ is the indicator function of event $A$. We are interested in the mean of this value, by assuming that the agent starts from node $i$, i.e. $\sum_{n=1}^{\infty}\E{I_{\left\{\tilde{X}_n =j\right\}}|\tilde{X}_0 =i}$. In order to compute this value we observe that the term of the sum is the probability $\Prob(\tilde{X}_n =j|\tilde{X}_0 =i)$ which, by our definition, is equal to $\Prob(\{X_n=j\}\cap\{U_1< U_2 < U_3 < \cdots < U_n\}|X_0 = i)$. This probability is exactly $(1/n!)\Pd^{(n)}(i,j)$, where $\Pd^{(n)}(i,j)$ is the probability of the transition from $i$ to $j$ through walks of length $n$. Therefore, the matrix $\textbf{W}$ considered in Eq.~\eqref{Eq:W} is a matrix of mean values associated to the ARW. The element $\textbf{W}(i,j)$ provides the mean number of visits that node $j$ receives when the agent starts at node $i$ and follows the ARW. 

Therefore, the probability of transition between any pair of nodes through ARW is given by
\begin{equation}\label{Eq:P}
\textbf{P} = \frac{\textbf{W}}{e}.
\end{equation}
Note that the matrix $\textbf{W}$ weighs all walks by the inverse of the factorial of lengths. Therefore, this definition penalizes longer walks, i.e., the shortest walks receive more weight than the longest ones. We define the generalized expression for the accessibility as \cite{deArruda2014}
\begin{equation}\label{Eq:acc}
\alpha(i) = \exp\left(-\sum_j \textbf{P}(i,j)\log \textbf{P}(i,j)\right),
\end{equation}
which we call \emph{generalized random walk accessibility}. Figure \ref{fig:ex_acc} illustrates this measure on the Zachary Karate-club network \cite{Zachary77}.

We note that the exponential matrix is also considered in the definition of the communicability \cite{Estrada08, Estrada2011}. The difference is that the accessibility is based on the concept of diversity \cite{Hill1973, Jost06} whereas communicability is associated with the communication between any pair of nodes \cite{Estrada2011}. Moreover, the former is related to the probability transition matrix, whereas the latter to the adjacency matrix. Thus, there is no trivial relationship between these two metrics in irregular graphs.

\subsubsection{Global measures} \label{sec:global}

Here we discuss some metrics that may be used to globally quantify a specific feature of our network. First of all, it can be observed that, in principle, any summary statistics\footnote{Summary statistics are used to summarize a set of observations. As examples, we can cite the average value or any statistical moment, the variation, the standard deviation, or any central moment.} might be applied to local features. Formally, the \emph{$m$-th moment} of any distribution is defined as
\begin{equation}
 \E{X^m} = \sum_i x_i^m \Prob (x_i),
\end{equation}
where $X$ is a random variable (with a countable set -- discrete case), whose outcomes are $x_i$ and its occurrence is given by $\Prob(x_i)$. As an example, the average degree quantifies the average number of connections of a node found in a given network and is calculated as $\E{k} = \sum_k k \Prob (k)$. Complementary, we may use the \emph{variance} to quantify the spreading out from their average value. The variance is defined by
\begin{equation}
 \mathbb{V}(X) = \E{X^2} - \E{X}^2,
\end{equation}
where $\E{X^2}$ is the second statistical moment and $\E{X}^2$ is the squared value of the first moment. Observe that the variance is measured in squared units. In order to express it in the same unit of the original data we might consider the \emph{standard deviation}, which is simply given as $\sigma(X) = \sqrt{\mathbb{V}(X)}$.

The degree-degree correlation among nodes can be calculated considering the \emph{Pearson correlation coefficient} on the extremes of each edge \cite{Newman02:PRL}. This correlation is calculated as
\begin{equation}
 \rho^P = \frac{(1/M)\sum_{j>i}k_ik_j A_{ij} - [(1/M)\sum_{j>i}(1/2)(k_i +k_j) A_{ij}]^2}
{(1/M)\sum_{j>i}(k_i^2 + k_j^2) A_{ij} - [(1/M)\sum_{j>i}(1/2)(k_i +k_j) A_{ij}]^2},
\end{equation}
where $M$ is the number of edges and the superscript $P$ in $\rho^P$ stands for the Pearson correlation coefficient. Alternatively, the Spearman correlation coefficient \cite{PhysRevE.87.022801} can also be used. If $\rho^P > 0$, then the network is assortative, indicating that nodes with similar degrees tend to connect among themselves. On the other hand, if $\rho^P < 0$, then the network is called disassortative, which indicates that nodes with higher degrees tend to connect with low degree nodes. The case $\rho^P = 0$ suggests that there is no degree correlation. Note that the Pearson correlation coefficient measures only the linear correlation.

\subsection{Network optimization: degree-degree correlation tuning} \label{sec:simulated_annealed}

Under some circumstances, it is interesting to tune/optimize a specific network measurement, which can be a structural or even dynamical feature of the system. For instance, in \cite{deArruda2016} the authors evaluated the impact of assortative/disassortative structures on epidemic spreading. Here, we describe an optimization method in terms of the assortative coefficient, however, we remark that it can be used for any global network property of interest. 

Thus, in order to control the level of degree-degree correlations in random networks, we consider a simulated annealing algorithm \cite{Kirkpatrick1983}. This algorithm is based on two functions, i.e., (i) the perturbation function, which changes the system configuration, and (ii) the energy function, which is minimized. In our case, the perturbation function is a rewiring procedure that preserves the degree distribution of the network but changes the large-scale degree-degree correlations. The energy function is defined as $E_t = c(\rho_t + 1)$, where $\rho_t$ is the network assortativity at time $t$ and $c$ is a constant related to the level of degree-degree correlation, i.e., $c = -1$ if the goal is to obtain an assortative network or $c = 1$ if the goal is a disassortative network.

Given an initial network configuration, an initial temperature $T$, and a cooling factor $\alpha$, the algorithm can be described by the following steps: (i) the energy function is initialized as $E_0$; (ii) while the number of iterations is less than a threshold or the optimal solution is not found (or good solution, given a tolerance) the following steps are performed: (iii) a rewiring preserving the degree distribution is executed, according to our perturbation function; (iv) the new energy function, $E_{t+1}$, is calculated; (v) if $E_t - E_{t+1} < 0$ or $\exp \left( \frac{-(E_t - E_{t+1})}{T} \right) < U(0,1)$, where $U(0,1)$ is a random number sampled from a uniform distribution in $[0,1]$, then the new solution is accepted; (vi) the temperature is updated, $T = \alpha T$; and (vii) increment the iteration counter. Observe that a worse state than the current one can be accepted with a probability $\exp \left( \frac{-(E_t - E_{t+1})}{T} \right)$. This mechanism allows the system to avoid local minima. Following this procedure, we can generate random networks with a defined level of degree-degree correlation.

\subsection{Network models} \label{sec:models}

Network models are stochastic methods to generate networks with different structures \cite{Boccaletti06:PR, Costa07:AP, Newman010:book, Barrat08:book}. The main goal of the theoretical approaches is to reproduce some features observed in real systems. One of the first network models was proposed by Erd\"os and R\'enyi (ER) \cite{erdos1959random}. Following this model, nodes are connected according to a uniform probability, without any preference. Despite its mathematical elegance, it cannot describe properties of real networks, as discussed in \cite{Newman03, Newman010:book}. Thus, many other models have been proposed in the literature to overcome the limitations of the ER model. In this section, we briefly describe some important models that can be used to either mathematically evaluate a given model or describe a real network structure.

\subsubsection{Erd\"os--R\'enyi model} \label{sec:er}

\begin{figure*}
\includegraphics[width=0.96\textwidth]{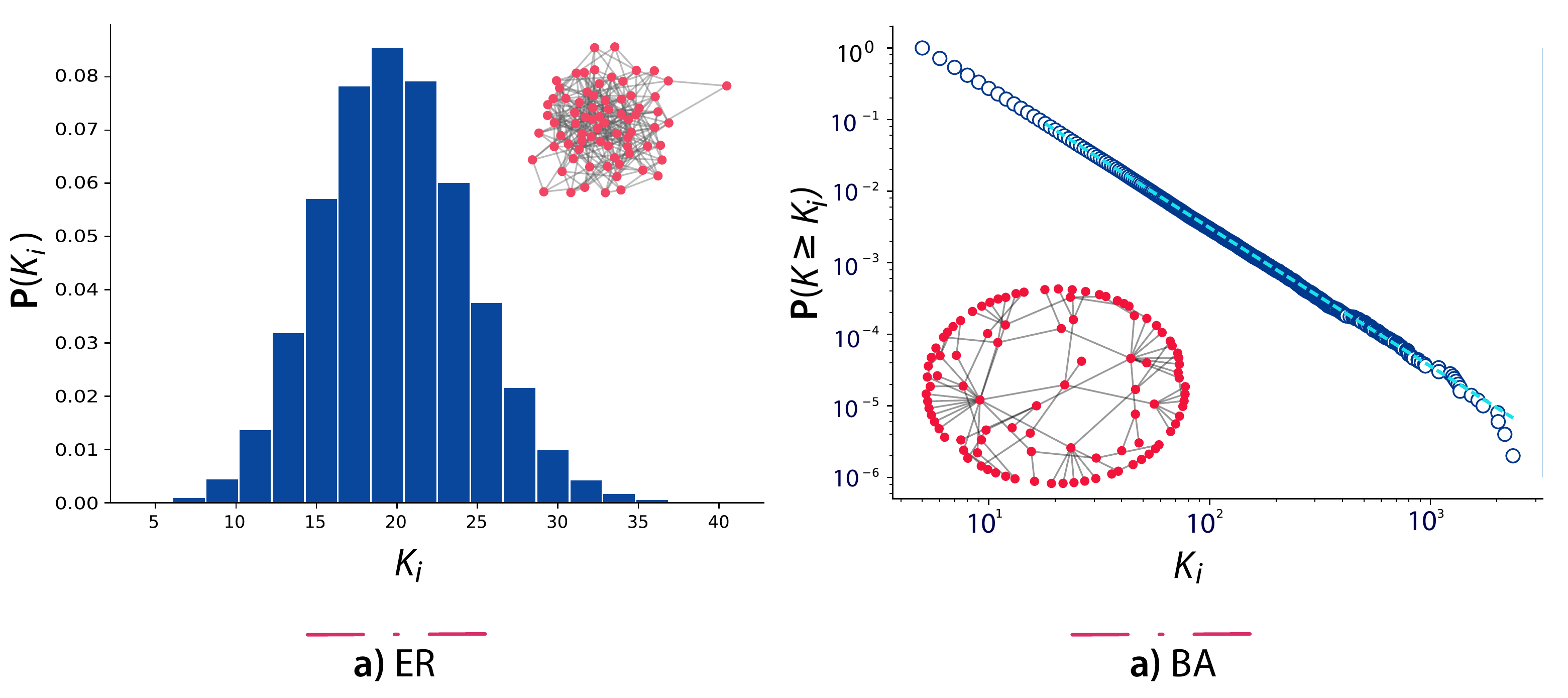}
\caption{In (a) we present the degree distribution of an Erd\"os and R\'enyi network with $N = 10^5$ and $\E{k} = 20$, while in the inset, a small ER network is shown for visualization purposes. In (b) we present the degree distribution of a Barab\'{a}si--Albert network with $N = 5 \times 10^5$ and $\E{k} = 10$, where the estimated $\zeta \approx 3.0$. In the inset, a small BA network is shown for visualization purposes. This figure and its approximation of $\zeta$ were generated using the code provided in \href{http://tuvalu.santafe.edu/~aaronc/powerlaws/}{http://tuvalu.santafe.edu/~aaronc/powerlaws/} (Accessed on 4th September, 2017).} \label{fig:deg_er_ba}
\end{figure*}

In 1959, the mathematicians Paul Erd\"os and Alfred Renyi proposed a random model \cite{erdos1959random}, where the probability of connection between two different nodes is the same. It is also called the ER model. This model was studied in-depth in \cite{bollobas2001random}. Besides its mathematical elegance, the structures built by this model do not represent real networks, as empirically verified by Newman in \cite{Newman03}.

Systematically, the algorithm that generates this model is described by the following rules: starting with a set of $N$ disconnected nodes, at each pair of nodes an edge is added with probability $p$ -- and not added with a complementary probability $(1-p)$. Thus, each edge on the graph has the same probability and the graph is completely homogeneous. Obviously, the average number of connections for each node is $k = p(N - 1)$. Note that its degree distribution is given by a binomial distribution since this is a Bernoulli process. Thus, for a large value of $N$, i.e. on the thermodynamic limit, and if the average degree is kept constant, the degree distribution tends to a Poisson distribution, as a consequence of the law of rare events \cite{bollobas2001random}. Figure \ref{fig:deg_er_ba} (a) shows an example of ER degree distribution and a small visualization of the network.

\subsubsection{Barab\'{a}si--Albert model} \label{sec:ba}

Barab\'{a}si and Albert proposed a model which considers growth and preferential attachment rules to generate scale-free networks \cite{Barabasi99}. It is also known as the BA model. In this case, a network is generated starting with a set of $m_0^*$ connected nodes.  After that, a new node with $m^*$ edges is included in the network. The probability of the new node $i$ to connect with an existing node $j$ in the network is proportional to the number of connections of $j$, i.e.,
\begin{equation}
\Prob(i,j) = \frac{k_j}{\sum_u k_u}.
\end{equation}
The most connected vertices have a greater probability of receiving new connections. Thus, networks generated by this model present a power-law degree distribution, $\Prob(k) = k^{-\zeta}$, where $\zeta = 3$ in the thermodynamic limit ($N\rightarrow \infty$) \cite{Barabasi99}. For the sake of an example, we show Figure \ref{fig:deg_er_ba} (b), where we present the degree distribution of a BA network and a small visualization of the network. Moreover, compared to Figure \ref{fig:deg_er_ba} (a) we can see the differences between homogeneous, ER, and heterogeneous, BA, degree distributions.

\subsubsection{Spatial models} \label{sec:spatial}

Regarding a homogeneous model, the model proposed by Waxman \cite{waxman1988routing} considers that nodes are uniformly distributed into a square of a unitary area. Each pair of nodes is connected according to a fixed probability, that depends on their distances, given as:
\begin{equation}
\Prob(i,j) = \eta_s \exp(-\eta_s d_{ij}), 
\end{equation}
where $\eta_s$ is a parameter that controls the average degree and $d_{ij}$ is the Euclidean distance between nodes $i$ and $j$. This model generates networks with an exponential degree distribution, which means that the probability of a node having a degree different than $\E{ k }$ decays exponentially. 

The model introduced by Barth{\'e}lemy \cite{Barthelemy2003:EPL, Barthelemy2011} produces scale-free networks embedded in space. This model is called SSF, which stands for spatial scale-free. Considering a regular $d$ dimensional lattice with length $L$, the algorithm has three main steps. Initially, $n_0$ initial active nodes are selected at random. Next, an inactive node $i$ is randomly selected, and connected to an active node $j$ with probability 
\begin{equation}
\Prob(i,j) \propto \frac{k_j+1}{\exp(d_{ij}/r_c)},
\end{equation}
where $k_j$ is the number of connections of node $j$, $r_c$ is a finite scale parameter and $d_{ij}$ is the Euclidean distance between nodes $i$ and $j$. Finally, the node $i$ becomes active and the second and third steps are repeated until all nodes are active. For each node, the second and third steps are repeated $m^*$ times in order to set the average connectivity as $\E{k} = 2m^*$ \cite{Barthelemy2003:EPL, Barthelemy2011}. The parameter $r_c$ controls the clustering coefficient \cite{watts:nature98} and assortativity \cite{Newman02:PRL} of the network.

\begin{figure*}
\includegraphics[width=0.96\textwidth]{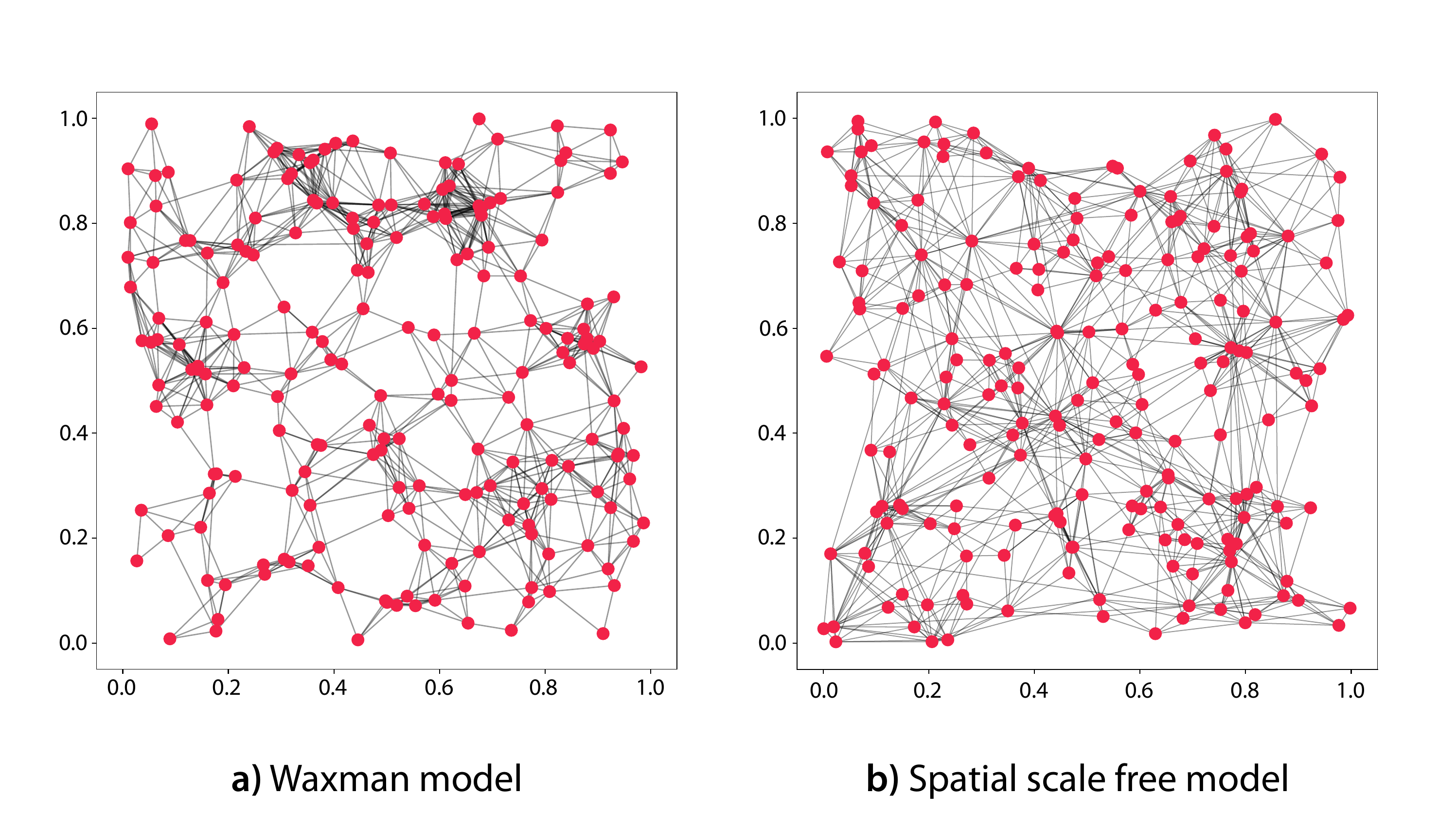}
\caption{In (a) we present a network obtained from the Waxman model (homogeneous), while in (b) the spatial scale free model proposed by Barth{\'e}lemy (heterogeneous). Each networks has $N = 200$. Observe that the network in (a) does not present hubs and long-range connections are rare, while in (b) those features are more often observed.} \label{fig:ws_ssf}
\end{figure*}

Figure \ref{fig:ws_ssf} shows an example of both models discussed in this sub-section. In (a) the Waxman model, which produces a homogeneous network, whilst in (b) the SSF model, producing a scale-free network that weighs two different mechanisms: (i) the spatial distribution of the nodes and (ii) the preferential attachment. Here we consider $r_c = 0.05$, $L = 1$ and $d = 2$. These values are similar to those used in the original review \cite{Barthelemy2003:EPL}.

\subsubsection{Configuration model} \label{sec:conf}

Empirical evidence from many areas shows that real networks present a heterogeneous pattern of connections \cite{Boccaletti06:PR, Costa07:AP, Newman010:book, Barrat08:book}. Many networks present a heavy-tailed degree distribution often approximated by a power-law of the form $\Prob(k) \sim k^{-\zeta}$. Thus, it is important to have an algorithm to generate a network with such a feature but random regarding other network properties. The BA model generates scale-free networks, but we cannot control the value of the exponent $\zeta$. On the other hand, the so-called configuration model can generate a network with any given degree distribution (more specifically, with a given degree sequence), enabling the control of $\zeta$ for networks presenting power-law degree distribution. The model is implemented as follows: (i) for each node we assign a degree $k_i$ based on the degree distribution $\Prob(k)$; (ii) the graph is constructed by randomly connecting pairs of nodes until every node is connected. Observe that there is no degree-degree correlation associated with this process since the edges are created without taking into account the degree of the nodes that are connected.

Regarding uncorrelated scale-free networks, i.e. $\Prob(k) \sim k^{-\zeta}$, there is an upper bound for the maximum degree of the network of $k_{max} = \frac{1}{\sqrt{N}}$ (imposing a structural cut-off, $k_{max} \sim N^{\frac{1}{2}}$), as shown in \cite{Catanzaro2005}. This upper bound applies to networks without self-loops and multiple edges. Besides, allowing a free value of $k_{max} = N$ would also introduce correlations in this kind of networks. In this case, for the natural cut-off, where $k_{max} \sim N^{\frac{1}{(\zeta-1)}}$ \cite{Barrat08:book}, it is impossible to construct uncorrelated power-law distributions with $2 < \zeta \leq 3$ without allowing multiple edges and self-loops \cite{Catanzaro2005}.

The configuration model can be adapted to produce random networks with a distribution of triangles, allowing us to control the clustering coefficient \cite{Newman2009}. This model can be used, for example, to quantify the impact of the triangles over a certain dynamical process. It is especially useful in social dynamics since this is a typical feature of this class of networks \cite{Boccaletti06:PR, Costa07:AP}.

\subsubsection{Random regular networks} \label{sec:rrn}

Random regular networks (RRN) are a class of homogeneous graphs in which every node has the same degree, $k_i = k$, but their connections are made randomly. Formally we have that $\Prob(k_i = k) = 1$, $\forall i \in [N]$. This model does not represent real systems, but it is helpful when analyzing and evaluating dynamical processes in networks. For instance, its homogeneity also implies a simplification of the mathematical modeling. Moreover, due to correlations, it is an interesting structural model in the evaluation of the accuracy of dynamical models, as seen in \cite{Mata2013}, where the authors have used this network to evaluate their prediction of the critical point in epidemic spreading. Furthermore, in Section \ref{sec:model_accuracy}, we also use this model when we discuss the precision of mean field approaches.

\section{Spectral characterization of networks} \label{sec:spectral_characterization}

We can describe the network topology in terms of its spectral properties~\cite{Mieghem:2011}. Here we focus on single-layer networks, but the generalization to multilayer networks is natural and straightforward, due to their matrix representation. We consider only undirected and weighted networks. In case we relax these constraints we will specify it in the context. Note that directed networks do not have symmetric associated matrices. Historically, Augustin--Louis Cauchy proved the spectral theorem for self-adjoint matrices (or Hermitian matrices), i.e., that every real, symmetric matrix is diagonalizable. Furthermore, the spectral decomposition, eigenvalue decomposition, or eigendecomposition is the decomposition of a matrix on the underlying vector space on which the operator acts. Mathematically such decomposition of $\M$ is given as
\begin{equation}
 \M = \V \D^* \V^{-1},
\end{equation}
where $\V$ is a $N \times N$ matrix, whose $i$-th column is the eigenvector $\V_i$ and $\D^*$ is a diagonal matrix whose elements are the associated eigenvalues, $\D^*_{ii} = \lambda_i$.

The first natural structure is the adjacency matrix since it completely describes a network. Moreover, this matrix and its spectra also appear in the modeling of spreading processes, as will be shown in Section \ref{sec:qmf}. A brief description of its spectral properties is shown in Section \ref{sec:A}. The next matrix studied is the Laplacian matrix (see Section \ref{sec:L}) which also appears in the analysis of dynamical processes in networks. However, its use is more frequent in synchronization processes \cite{Arenas2008} and community organization of networks \cite{Fortunato2010}. In Section \ref{sec:Pd}, we discuss the probability transition matrix, which is closely connected to random walks and even epidemic processes in networks \cite{Gomez2010, Arruda2017}. Finally, in Section \ref{sec:spectral_characterization_B} we discuss the non-backtracking matrix which brings some interesting structural properties, such as communities \cite{Krzakala2013} and their relation to spreading processes \cite{Shrestha2015}.

\subsection{Adjacency matrix} \label{sec:A}

\begin{figure}[!t]
\begin{center}\centering
\includegraphics[width=0.75\columnwidth]{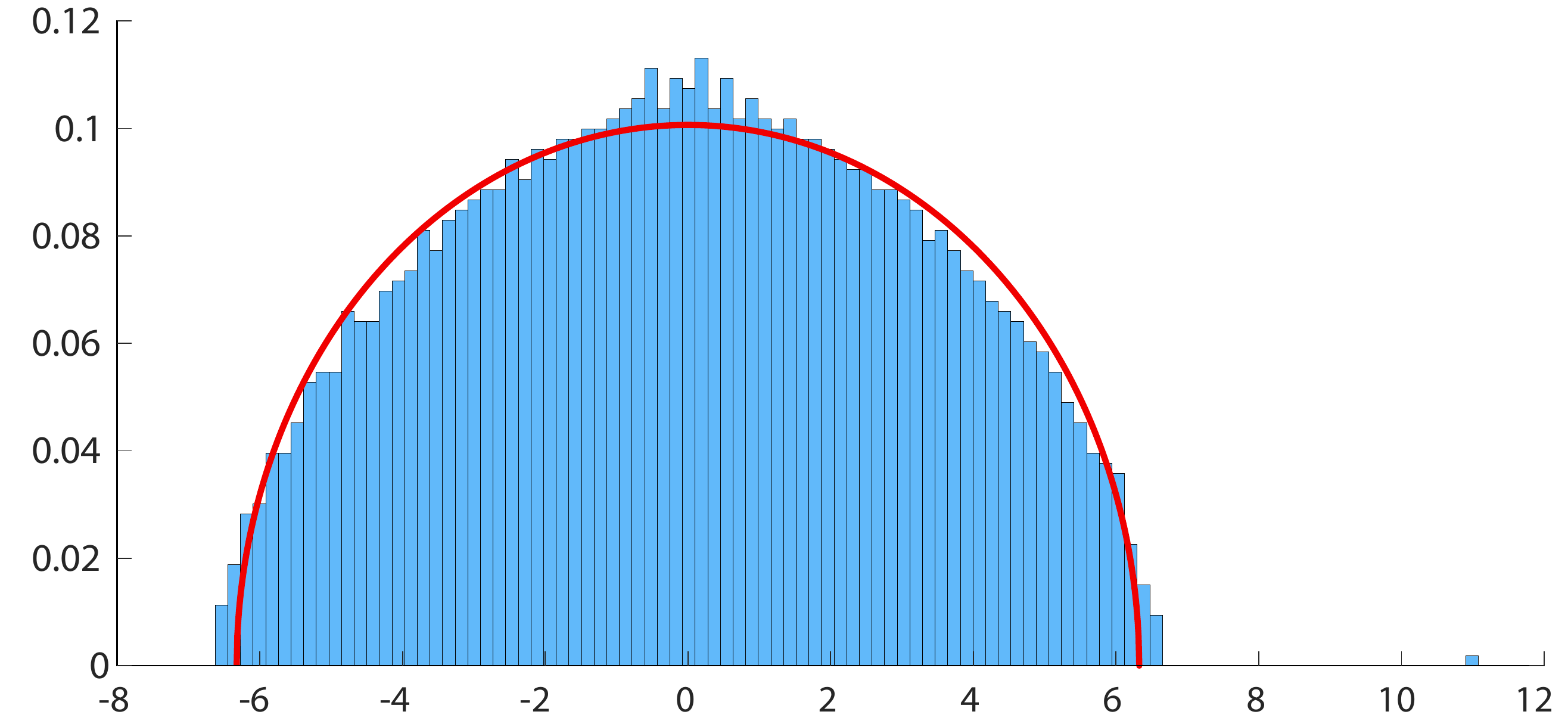}
\caption{Spectrum of the adjacency matrix of an Erd\"os and R\'enyi network with $n=3000$ and $\E{k} = 10$. The red line represents the spectral distribution given by the Wigner's semicircle law.} \label{fig:Spec_A}
\end{center}
\end{figure}

The adjacency matrix completely describes a graph or network. As previously mentioned, it is a binary matrix, whose elements $i$ and $j$ are one if there is an edge between these nodes, i.e.,
\begin{equation}
 \A_{ij} = \begin{cases}
            1 \hspace{1cm} \text{if there is an edge connecting nodes $i$ and $j$}\\
            0 \hspace{1cm} \text{otherwise}.
           \end{cases}
\end{equation}

As commented in \cite{Krzakala2013}, for sufficient dense Erd\"os and R\'enyi networks, the Wigner's semicircle law \cite{Wigner1958} can be applied. Formally, it states that the bulk of eigenvalues is distributed as  
\begin{equation}
 \Prob (\lambda) = \frac{\sqrt{4c - \lambda^2}}{2 \pi c},
\end{equation}
thus, the bulk of the spectrum lies in $[-2\sqrt{c}, 2\sqrt{c}]$ and $c$ is the average degree $c = \E{k}$, where we used the same notation as the original reviews. Figure \ref{fig:Spec_A} shows an example of the spectral distribution of an Erd\"os and R\'enyi network with $n=3000$ and $\E{k} = 10$ and its prediction from Wigner's semicircle law.

In the case of power-law degree distributions, $P(k) \sim k^{-\zeta}$, we can use previous results about the spectrum of graphs \cite{ChungPNAS2003} and obtain the following expression for the leading eigenvalue of $\A$
\begin{equation}
 \lambda_{max} \simeq 
 \begin{cases}
  \sqrt{k_{max}} &\hspace{1cm} \zeta > \frac{5}{2} \\
  \frac{\E{k^2}}{\E{k}} &\hspace{1cm} 2 <\zeta < \frac{5}{2}
 \end{cases}
\end{equation}
where $k_{max}$ is the maximum degree of the network. Note that, in the thermodynamic limit, the maximum degree is a growing function of the network size.

The leading eigenvalue is connected to many dynamical processes in networks (see Section \ref{sec:qmf} and~\cite{Boccaletti06:PR}). We can guarantee, using the Perron-Frobenius theorem, that any adjacency matrix has a largest real eigenvalue and that the corresponding eigenvector can be chosen to have strictly positive components.

\subsection{Laplacian matrix} \label{sec:L}

\begin{figure}[!t]
\begin{center}
\includegraphics[width=0.65\columnwidth]{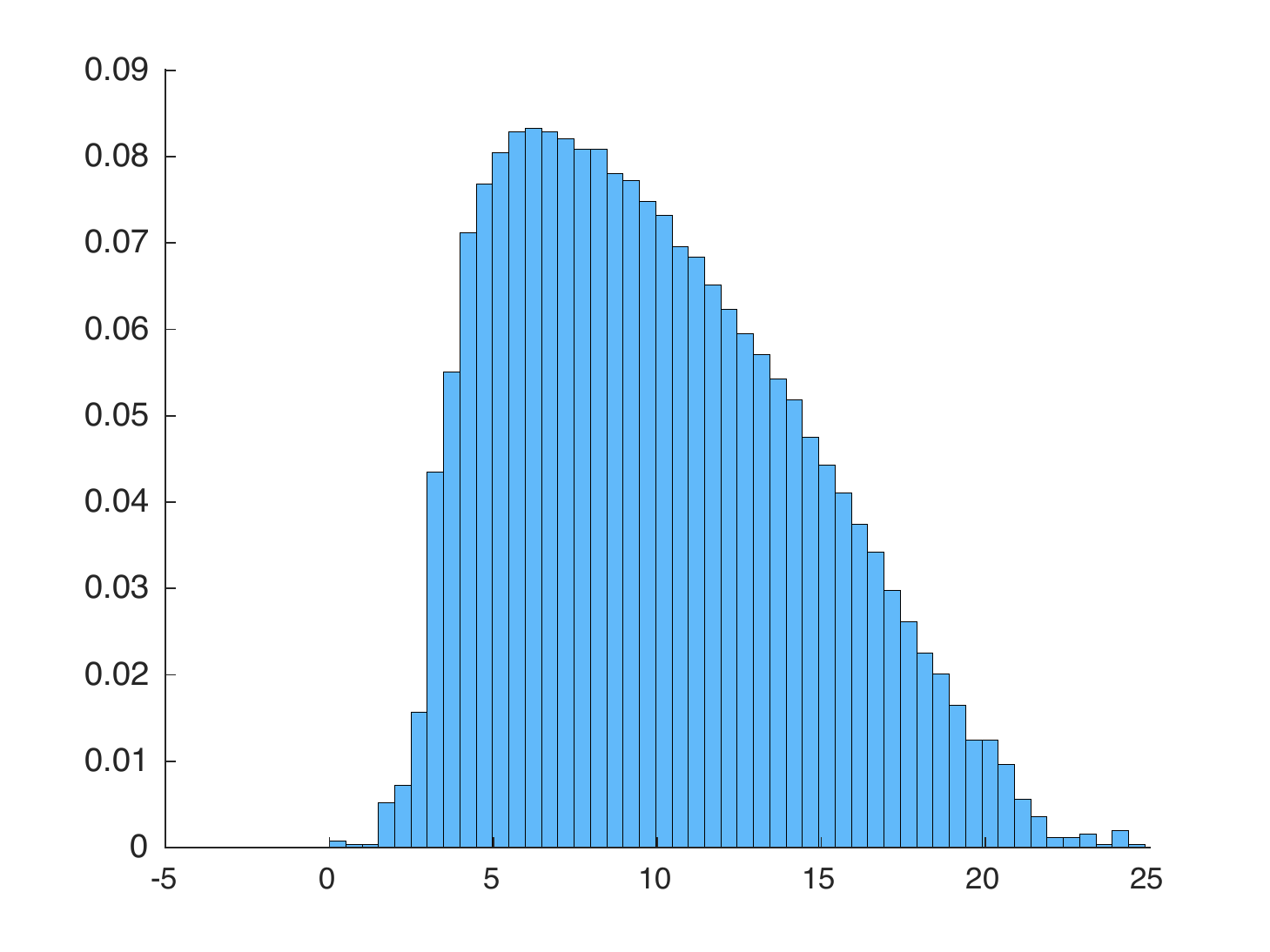}
\caption{Spectrum of the Laplacian matrix of an Erd\"os and R\'enyi network with $n=3000$ and $\E{k} = 10$. } \label{fig:Spec_L}
\end{center}
\end{figure}

In order to introduce the Laplacian matrix, we show it in terms of a diffusion process. Assuming that the flux of a given quantity is proportional to a control parameter times the flux difference between nodes, we have the following master equation
\begin{equation}
 \frac{d \psi_i}{dt} = C \sum_j \A_{ij} (\psi_j - \psi_i),
\end{equation}
where $\psi_i$ is the quantity under study and $C$ is the control parameter, which can be understood also as a coupling parameter. Thus, this equation can be rewritten in its matrix form as
\begin{equation}
 \frac{d \psi}{dt} + C \Lap \psi = 0,
\end{equation}
where $\psi$ is a vector, $\psi = \{\psi_1, \ldots, \psi_N \}$. Notice that this differential equation has a similar form to the gas diffusion equation, except that the Laplacian operator $\nabla^2$ has been replaced by the Laplacian matrix $\Lap$. For this reason, we call it the Laplacian graph \cite{Newman010:book}. Thus, the Laplacian matrix can be formally defined as 
\begin{equation} \label{eq:laplacian_def}
 \Lap = \D - \A,
\end{equation}
where $\D$ is a diagonal matrix comprising the degree of each node, $\D_{ii} = k_i$ and $\D_{ij} = 0$ for $i\neq j$, and $\A$ is the adjacency matrix. Observe that the sum of the rows of this matrix is zero for all of its rows. This also implies that zero is an eigenvalue of $\Lap$. Furthermore, the Laplacian matrix is positive semi-definite and its smallest eigenvalue is zero. Therefore, our main interest lies in the second smallest eigenvalue, which is also called \emph{algebraic connectivity}, and its corresponding eigenvector is called the \emph{Fiedler vector}. These two objects are directly connected to the community structure \cite{Shi2000, Fortunato2010} and dynamical processes on networks \cite{Newman010:book}. Moreover, for the sake of an example, in Figure \ref{fig:Spec_L} we show the spectral distribution of the Laplacian matrix. Note that all the eigenvalues are positive and zero is also an eigenvalue.

\subsection{Probability transition matrix} \label{sec:Pd}

The probability transition matrix was already presented in Section \ref{Sec:Acc}. It was defined as $\Pd_{ij} = \frac{\A_{ij}}{k_j}$. However, we can also define it in its matrix form as
\begin{equation}
 \Pd = {\D}^{-1}\A,
\end{equation}
where $\D$ is a diagonal matrix comprising the degree of each node, $\D_{ii} = k_i$. This matrix is closely related to random walks \cite{Meila01, Travenccolo2008:PLA, Travencolo2009:NJP, deArruda2014} and spreading processes in networks \cite{Gomez2010, Arruda2017}.

In \cite{Meila01}, the authors show that the spectral properties of $\Pd$ are related to the problem of clustering proposed in \cite{Shi2000}. In \cite{Meila01} the authors propose to cut a graph into pieces using the following relation with the Laplacian matrix,
\begin{equation} \label{eq:Lap_gen}
 \Lap v = \mu {\D} v.
\end{equation}
This model proved to be efficient in clustering tasks. In \cite{Meila01} it was shown that $\mu = (1 - \lambda)$, where $\lambda$ are solutions of
\begin{equation}
 \Pd v = \lambda v.
\end{equation}
This result can be easily obtained by multiplying Equation \ref{eq:Lap_gen} on the left by ${\D}^{-1}$. The interest in this relationship lies in the fact that the probability matrix shows a strong relation with the group structure in the network and also with the Laplacian matrix.

\subsection{Non-backtracking matrix}  \label{sec:spectral_characterization_B}

\begin{figure}[!t]
\begin{center}
\includegraphics[width=0.75\columnwidth]{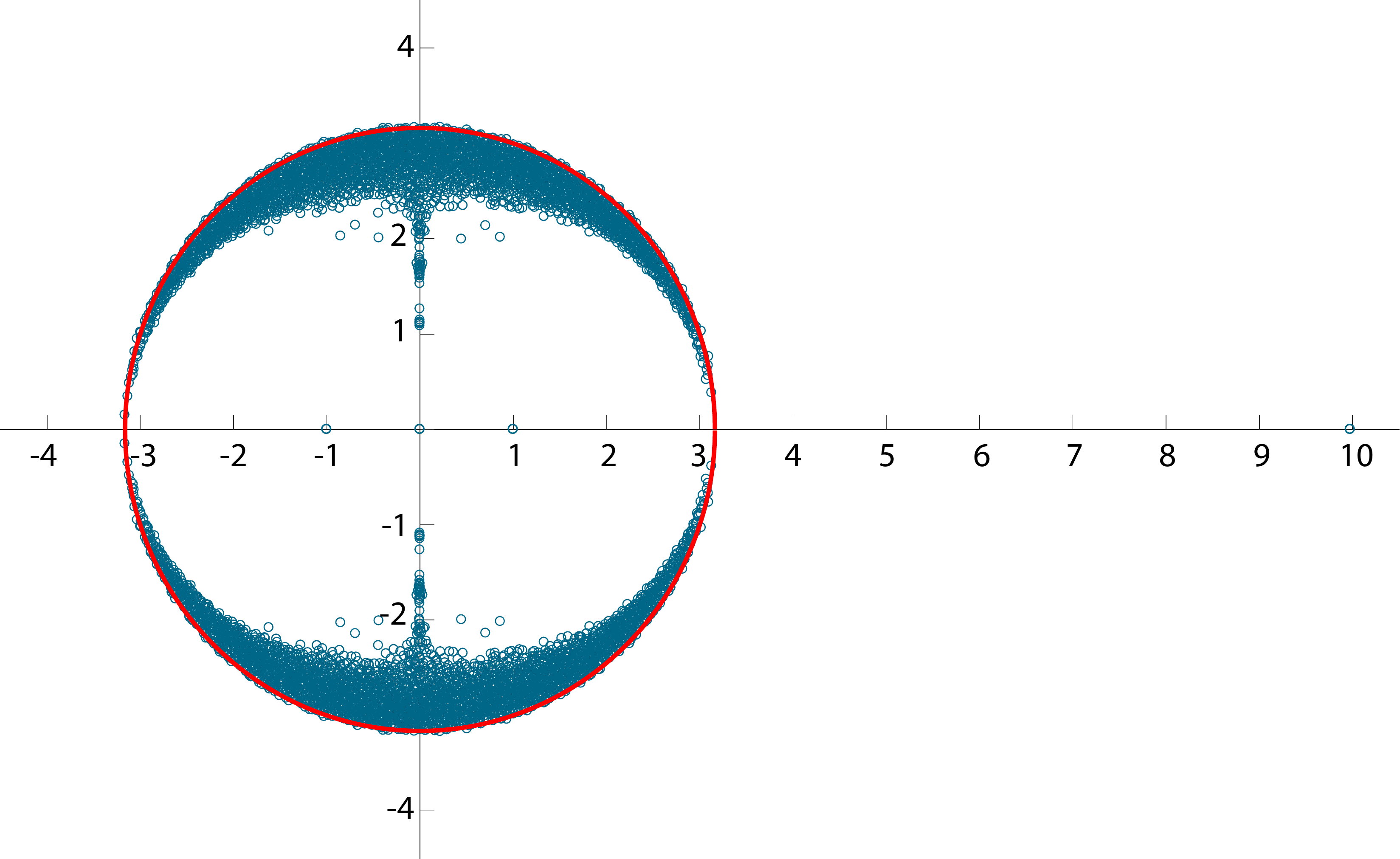} 
\caption{Spectrum of the Non-backtracking matrix of an Erd\"os and R\'enyi network with $n=3000$ and $\E{k} = 10$.} \label{fig:non-backtracking}
\end{center}
\end{figure}

The non-backtracking matrix is a $M \times M$ matrix defined in terms of  directed edges of the graph. Formally it is given as \cite{Krzakala2013}
\begin{equation}
 \B_{(i\rightarrow j, k\rightarrow l)} = \begin{cases}
            1 \hspace{1cm} \text{if $j = k$ and $i \neq l$}\\
            0 \hspace{1cm} \text{otherwise}.
           \end{cases}
\end{equation}

As described in \cite{Krzakala2013}, the spectrum of $\B$ is less sensitive to high degree nodes. Furthermore, trees dangling off or disconnected from the graph simply contribute to zero eigenvalues. Similarly, it can be observed that unicyclic components yield $0$, $1$ and $-1$ eigenvalues. Besides these particularities, $\B$'s bulk is confined to a disk in the complex plane, whose radius is $\sqrt{c}$ \cite{Krzakala2013}. Additionally, similarly to the adjacency matrix, using the Perron-Frobenius theorem we can guarantee that the non-backtracking matrix has a largest real eigenvalue and that the corresponding eigenvector can be chosen to have strictly positive components. In order to exemplify these concepts, in Figure \ref{fig:non-backtracking} we show the spectrum of the non-backtracking matrix for an Erd\"os and R\'enyi network with $n=3000$ and $\E{k} = 10$.

As a result of the previously mentioned properties, $\B$'s spectrum can be reduced (the important part) to the spectrum of \cite{Hashimoto1989, Bass1992, Angel2007, Krzakala2013}
\begin{equation}
\B' =
 \begin{pmatrix}
  0 & \D - \I \\
  -\I & \A
 \end{pmatrix},
\end{equation}
which is a $2N \times 2N$ matrix. This observation drastically reduces the computational complexity. Note that in dense networks $M \in \bigO{N^2}$. Moreover, we can re-write the eigenvalue problem in terms of the block matrices as
\begin{equation}
 \begin{pmatrix}
  0 & \D - \I \\
  -\I & \A
 \end{pmatrix}
 \begin{pmatrix}
  v_1 \\
  v_2
 \end{pmatrix}
 = \lambda
 \begin{pmatrix}
  v_1 \\
  v_2
 \end{pmatrix},
\end{equation}
yielding to the following quadratic eigenvalue problem,
\begin{equation}
 \left( \lambda^2 \I - \lambda \A + \D-\I \right) v_2 = 0.
\end{equation}
Thus, the characteristic polynomial is given as
\begin{equation}
 \text{det} \left[ \lambda^2 \I - \lambda \A + (\D-\I) \right] = 0,
\end{equation}
whose solutions are the eigenvalues of $\B'$ and, therefore, also $\B$. It accounts for the $2N$ of $\B$'s eigenvalues, while the other $2(M-N)$ are $\pm1$. This equation is well-known in the theory of graph zeta functions \cite{Hashimoto1989, Bass1992, Angel2007, Krzakala2013}.

\section{Multilayer network theory} \label{sec:multilayer_structure}

\begin{figure*}[!t]
\begin{center}
\includegraphics[width=\linewidth]{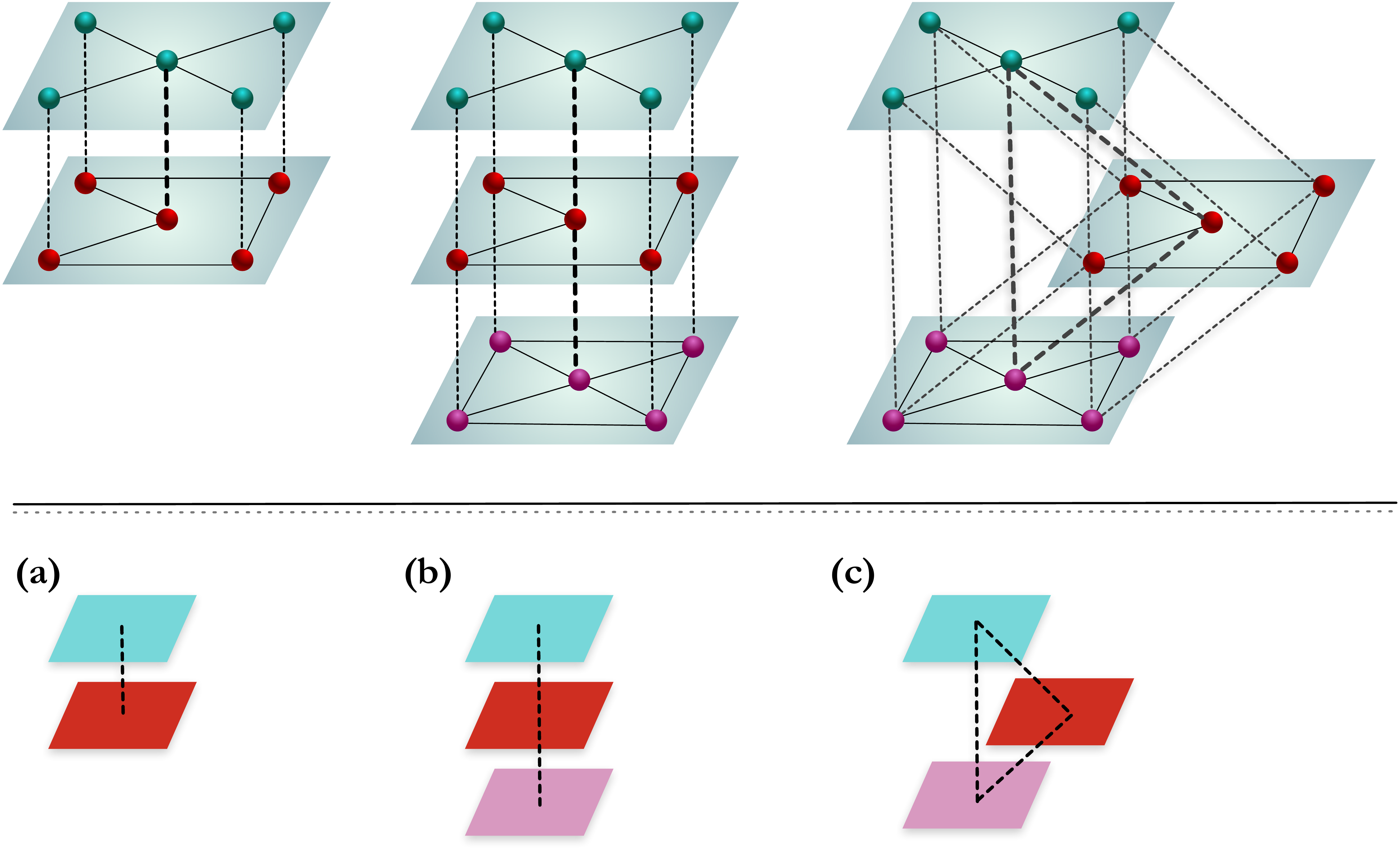}
\caption{Illustration of 3 multilayer networks. (a) and (c) represent 2 or 3-layer multiplex networks, while (b) represents an interconnected network.} \label{fig:schematic}
\end{center}
\end{figure*}

Multilayer networks are constituted by interacting layers, representing different kinds of contact. Consequently, layers comprise nodes and edges, modeling the interactions on the same kind. Here we focus on two specific cases, multiplex and  interconnected networks. In the multiplex case, the network of layers is a complete graph, i.e. each node has its counterpart in all layers. It is worth mentioning that one might also consider a multiplex with a different number of nodes in different layers. In the interconnected case, we can have a network of layers with irregular structure and also sparsity on the inter-layer edges. However, we focus here on the one-to-one case, where if there is a connection between two layers, all nodes on one layer have a counterpart on the other layer. Social networks represent a possible example of this system. The same user can have an account in different social networks, like Facebook and Twitter, which represent the layers. The connections between these layers are defined between the users that they share. We show an illustration of these multilayer networks in Figure \ref{fig:schematic}, where (a) and (c) represent 2 or 3-layer multiplex networks, while (b) represents an interconnected network. For more definitions, we refer the reader to \cite{Kivela2014}.

In this section, we focus on the basic concepts of multilayer representations. Firstly, in Section \ref{sec:mux_mat}, we focus on multiplex networks, defining them in terms of their supra-adjacency matrix. Note that the generalization to the multilayer case is natural, however, the symmetries introduced in the multiplex case allow us to extract some spectral properties of this system. Next, in Section \ref{sec:tensorial_notation}, we describe the tensorial notation, showing its projections and generalizing some concepts of the previous section to this formalism. It is important to emphasize that both formalisms are equally valid and choosing one over the other might depend on the application or even being a matter of taste. It is worth mentioning that the matrix notation provides a more straightforward notation for sparse coupling matrices as it allows a different number of nodes in each layer. This case, on a tensorial notation, implies in having many zeros in the tensor, which can be easily treated. On the other hand, the main advantage of the tensorial notation is the compactness of the equations and its projections, which appear naturally and also have physical meanings.

\subsection{Matrix representation} \label{sec:mux_mat}

In this section, we focus on the matrix representation of multiplex networks, which are probably the most natural extension of the single-layer theory to the multilayer formalism. In the following section, we formally define our mathematical objects, followed by the concepts of quotient graphs and coarse-grained versions of the original multiplex. Finally, we introduce the eigenvalue interlacing between the original multiplex network and its coarse-grained versions, which are the main results of this section. These concepts were formerly defined and shown in \cite{Cozzo2015}.

\subsubsection{Definitions}

\begin{figure*}
\includegraphics[width=0.96\textwidth]{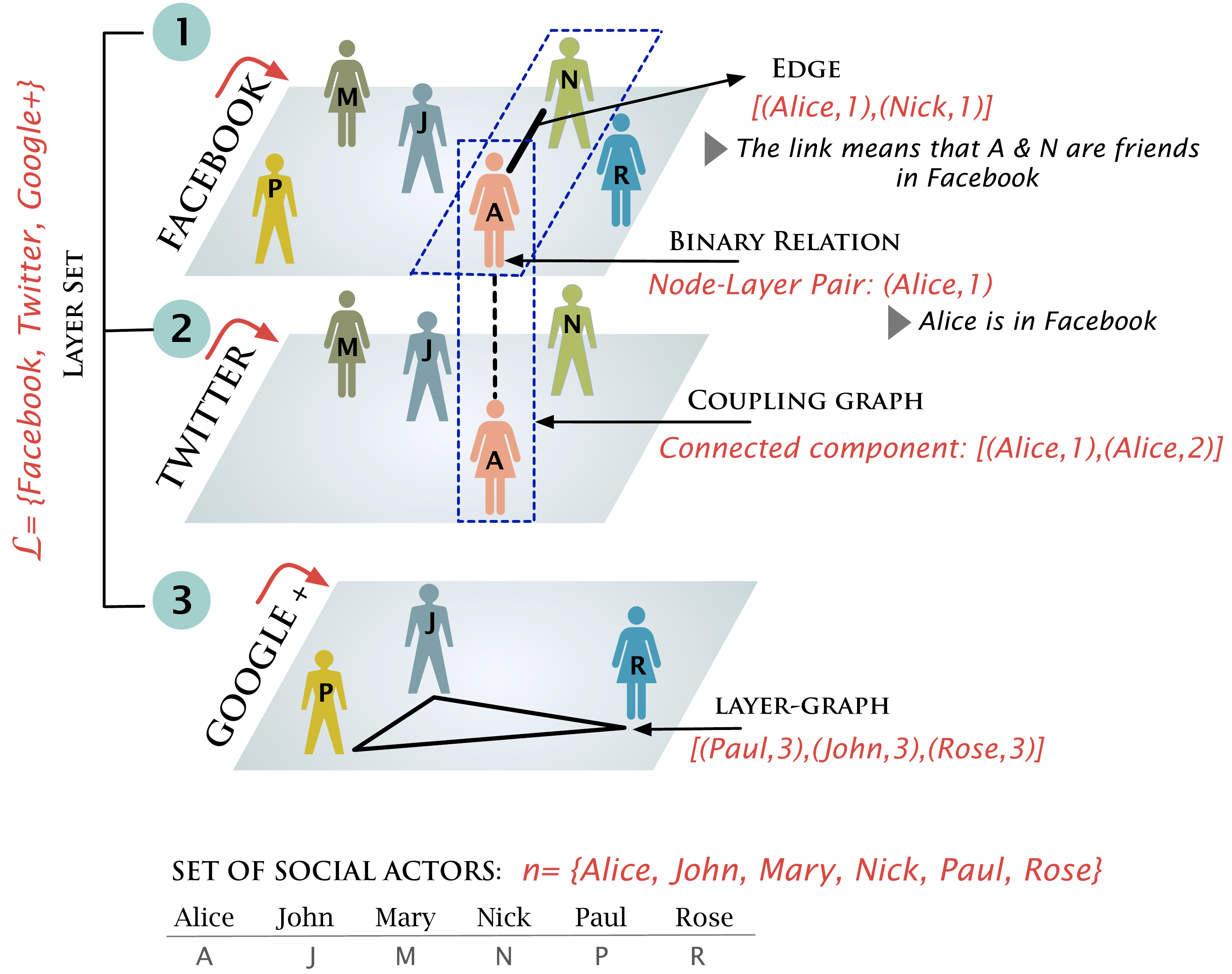}
\caption{Schematic representation of a multiplex network composed by on-line social networks (OSN) illustrating the definitions presented on the main text.} \label{fig:OSNs}
\end{figure*}

A multiplex network is a quadruple $\mathcal{M}=(\mathfrak{L},\mathfrak{n},\mathfrak{P},\mathfrak{M})$. $\mathfrak{L}=\{1,\dots,m\}$ is an index set that we call the layer set. Here we have assumed $\mathfrak{L}\subset\mathcal{N}$ for practical reasons and without loss of generality. We indicate the general element of $\mathfrak{L}$ with Greek lower case letters. Moreover, $\mathfrak{n}$ is a set of nodes and $\mathfrak{P}=(\mathfrak{n},\mathfrak{L},\mathfrak{N})$,  $\mathfrak{N}\subseteq\mathfrak{n}\times\mathfrak{L}$ is a binary relation. Finally, the statement $(n,\alpha)\in \mathfrak{N}$ is read \textit{node n participates in layer $\alpha$}. We call the ordered pair $(n,\alpha)\in\mathfrak{N}$ a node-layer pair, and we say that the node-layer pair $(n,\alpha)$ is the representative of node $n$ in layer $\alpha$.

On the other hand, $\mathfrak{M}=\{G_\alpha\}_{\alpha\in\mathfrak{L}}$ is a set of graphs, that we call layer-graphs, indexed by means of $\mathfrak{L}$. The node set of a layer-graph $G_\beta\in \mathfrak{M}$ is a sub-set $\mathfrak{n}_\beta\subset\mathfrak{N}$ such that $\mathfrak{n}_\beta=\{(n,\alpha)\in\mathfrak{P}\mid \alpha=\beta\}$, so the nodes of $G_\beta$ are node-layer pairs; therefore we say that node-layer pairs represent nodes in layers. The edge set of a graph $G_\alpha\in\mathfrak{M}$ is $\mathfrak{E}_\beta\subseteq \mathfrak{n}_\beta\times \mathfrak{n}_\beta$. Additionally, the binary relation $\mathfrak{P}$ can be identified with its graph $G_\mathfrak{P}$. $G_\mathfrak{P}$ has node set given by $\mathfrak{n}\cup \mathfrak{L}$, and edge set $\mathfrak{E}_\mathfrak{P}=\mathfrak{N}$, and we call it the \textit{participation graph}.

Consider the graph $G_\mathfrak{C}$ on $\mathfrak{N}$ in which there is an edge between two node-layer pairs $(n,\alpha)$ and $(m,\beta)$ only if $n=m$; that is, only if the two edges in the graph $G_\mathfrak{P}$ are incident on the same node $n\in\mathfrak{n}$, which means that the two node-layer pairs represent the same node in different layers. We call $G_\mathfrak{C}$ the coupling graph. It is easy to note that the coupling graph is composed of $n=\mid\mathfrak{n}\mid$ disconnected components that are cliques or isolated nodes. Each clique is formed by all the representatives of a node in the layers, we call the components of $G_\mathfrak{C}$ \textit{supra-nodes}.

Let us now also consider the graph $G_\mathfrak{l}$ on the same node set $\mathfrak{N}$, where there is an edge between two node-layer pairs $(n,\alpha)$, $(m,\beta)$ only if $\alpha=\beta$; that is, only if the two edges in graph $G_\mathfrak{P}$ are incident on the same node $\alpha\in\mathfrak{L}$. We call $G_\mathfrak{l}$ the layer graph. It can be observed that this graph is formed by $m=\mid\mathfrak{L}\mid$ disconnected components that are called cliques.

Finally, we can define the \textit{supra-graph} $G_\mathcal{M}$ as the union of the layer-graphs with the coupling graph: $G_\mathcal{C}\cup \mathfrak{M}$. $G_\mathcal{M}$ has node set $\mathfrak{N}$ and edge set $\bigcup_\alpha \mathfrak{E}_\alpha \cup \mathfrak{E}_\mathfrak{C}$. $G_\mathcal{M}$ is a synthetic representation of the Multiplex Network $\mathcal{M}$. It results that each layer-graph $G_\alpha$ is a sub-graph of $G_\mathcal{M}$ induced by $\mathfrak{n}_\alpha$. Furthermore, when all nodes participate in all layer-graphs, the Multiplex Network is said to be fully aligned \cite{Kivela2014} and the coupling graph is made of $n$ complete graphs of $m$ nodes.

It is useful to come back to our system of social agents as a paradigmatic multiplex network to make sense of the previous definitions. The layer set is the list of OSNs, for example, $\mathfrak{L}=\{Facebook, Twitter, Google+\}$. Since for practical purposes we want a set of indexes that are natural numbers, we may say that: Facebook is $1$, Twitter is $2$, and $Google+$ is 3. The set of nodes is the set of social actors, for example, $n=\{Marc, Alice, Bifi, Nick, Rose\}$. The binary relations represent the participation of each of these agents in some OSNs, thus we have that a statement of the type \textit{Alice has a Facebook account} is represented by the pair $(Alice,1)$, that is a node-layer pair. Each set of relationships in each OSN is represented by a graph, for example, the link $[(Alice,1),(Nick,1)]$ means that Alice and Nick are friends on Facebook. If Alice has a Facebook account and a Twitter account, but not a Google+ account, in the coupling graph we will have the connected component $[(Alice,1),(Alice,2)]$ that is the supra-node related to Alice. If only the Bifi, Nick, and Rose have Google+ accounts, in the layer graph we will have the connected component $[(Bifi,3),(Nick,3),(Rose,3)]$. See Fig. \ref{fig:OSNs} for a schematic representation of several of the definitions given above.

\subsubsection{The supra-adjacency and supra-Laplacian matrices} \label{sec:supradjacency}

The supra-adjacency matrix is the adjacency matrix of the supra-graph $G_\mathcal{M}$. Just as $G_\mathcal{M}$, $\A$ is a synthetic representation of the whole multiplex $\mathcal{M}$. By definition, it can be obtained from the intra-layer adjacency matrices and the coupling matrix in the following way:
\begin{equation}
 \A =\bigoplus_\alpha \mathbf{A}^\alpha + \mathcal{C},
\label{supradjacency}
\end{equation}  
where the same consideration as in $\mathcal{C}$ applies for the indices. We also define $\bar{\A}=\bigoplus \mathbf{A}^\alpha$, and we call it the intra-layer adjacency matrix. Furthermore, one might weight the inter and intra-layer interactions differently. Thus, Equation \ref{supradjacency} can be re-written as
\begin{equation}
\A = \bigoplus_{m=1}^M \A^{(m)} + p \Cuop, 
\end{equation}
where $p$ is a parameter that defines the strength of the inter-layer edges and $\Cuop$  is the interlayer coupling matrix, whose elements represent the relations between nodes in different layers, thus implicitly containing the information of a network of layers \cite{Garcia2014}. 

Similarly to single-layer networks, the supra-Laplacian is also defined by equation \ref{eq:laplacian_def}, which can also be written as
\begin{equation}
\Lap =\bigoplus_\alpha \Lap^{(\alpha)} + \Lap_C,
\label{formalism:laplacian2}
\end{equation}
where $\Lap^{(\alpha)}$ is the Laplacian matrix of the layer $\alpha$ and $\Lap_C$ is the contribution of the interlayer coupling matrix, $\Cuop$. For the sake of an example we present the block matrices of the supra-adjacency for the 2-Layer node-aligned multiplex case, which have the form
\begin{equation}
 \A = 
  \begin{bmatrix}
  \A^{(1)} & p\I \\
  p\I & \A^{(2)} 
 \end{bmatrix},
\end{equation} 
while its respective supra-Laplacian matrix follows
\begin{equation} 
 \Lap = \D - \A = 
  \begin{bmatrix}
  p\I + \Lap^{(1)} & -p\I \\
  -p\I & p\I + \Lap^{(2)}
 \end{bmatrix}.
\end{equation}

\subsubsection{Quotient graphs}

We next briefly introduce the notion of network quotient associated to a partition of the node set. Suppose that ${V_1, . . . ,V_m}$ is a partition of the node set of a network $G$ with adjacency matrix $A$, and write $n_i = |V_i|$. The quotient network $Q$ of $G$ is a coarse-grained representation of the network with respect to the partition. It has one node per cluster $V_i$ and an edge from $V_i$ to $V_j$ weighted by an average connectivity from $V_i$ to $V_j$
\begin{equation}
b_{ij}=\frac{1}{\sigma}\sum_{\substack{k\in V_i\\l\in V_j}}a_{kl}.
\label{quotientelement}
\end{equation}
Different choices are possible for the normalization parameter $\sigma$: $\sigma_i=n_i,\ \sigma_j=n_j$ or $\sigma_{ij}=\sqrt{n_in_j}$. Depending on the choice for $\sigma$ we call the resulting quotient respectively: left, right or symmetric quotient. We can express the left quotient $Q_l(A)$ in matrix form. Consider the $n\times m$ characteristic matrix of the partition $S=s_{ij}$, with $s_{ij}=1$ if $i\in V_j$ and zero otherwise. Then
\begin{equation}
Q_l(A)=\Lambda^{-1}S^TAS,
\label{quotientadjacency}
\end{equation} 
where $\Lambda=diag\{n_1,\dots,n_m\}$.

\subsubsection{Aggregate Network and Network of Layers of a Multiplex Network}

In the context of Multiplex Networks, two quotient graphs arise naturally \cite{Garcia2014, Cozzo2015} by considering coupled node-layer pairs and layers. Supra-nodes partition the supra-graph, and the supra-nodes characteristic matrix $S_n$ is the associated characteristic matrix. Then, we define the aggregate network of the multiplex network as the quotient associated to that partition:
\begin{equation}
\tilde{\mathbf{A}}=\Lambda^{-1}\mathcal{S}_n^T\bar{\mathcal{A}}\mathcal{S}_n,
\label{aggregate}
\end{equation}  
where $\Lambda=diag\{\kappa_1,\dots,\kappa_n\}$ is the multiplexity degree matrix.
Since, the Laplacian of the quotient is equal to the quotient of the Laplacian, the Laplacian of the aggregate network is given by:
\begin{equation}
\tilde{\mathbf{L}}=\Lambda^{-1}\mathcal{S}_n^T\bar{\mathcal{L}}\mathcal{S}_n.
\label{aggregateLaplacian}
\end{equation}
In the same way, layers partition the supra-graph, thus the network of layers is defined by
\begin{equation} \label{network_of_layers}
\tilde{\mathbf{A}}_\mathfrak{l}=\Lambda^{-1}\mathcal{S}_\mathfrak{l}^T\bar{\mathcal{A}}\mathcal{S}_\mathfrak{l},
\end{equation}
and its Laplacian is given by
\begin{equation}
\mathbf{\tilde{L}_\mathfrak{l}}=\Lambda^{-1}S_\mathfrak{l}^T\bar{L}S_\mathfrak{l}.
\label{nolLaplacian}
\end{equation}

\subsubsection{Spectral relations between \textit{supra} and coarse-grained representations}

The fundamental spectral result related to a quotient network is that adjacency eigenvalues of a quotient network interlace the adjacency eigenvalues of the parent network. That is, if $\mu_i,\dots,\mu_m$ are the adjacency eigenvalues of the quotient network, and $\lambda_i,\dots,\lambda_n$ are the adjacency eigenvalues of the parent network, the result is that
\begin{equation}
\lambda_i\leq\mu_i\leq\lambda_{i+n-m}.
\label{interlacing}
\end{equation}
The same result applies for Laplacian eigenvalues. We can derive directly from that result a list of bounds for the supra-adjacency and the supra-Laplacian in terms of the aggregate network and of the network of layers \cite{Garcia2014, Cozzo2015}. Besides, in the case of node aligned multiplex networks, we have that the eigenvalues of the Laplacian of the network of layers are a subset of the spectrum of the supra-Laplacian. This result is of special relevance in studying the structural properties of a multiplex network since it states that the adjacency (Laplacian) eigenvalues of the cross-grained representation of a multiplex interlace the adjacency (Laplacian) eigenvalues of the parent. In the case of a node-aligned multiplex, the Laplacian eigenvalues of the network of layers are a sub-set of the Laplacian eigenvalues of the parent multiplex network.

\subsection{Tensorial representation} \label{sec:tensorial_notation}

\subsubsection{Definitions} \label{sec:tensor_rep}

Here we use the representation formerly presented in \cite{DeDomenico2013}. We also adopt the Einstein summation convention, in order to have more compact equations: if two indices are repeated, where one is a superscript and the other a subscript, then this operation implies a summation. Moreover, the result is a tensor whose rank lowers by 2. For instance, $A^{\alpha}_{\beta} A_{\alpha}^{\gamma} = \sum_\alpha A^{\alpha}_{\beta} A_{\alpha}^{\gamma}$. In our notation, we use Greek letters to indicate the components of a tensor. In addition, we use a tilde ($\tilde{\cdotp}$) to denote the components related to the layers, with dimension $m$, while the components without a tilde present dimension $n$ and are related to the nodes.

A multilayer network is represented as the fourth-order adjacency tensor
$M \in \mathbb{R}^{n \times n \times m \times m}$, which can represent various relations among nodes \cite{DeDomenico2013}. Formally we have
\begin{equation}
 M_{\beta \tilde{\gamma}}^{\alpha \tilde{\delta}} = \sum_{\tilde{h},\tilde{k} = 1}^m C^{\alpha}_{\beta}(\tilde{h}\tilde{k}) E^{\tilde{\delta}}_{\tilde{\gamma}}(\tilde{h}\tilde{k}) = \sum_{\tilde{h},\tilde{k} = 1}^m \sum_{i,j = 1}^n w_{ij} (\tilde{h}\tilde{k}) \mathcal{E}_{\beta \tilde{\gamma}}^{\alpha \tilde{\delta}}(ij\tilde{h}\tilde{k}),
\end{equation}
where $E_{\tilde{\delta}}^{\tilde{\gamma}}(\tilde{h}\tilde{k}) \in
\mathbb{R}^{m \times m}$ and $\mathcal{E}_{\beta \tilde{\gamma}}^{\alpha \tilde{\delta}}(ij\tilde{h}\tilde{k}) \in
\mathbb{R}^{n \times n \times m \times m}$ indicate the tensor in its respective canonical basis. 

In addition to the tensor $M \in \mathbb{R}^{n \times n \times m \times m}$, we are usually interested in the weighted tensor, where the inter and intra layer edges have different weights. This tensor is denoted as
\begin{equation} \label{eq:adj_tensor}
\mathcal{R}_{\beta \tilde{\delta}}^{\alpha \tilde{\gamma}}(\lambda, \eta) = M_{\beta \tilde{\sigma}}^{\alpha \tilde{\eta}} E^{\tilde{\sigma}}_{\tilde{\eta}}(\tilde{\gamma} \tilde{\delta})  \delta^{\tilde{\gamma}}_{\tilde{\delta}} +
\frac{\eta}{\lambda} M_{\beta \tilde{\sigma}}^{\alpha \tilde{\eta}} E^{\tilde{\sigma}}_{\tilde{\eta}}(\tilde{\gamma} \tilde{\delta}) (U^{\tilde{\gamma}}_{\tilde{\delta}} - \delta^{\tilde{\gamma}}_{\tilde{\delta}})
\end{equation}
and it is called supra contact tensor, whose name comes from its definition on epidemic spreading \cite{Arruda2017}. Note that the intra-layer edges are weighted by $\eta$, while the inter-layer edges are weighted by $\lambda$. Furthermore, since a scalar does not change the spectral properties of our tensor, we divide it by $\lambda$, which leave us with just one parameter, the so-called coupling parameter, $\frac{\eta}{\lambda}$.

\subsubsection{Projections} \label{sec:tensor_projection}

One of the advantages of the tensorial representation is the possibility to define different projections, which allow us to have very compact equations. In the context of multilayer networks, these projections often present a physical meaning, allowing us to characterize different levels of the system. First of all, observe that we can extract one layer by projecting the tensor $M_{\beta \tilde{\gamma}}^{\alpha \tilde{\delta}}$
to the canonical tensor
$E_{\tilde{\delta}}^{\tilde{\gamma}}(\tilde{r}\tilde{r})$. Formally,
from \cite{DeDomenico2013} we have
\begin{equation} \label{eq:extraction}
 M_{\beta \tilde{\gamma}}^{\alpha \tilde{\delta}} E_{\tilde{\delta}}^{\tilde{\gamma}}(\tilde{r}\tilde{r}) = C^{\alpha}_{\beta}(\tilde{r}\tilde{r}) = A^{\alpha}_{\beta}(\tilde{r}),
\end{equation}
where $\tilde{r} \in \{ 1, 2, ..., m \}$ is the selected layer and $A^{\alpha}_{\beta}(\tilde{r})$ is the adjacency matrix (rank-2 tensor). Moreover, aiming at having more compact and clear equations, we define the all-one tensors $u_\alpha \in \mathbb{R}^n$ and $U^{\beta \tilde{\delta}} \in \mathbb{R}^{n \times m}$. Here, we restrict our analysis to multilayer networks with a diagonal coupling \cite{Kivela2014}. In other words, each node can have at most one counterpart on the other layers. In addition, for the sake of simplicity, we focus on unweighted and undirected connected networks, in which there is a path from each node to all other nodes.

Besides the adjacency tensor presented above, the network of layers \cite{Garcia2014} also characterizes the topology of the system. In this reduced network representation, each node represents one layer and the edges between them codify the number of edges connecting those two layers. Formally we have,
\begin{equation} \label{eq:net_layers}
  \Psi_{\tilde{\delta}}^{\tilde{\gamma}} = M_{\beta \tilde{\delta}}^{\alpha \tilde{\gamma}} U^{\beta}_{\alpha},
\end{equation}
where $\Psi_{\tilde{\delta}}^{\tilde{\gamma}} \in \mathbb{R}^{m \times m}$. Note that this network presents self-loops, which are weighted by the number of edges on the layer. Additionally, since we assume that the layers have the same number of nodes, the edges of the network of layers have weights equal to the number of nodes $n$.

Another important reduction of the multilayer network is the so-called projection \cite{DeDomenico2013}. This network aggregates all the information into one layer, including self-loops that stand for the number of layers in which a node appears. Mathematically, we have
\begin{equation} \label{eq:agregated}
 P_\beta^\alpha = M_{\beta \tilde{\gamma}}^{\alpha \tilde{\delta}} U_{\tilde{\delta}}^{\tilde{\gamma}},
\end{equation}
where $P_\beta^\alpha \in \mathbb{R}^{n \times n}$. Complementary, a version of the projection without self-edges is called the overlay network and is given as the contraction over the layers \cite{DeDomenico2013}, i.e.,
\begin{equation} \label{eq:overlay}
O_\beta^\alpha = M_{\beta \tilde{\gamma}}^{\alpha \tilde{\gamma}}.
\end{equation}
Observe that the overlay network does not consider the contribution of the interlayer connections, whereas the projection does. Comparisons between the assortativity of these two different representations of the system reveal the key role of such inter-links (for more details, please refer to \cite{deArruda2016}).

\subsubsection{Eigentensor}

Here we describe the eigenvalue problem considering the tensorial representation. This eigenvalue problem can be generalized to the case of a rank-4 tensor leading to
\begin{equation}
 \mathcal{R}_{\beta \tilde{\delta}}^{\alpha \tilde{\gamma}}  f_{\alpha \tilde{\gamma}}(\Lambda) = \Lambda f_{\beta \tilde{\delta}}(\Lambda),
\end{equation}
where $\Lambda$ is an eigenvalue and $f_{\beta \tilde{\delta}}(\Lambda)$ is the corresponding eigentensor. In addition, we are assuming that the eigentensors form an orthonormal basis. Importantly, the supra-contact matrix, $R$, in \cite{Cozzo:2013} can be understood as a flattened version of the tensor $\mathcal{R}_{\beta \tilde{\delta}}^{\alpha \tilde{\gamma}}(\lambda, \eta)$. Consequently, all the results for $R$ also apply to the tensor $\mathcal{R}$. As argued in \cite{DeDomenico2013}, this supra-adjacency matrix corresponds to unique unfolding of the fourth-order tensor $M$ yielding square matrices. Following this unique mapping, we have the correspondence of the eigensystems.

\subsubsection{Interlacing properties} \label{sec:Spectra_interlacing}

\begin{table}[t] 
\begin{center}
\caption{Structure and spectra of the normalized network of layers $\Phi_{\tilde{\delta}}^{\tilde{\gamma}}(\lambda, \eta)$. The eigenvalues assume that the average degree of each layer, $\E{ k^{l} }$, is the same, i.e. $\E{ k^{l}  } = \E{ k }, \forall l$.}
\begin{tabular}{c|c|c}
\hline
Network & $\Phi_{\tilde{\delta}}^{\tilde{\gamma}}(\lambda, \eta)$ & Eigenvalues \\
\hline
\multirow{3}{*}{Line with 2 nodes} &

\multirow{3}{*}{$\begin{bmatrix}
 \E{ k^{l = 1} } & \frac{\eta}{\lambda} \\
 \frac{\eta}{\lambda} & \E{ k^{l = 2} } 
\end{bmatrix}$} & $\E{ k } - \frac{\eta}{\lambda}$\\
    & & $\E{ k } + \frac{\eta}{\lambda}$ \\
    & & \\
\hline
\multirow{4}{*}{Line with 3 nodes} &

\multirow{4}{*}{$\begin{bmatrix}
 \E{ k^{l = 1} } & \frac{\eta}{\lambda}  & 0 \\
 \frac{\eta}{\lambda} & \E{ k^{l = 2} }  & \frac{\eta}{\lambda} \\
 0 & \frac{\eta}{\lambda} & \E{ k^{l = 3} }
\end{bmatrix}$} & $\E{ k }$\\
 & & $\E{ k } - \sqrt{2} \frac{\eta}{\lambda}$ \\
 & & $\E{ k } + \sqrt{2} \frac{\eta}{\lambda}$ \\
 & & \\
\hline
\multirow{4}{*}{Multiplex} &

\multirow{4}{*}{$\begin{bmatrix}
 \E{ k^{l = 1} } & \frac{\eta}{\lambda}  & \frac{\eta}{\lambda} \\
 \frac{\eta}{\lambda} & \E{ k^{l = 2} }  & \frac{\eta}{\lambda} \\
 \frac{\eta}{\lambda} & \frac{\eta}{\lambda} & \E{ k^{l = 3} }
\end{bmatrix}$} &  $\E{ k } - \frac{\eta}{\lambda}$\\
& & $\E{ k } - \frac{\eta}{\lambda}$\\
& & $\E{ k } + 2 \frac{\eta}{\lambda}$\\
& & \\
\hline
\end{tabular}
\label{tab:spec_net_net}
\end{center} 
\end{table}

Invoking the unique mapping presented in the previous sub-section and considering the results of \cite{Garcia2014,Cozzo2015}, we can use the interlacing properties to relate the spectra of the multilayer network with the spectra of the network of layers. We define the normalized network of layers (for more see Section \ref{sec:tensor_projection}) in terms of the supra contact tensor as 
\begin{equation} \label{eq:Net_net}
 \Phi_{\tilde{\delta}}^{\tilde{\gamma}}(\lambda, \eta) = \frac{1}{n} \mathcal{R}_{\beta \tilde{\delta}}^{\alpha \tilde{\gamma}}(\lambda, \eta) U^{\beta}_{\alpha},
\end{equation}
where we are implicitly assuming a multilayer network in which the layers have the same number of nodes and a dependency on the coupling parameters (the demonstration that this tensor is an unfolding of the matrix in \cite{Garcia2014} is shown in \cite{Arruda2017}). Additionally, let us denote by $\mu_1 \geq \mu_2 \geq ... \geq \mu_{m}$ the ordered eigenvalues of $\Phi_{\tilde{\delta}}^{\tilde{\gamma}}(\lambda, \eta)$. Following \cite{Garcia2014}, the interlacing properties imply 
\begin{equation}
 \Lambda_{nm-m+j} \leq \mu_{j} \leq \Lambda_{j},
\end{equation}
for $j = m, ..., 1$. As examples, Table \ref{tab:spec_net_net} shows the spectrum of three simple networks of layers that can be computed analytically: a line with two and three nodes and a triangle. Figure \ref{fig:schematic} shows a schematic illustration of these 3 multilayer networks. 

Furthermore, using similar arguments we can also obtain results for the normalized projection, formally given as
\begin{equation} \label{eq:norm_agregated}
 \mathbf{P}_\beta^\alpha = \frac{1}{m} \mathcal{R}_{\beta \tilde{\delta}}^{\alpha \tilde{\gamma}}(\lambda, \eta) U^{\tilde{\delta}}_{\tilde{\gamma}},
\end{equation}
whose ordered eigenvalues, denoted by $\nu_1 \geq \nu_2 \geq ... \geq \nu_{m}$, also interlace with the supra contact tensor satisfying
\begin{equation}
 \Lambda_{nm-n+j} \leq \nu_{j} \leq \Lambda_{j},
\end{equation}
for $j = n, ..., 1$. Finally, the adjacency tensor of an extracted layer also interlaces, yielding
\begin{equation}
 \Lambda_{nm-n+j} \leq \Lambda_{j}^l \leq \Lambda_{j},
\end{equation}
for $j = n, ..., 1$. These results show that the eigenvalue of the multilayer adjacency tensor is always larger than or equal to all the eigenvalues of the individual isolated layers as well as the network of layers.

Once we have briefly summarized the basic representation and metrics of both single-layer and multilayer networks, we are ready to move to the main focus of this review: the theoretical and numerical characterization of spreading processes.

\section{Epidemic spreading} \label{sec:epidemic}

Let us denote by $X$ those individuals that are susceptible and can be infected, by $Y$ those individuals that are infected and can spread the disease to their neighbors and, finally, by $Z$ the recovered or removed individuals, which do not take part on the spreading process. Here we define our local rules using the following notation, $Y + X  \xrightarrow{\lambda}  Y + Y$, where two individuals in states $X$ and $Y$ perform a contact with a rate (or probability, in the discrete time case) $\lambda$, turning $X$ into the state $Y$. We must emphasize that our notion of time is still not yet defined, therefore while defining our set of local rules we do not distinguish between rates and probabilities. This distinction is discussed throughout this review, especially in the mean field approximations, considering the continuous time approaches with the cellular automata formalism, in Sections \ref{sec:mfa} and \ref{sec:comp}. The first local rule presented here comprises the \emph{Susceptible-Infected} (SI) model. It is modeled as \cite{AnderssonBritton2000, Barrat08:book, Satorras015}
\begin{equation*}
\begin{array}{rcl}
Y + X & \xrightarrow{\lambda} & Y + Y,\\[.2cm]
\end{array}
\end{equation*}
where the disease is spread through the contact from an infected to a susceptible individual. Note that this process presents a single absorbing state\footnote{An absorbing state is defined as a state from which is impossible to make a transition to other states. Formally, consider the state $i$, then $\Prob(X_{t+1}=i | X_{t}=i) = p_{ii} = 1$ and $\Prob(X_{t+1}=j | X_{t}=i) = p_{ij} = 0$ $\forall j$.}. For a non-zero initial condition, this rule will always drive the system to the all infected state, where every connected individual of the population is infected. 

Although several diseases could be modeled using this rule, most diseases present a recovering mechanism. One way to model the latter is to consider the \emph{Susceptible-Infected-Susceptible} (SIS) model, which is given by the following set of local rules \cite{AnderssonBritton2000, Barrat08:book, Satorras015}
\begin{equation*}
\begin{array}{rcl}
 Y + X & \xrightarrow{\lambda} & Y + Y,\\[.2cm]
 Y &\xrightarrow{\delta} &X,\\[.2cm]
\end{array}
\end{equation*}
where the spreading mechanism is the same as the SI model. Note that infections do not consider immunization upon recovery from infection, and individuals become susceptible again. Moreover, regarding a finite population, the only absorbing state is the inactive state, where every individual in the population is healthy (state $X$). As discussed in section \ref{sec:exact}, this model also presents a meta-state \footnote{In our context, a meta-state is configured when there is a set of micro-states which has a finite probability larger than zero and the system remains ``traped'' in this set. In other words, there are many different configurations for which the fraction of infected nodes is the same and the system fluctuates around this average value. In the Markov sense the average value is not an absorbing state, but the dynamics fluctuates around this average value.}, in which the disease can be sustained for long periods of time. Furthermore, in the thermodynamic limit, infinite system size, the system stays trapped in this state. The SIS dynamics captures some of the most important features of real diseases while being simple enough. 

Whilst the SIS model could represent many diseases, there are many others where the individual acquires permanent immunity after being cured. For such cases we can invoke the \emph{Susceptible-Infected-Recovered} (SIR) model. In this scenario, the local rule that accounts for the recovering must be changed \cite{AnderssonBritton2000, Barrat08:book, Satorras015}. Formally, we have the SIR set of local rules defined as
\begin{equation*}
\begin{array}{rcl}
 Y + X & \xrightarrow{\lambda} & Y + Y,\\[.2cm]
 Y &\xrightarrow{\delta} &Z,\\[.2cm]
\end{array}
\end{equation*}
where the last rule imples that recovered individuals are removed from the dynamics. In contrast to the SIS, here we have infinitely many absorbing states in the thermodynamic limit.

The last disease model presented here is the \emph{Susceptible-Infected-Recovered-Susceptible} (SIRS) model, defined by the following set of local rules \cite{AnderssonBritton2000, Barrat08:book, Satorras015}
\begin{equation*}
\begin{array}{rcl}
 Y + X & \xrightarrow{\lambda} & Y + Y,\\[.2cm]
 Y &\xrightarrow{\delta} &Z,\\[.2cm]
 Z &\xrightarrow{\gamma} &X,
\end{array}
\end{equation*}
where we have two spontaneous transitions: one from $Y$ to $Z$ and from $Z$ to $X$. Note that this model also presents just one absorbing state for finite populations, similarly to the SIS dynamics. However, it has not been explored in the network literature as intensively as the SIS and SIR models due to its higher mathematical complexity.

In order to formally describe the evolution of an epidemic spreading process on networks, we construct a Markov chain $(\xi_t)_{t\geq 0}$ with state space\footnote{Note that for the SIS dynamics we only need one random variable, however, here we present the most general case.} $\mathcal{S}=\{(1,0,0),(0,1,0),(0,0,1)\}^{[N]}$, where $[N]$ is the set of nodes. More precisely, we define $\xi_t :=\{(X_i(t),Y_i(t),Z_i(t)): i\in [N]\}$, where $X_i(t)$, $Y_i(t)$ and $Z_i(t)$ are Bernoulli random variables indicating whether the node $i\in[N]$ is susceptible, infected, or recovered at time $t$, respectively. Therefore, $(1,0,0),(0,1,0)$ and $(0,0,1)$ represent the states $X$, $Y$ and $Z$, respectively. Each point $\xi \in \mathcal{S}$ is called a configuration. To construct the Markov chain, we consider random objects defined in the same suitable probability space $(\Omega, \mathcal{F}, \mathbb{P})$, where $\Omega$ is the sample space,  $\mathcal{F}$ is a $\sigma$-algebra of subsets of $\Omega$, i.e., the set of configurations, and $\mathbb{P}$ is a probability measure function. 

Furthermore, observe that, regarding an epidemic spreading process in continuous time, it is modeled as a collection of independent Poisson processes with rates $\lambda$ and $\delta$. Thus, one might rescale time, obtaining the effective spreading rate $\tau = \frac{\lambda}{\delta}$, which is a suitable control parameter for our system, simplifying our analysis. However, in the discrete time case, $\lambda$ and $\delta$ are probabilities and this rescaling is not always possible, as can be seen in the analysis of the cellular automata case, Section \ref{sec:dtmc}.

\section{Rumor spreading}

Along the same lines as in the previous section, here we initially present the set of local rules, then we perform a formal description of our model. The two most common rumor spreading models were introduced by Daley and Kendall (DK)~\cite{daleykendall1964} and Maki and Thompson (MT)~\cite{makithompson1973}. Let us denote by $Y$ the ignorants, i.e., individuals who are not aware of a rumor/information, by $X$ the spreaders, who know the rumor and are willing to spread it and, finally, $Z$ the stiflers, individuals who are aware of the rumor/information, but are unwilling to spread it (they lost interest in it). Chronologically, beginning with the DK model \cite{daleykendall1964, daleykendall1965}, their rules are defined as 
\begin{equation*}
\begin{array}{rcl}
 Y + X & \xrightarrow{\lambda} & Y + Y,\\[.2cm]
 Y + Y &\xrightarrow{\alpha} &Z + Z,\\[.2cm]
 Y + Z &\xrightarrow{\alpha} &Z + Z,\\[.2cm]
\end{array}
\end{equation*}
where both, the spreading and annihilation mechanisms are based on the contact, which contrasts with the epidemic spreading models previously described. Besides, observe that such mechanism is undirected, since on the contact of two spreaders both turn into stiflers. In fact this is the difference between the DK model and the Maki--Thompson (MT) model \cite{makithompson1973}. In the latter, the contacts are assumed to be directed, yielding to
\begin{equation*}
\begin{array}{rcl}
 Y + X & \xrightarrow{\lambda} & Y + Y,\\[.2cm]
 Y + Y &\xrightarrow{\alpha} &Z + Y,\\[.2cm]
 Y + Z &\xrightarrow{\alpha} &Z + Z.\\[.2cm]
\end{array}
\end{equation*}
In \cite{Lebensztayn2011}, the authors re-parametrized the traditional rumor models of Daley and Kendal \cite{daleykendall1964, daleykendall1965} and of Maki--Thompson \cite{makithompson1973} and considered a new class of individuals, called uninterested, studying the final state and variance of the final fraction of the stiflers on finite populations. 

Furthermore, Gonz\'alez-Bail\'on et al.~\cite{Gonzalez2011} analyzed the dynamics of protest recruitment using Twitter data from the mobilizations in Spain during 2011. They showed the existence of influential spreaders, similarly to what was found in the SIR epidemic spreading model \cite{Kitsak10:NP}. On the other hand, in \cite{Borge12} the authors found evidence of  an absence of influential spreaders in complex networks. These findings motivated Borge-Holthoefer et al. to propose two different models presenting this feature \cite{Borge-Holthoefer2012}. The first considers an activity parameter, while the second aims to model the apathy, in which a node can receive the information, but is not interested in spreading it. Thus, an individual can be a stifler without passing through the spreader state. This model considers that after a contact, an ignorant turns into a spreader with a rate/probability $\lambda \eta$, while it can turn into a stifler directly with rate/probability $\lambda (1 - \eta)$. Formally, their model's set of local rules are given as
\begin{equation*}
\begin{array}{rcl}
 Y + X & \xrightarrow{\lambda \eta} & Y + Y,\\[.2cm]
 Y + X & \xrightarrow{\lambda (1-\eta)} &Y + Z,\\[.2cm]
 Y + Y &\xrightarrow{\alpha} &Z + Y,\\[.2cm]
 Y + Z &\xrightarrow{\alpha} &Z + Z.\\[.2cm]
\end{array}
\end{equation*}

In a similar way as done for the epidemic spreading in the previous section, we can also describe the evolution of this process constructing a Markov chain $(\xi_t)_{t\geq 0}$ with the same state space $\mathcal{S}=\{(1,0,0),(0,1,0),(0,0,1)\}^{[N]}$ as previously defined. Furthermore, to construct the Markov chain, we consider random objects defined in the same suitable probability space $(\Omega, \mathcal{F}, \mathbb{P})$, where $\Omega$ is the sample space,  $\mathcal{F}$ is a $\sigma$-algebra of subsets of $\Omega$, i.e., the set of configurations, and $\mathbb{P}$ is a probability measure function. Regarding time, we should mention that most of the approaches in the literature consider continuous time approaches \cite{daleykendall1964, daleykendall1965, makithompson1973, Lebensztayn2011, Borge-Holthoefer2012}. In other words, it implies that they are modeled with a collection of independent Poisson processes with rates $\lambda$ and $\alpha$. 

\begin{table*}[!th]
\centering
\begin{adjustbox}{max width=0.95\textwidth}
\begin{threeparttable}[b]
\caption{A brief literature review: A summary of previous epidemic and rumor spreading models. The states are (i) susceptible or ignorant ($X$), (ii) infected or spreader ($Y$) and (iii) recovered or stifler ($Z$).}
\begin{tabular}{c|l|c|c|c|c}
\hline
& Model & Interactions & Threshold & Networks & Comments \\
\hline
\hline
\multirow{9}{*}{\vspace{-1cm}\hspace{0.2cm}\begin{rotate}{90}
\hbox{Epidemic}
\end{rotate}\hspace{0.2cm}} 
& SI~\tnote{1}   & $Y+X \xrightarrow{\lambda} Y+Y$ & \hspace{.7cm} -- \hspace{.7cm} & Yes & Only two fixed points.\\
\cline{2-6}
& \multirow{2}{*}{SIR}  & $Y+X \xrightarrow{\lambda} Y+Y$ & \multirow{2}{*}{$\lambda > \frac{\delta}{\Lambda_{max}}$}~\tnote{2} & \multirow{2}{*}{Yes} & Absorbing state, $Z$;\\
&      & $Y \xrightarrow{\delta} Z$      & &  & Presence of influential spreaders. \\
\cline{2-6}
& \multirow{2}{*}{SIS}  & $Y+X \xrightarrow{\lambda} Y+Y$ & \multirow{2}{*}{$\lambda > \frac{\delta}{\Lambda_{max}}$}\tnote{3} & \multirow{2}{*}{Yes} & Presents an active steady state;  \\
&      & $Y \xrightarrow{\delta} X$      & &  & Discrete and continuous time. \\
\cline{2-6}
& \multirow{3}{*}{SIRS} & $Y+X \xrightarrow{\lambda} Y+Y$ & \multirow{3}{*}{$\lambda > \delta \frac{\E{ k }}{\E{ k^2 }}$} & \multirow{3}{*}{Yes} & Presents an active steady state; \\
&              & $Y \xrightarrow{\delta} Z$    & & & Short-term immunity. \\
&              & $Z \xrightarrow{\gamma} X$    & & &  \\
\hline
\hline
\multirow{23}{*}{\vspace{-1cm}\hspace{0.2cm}\begin{rotate}{90}
\hbox{Rumor}
\end{rotate}\hspace{0.2cm}}
& \multirow{3}{*}{Daley--Kendall} & $Y+X \xrightarrow{\lambda} Y+Y$ & \multirow{3}{*}{$\frac{\lambda}{\alpha} > 0$} & \multirow{3}{*}{Yes} & Absorbing state, $R$; \\
&                 & $Y+Y \xrightarrow{\alpha} Z+Z$  & &               & Undirected contact; \\
&                 & $Y+Z \xrightarrow{\alpha} Z+Z$  & &               &  \\
\cline{2-6}
& \multirow{3}{*}{Maki--Thompson} & $Y+X \xrightarrow{\lambda} Y+Y$ & \multirow{3}{*}{$\frac{\lambda}{\alpha} > 0$} & \multirow{3}{*}{Yes} & Absorbing state, $Z$; \\
&                 & $Y+Y \xrightarrow{\alpha} Z+Y$  & &               & Directed contact.  \\
&                 & $Y+Z \xrightarrow{\alpha} Z+Z$  & &                 \\
\cline{2-6}
& \multirow{4}{*}{Nekovee et al.} & $Y+X \xrightarrow{\lambda} Y+Y$ & \multirow{4}{*}{$\frac{\lambda}{\delta} \geq \frac{\E{ k }}{\E{ k^2 }}$} & \multirow{4}{*}{Yes} & Absorbing state, $Z$; \\
&                 & $Y+Y \xrightarrow{\alpha} Z+Y$  & &               & Presents a   \\
&                 & $Y+Z \xrightarrow{\alpha} Z+Z$  & &               & ``lost of interest'' mechanism.  \\
&                 & $Y \xrightarrow{\delta} Z$  & &               &  \\
\cline{2-6}
& \multirow{4}{*}{Borge et al.~\tnote{4}} \hspace{0.1cm} & $Y+X \xrightarrow{\lambda \eta} Y+Y$ & \multirow{4}{*}{--} & \multirow{4}{*}{Yes} & Absorbing state, $Z$;\\
&                 & $Y+X \xrightarrow{\lambda (1-\eta)} Z+Y$  & &           & Presents an  \\
&                 & $Y+Y \xrightarrow{\alpha} Z+Z$  & &               & apathy mechanism;  \\
&                 & $Y+Z \xrightarrow{\alpha} Z+Z$  & &               & Models activity.  \\
\cline{2-6}
& \multirow{9}{*}{Kawachi et al.~\tnote{5}} & $Y+X \xrightarrow{\alpha_{xy} \theta_{xy}} Y+Y$ & \multirow{9}{*}{--} & \multirow{9}{*}{No}   \\
& & $Y+X \xrightarrow{\alpha_{xy} (1 - \theta_{xy})} Y+Z$ & & &  \\

& & $Z+X \xrightarrow{\alpha_{xz} \theta_{xz}} Z+Y$ & & &  \\
& & $Z+X \xrightarrow{\alpha_{xz} (1 - \theta_{xz})} Z+Z$ & & & Presents an active steady state.  \\

& & $Y+Y \xrightarrow{\beta} Y+Z$ & & &   \\

& & $Z+Y \xrightarrow{\gamma} Z+Z$ & & &  \\

& & $Z+Z \xrightarrow{\lambda p} Z+X$& & &  \\

& & $Y \xrightarrow{\eta_y} X$& & &  \\
& & $Z \xrightarrow{\eta_z} X$& & &  \\
\hline
\end{tabular}
\label{tab:literature}
\begin{tablenotes}
\item[1]{This models is similar to the Eden model \cite{eden1961}.}
\item[2]{This expression is obtained in detail in \cite{Newman010:book}, assuming uncorrelated networks. In \cite{Barrat08:book}, the authors show another compartmental based approach, which yields $\frac{\lambda}{\delta} > \frac{\E{ k }}{\E{ k^2 } - \E{ k }}$. The first follows a quenched mean field (QMF) approach, where the process takes place in a fixed network, while the second expression is obtained considering the heterogeneous mean field (HMF) approach, where we assume that every node with the same degree is statistically equivalent.}
\item[3]{This expression is obtained in detail in \cite{Newman010:book}, assuming uncorrelated networks. In \cite{Barrat08:book} the authors show that $\frac{\lambda}{\delta} > \frac{\E{ k }}{\E{ k^2 }}$, that is completely analogous to the expression in the table, since $\Lambda_{max} = \frac{\E{ k^2 }}{\E{ k }}$ for a random network generated by the configuration model \cite{ChungPNAS2003} considering scale-free networks in which $P(k) \sim k^{-\zeta}$ and $2 < \zeta < \frac{2}{5}$. Observe that these expressions are valid in finite networks and the threshold tends to zero in the thermodynamic limit, since $\E{ k^2 } \rightarrow \infty$.}
\item[4]{Considering the Model II in \cite{Borge-Holthoefer2012}, which takes into account the apathy of the individuals.}
\item[5]{The authors considered even more interactions, but did not evaluate most of the possibilities numerically. In addition, the transitions follow the notation used in \cite{Kawachi200855} and are rates, not necessarily probabilities, as the authors follow a continuous time approach.}
\end{tablenotes}
\end{threeparttable}
\end{adjustbox}
\end{table*}

For the sake of generality, and given that we shall aim to develop a general model for spreading processes, no matter if it is a disease or a rumor, we consider throughout this review the following states: (i) susceptible or ignorant, which we denote by state $X$, (ii) infected or spreader, which is represented by state $Y$; and (iii) recovered or stifler, corresponding to state $Z$. The reader is also referred to Table \ref{tab:literature}, where we summarize the main rumor and epidemic models studied so far. The transitions between states in each model and the thresholds, when available, are also shown for completeness.

\section{Exact SIS Markov chain} \label{sec:exact}

The exact SIS Markov chain model is formulated in Section \ref{sec:epidemic}. Here we also assume a Markovian process given as a collection of independent Poisson processes. In other words, to each directed edge emanating from an infected individual, $i$, we associate a Poisson process with parameter $\lambda$, $N^\lambda_{i \rightarrow j}$, modeling the spread of the disease. Additionally, in order to model the recovery of infected individuals, we associate a Poisson process with parameter $\delta$, $N^\delta_i$, to each node. We also consider that our population interacts in a network represented by the the adjacency matrix $\A$. For simplicity, throughout the review we consider an undirected and connected\footnote{A connected network is a network where there is a path from each node $i$ to each node $j$ in the network. In other words, it has only one component.} network. Based on this simple set of assumptions, the time evolution of the probability that node $i$ is infected, $\E{Y_i}$, is given by
\begin{equation} \label{eq:sis_exact}
 \frac{d\E{Y_i}}{dt} = \E {-\delta Y_i + \lambda \sum_{k=1}^N \A_{ki} X_i Y_k },
\end{equation}
where $\E{X_i} = 1 - \E{Y_i}$ is the probability that node $i$ is susceptible. Besides, we might also define the effective spreading rate $\tau = \frac{\lambda}{\delta}$, where we can rescale time in terms of $\delta$ enabling us to have just one control parameter, $\tau$. In fact, Equation \ref{eq:sis_exact} is exact, but it relies on the solution of $\E{Y_i Y_j}$, which can be obtained through the following formal definition \cite{Mieghem2014}
\begin{equation} \label{eq:dif_E}
 \frac{d \E{Y_i Y_j}}{dt} = \E {  Y_j \frac{d Y_i}{dt} + Y_i \frac{d Y_j}{dt} }.
\end{equation}
Note that by applying Equation \ref{eq:dif_E} we will have a dependency on the product of three random variables. Such formula can be applied up to the product of $N$ random variables. Note that terms such as $\E{Y_i Y_i}$ also appear naturally and it follows that $\E{Y_i Y_i} = \E{Y_i}$, as we are dealing with Bernoulli random variables. The same also applies to $\E{X_i Y_i} = 0$. Furthermore, as it can be easily seen, for the product of two random variables we have $\binom{N}{2}$, while for three we have $\binom{N}{3}$ and so forth, yielding to $\sum_{k=1}^N \binom{N}{k} = 2^N -1$ \cite{Mieghem2014}. Consequently, the exact solution can be obtained by solving the set of $2^N$ equations, which is prohibitive in practice\footnote{For instance, observe that for a network with $N = 30$, we have $2^N = 1073741824$ equations!}. Thus, the main importance of this formulation comes from the physical insights it provides and on possible approximations, as discussed in Section \ref{sec:mfa}. 

Aiming to extract useful information from our model we should define an order parameter\footnote{Following the Ehrenfest classification, the order parameter is defined as the first derivative of the free energy with respect to the external field. If it is continuous across the transition it is called a second order phase transition, which is the case of our classical SIS models. On the other hand, if it is discontinuous it is called a first order transition.}, which, in our case, must be a macro variable that captures the behavior of the whole system or a variable of interest. The order parameter is assumed to be the fraction of infected individuals, formally defined as
\begin{equation}
 \rho = \frac{1}{N} \sum_i^N \E{Y_i}.
\end{equation}

We next formally define what is meant by phase transition and epidemic threshold or, more generally, critical point. As we deal with a process with an absorbing state, we consider two different phases: (i) the inactive phase, also called the healthy phase and the (ii) active phase or endemic state. In the inactive phase, the order parameter goes to zero, i.e., for $\tau < \tau_c$, $\rho = 0$, while beyond the critical point the order parameter is larger than zero, i.e., for $\tau > \tau_c$ we have $\rho > 0$. It is worth emphasizing that the phase transition will only appear in the thermodynamic limit. In this case, we have an infinite number of micro-states \footnote{A micro-state in our context is the configuration of each node of the network, characterizing microscopically the dynamics. On the contrary, the macro-state characterize the global behavior of the dynamics.} (for the same order parameter) and an infinite number of infected individuals allowing the dynamics to be trapped in that state.

Then, observing that $X_i \geq 0$ in Equation \ref{eq:sis_exact} we have that~\cite{Mieghem2016}
\begin{equation} \label{eq:sis_exact2}
 \frac{d w}{dt} \leq \left(\tau \A -\I \right) \E{w(t)},
\end{equation}
where $w = [y_1, y_2, ..., y_N]$ is a binary vector containing the state of each node and whose solution is given by
\begin{equation} \label{eq:sis_exact3}
 \E{w(t)} \leq \exp{\left(\tau \A -\I \right)t} \E{w(0)},
\end{equation}
with
\begin{equation}
 \expm{\left(\tau \A -\I \right)t} = \sum_j^m \Bigg \{ \sum_k^{m_j} Y_{mk} t^{k-1} \Bigg\} e^{\Lambda_{m}t},
\end{equation}
where $Y_{mk}$ are linearly independent constant matrices and $\Lambda_{m}$ are eigenvalues of the adjacency matrix. Then, as reported in \cite{Mieghem2016}, before the critical point, the disease must die exponentially fast when $\tau < \frac{1}{\Lambda_1}$. Note that this is the first time in the review where we relate the spectral properties of our graph with dynamical properties of our system. This will be common in the next part of our text.

Additionally, an interesting connection between the spectral properties of the Laplacian matrix (see Section \ref{sec:spectral_characterization}) and the epidemic spreading might be obtained from the definition of the order parameter. From Equation \ref{eq:sis_exact} and after some algebra \cite{Mieghem2016}, we obtain
\begin{equation} \label{eq:drhodt}
 \frac{d\rho(t)}{dt} = \frac{1}{ N} \E {w^T(t) \left(\La - \tau\I \right) w(t) },
\end{equation}
where $w = [y_1, y_2, ..., y_N]$ is a binary vector containing the state of each node. In \cite{Mieghem2016} the author proposed an approximate formula to describe the time evolution of the SIS process using just a single non-linear equation. Moreover, Equation \ref{eq:drhodt} clarifies the connection of the Laplacian matrix with this process, which is most often analyzed using the adjacency matrix. However, it is also true that this connection is not as natural as with the adjacency matrix.

\begin{figure*}
\includegraphics[width=0.96\textwidth]{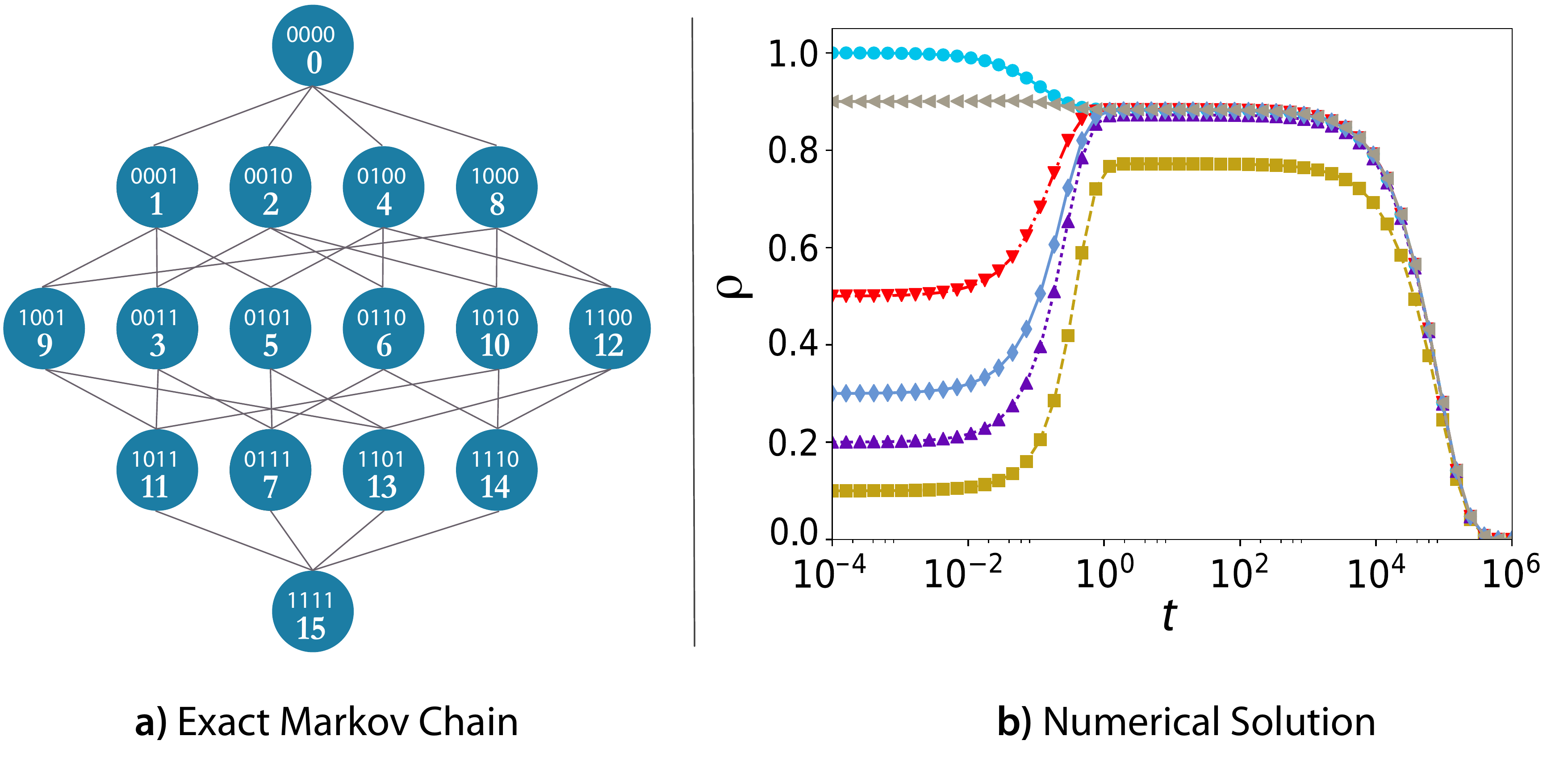}
\caption{In (a) the graph representation of the exact Markov chain, where each node is a state, defined by its binary and integer representation. In (b) the numerical solution of the exact model for a small Erd\"os and R\'enyi network with $N = 10$, $\E{k} = 5$, $\delta = 1$ and $\lambda = 2$ and different curves represent different initial conditions. Time is measured in $\frac{1}{\delta}$ units.} \label{fig:sis_exact}
\end{figure*}

There are other forms to express the exact processes described so far. Next, we use the infinitesimal generator approach to derive a set of equations that describes our dynamics. Denoting the micro-state of our Markov chain by the binary vector $S_i = [y_N, y_{N-1}, ..., y_1]$\footnote{The reversed order is important!}, we can represent it as the integer $i = \sum_{k=1}^N y_k(i) 2^{k-1}$ \cite{Mieghem09, Mieghem2014}. Figure \ref{fig:sis_exact} (a) shows a small example of this chain using its graph representation. Note that in state $i = 0$, all bits of our micro-state vector are zero, implying that none of the individuals is infected. On the other hand, the micro-state $i = 2^N -1$, implies all one vector, which means that all the individuals are infected. In other words, we map our intuitive binary vector that contains the Bernoulli random variables previously defined, keeping the micro-state of every node of our population, into an integer number. This procedure allows us to properly write the probability transitions between each micro-state. Since our process presents $2^N$ micro-states, the infinitesimal generator, $Q$, can be written as
\begin{equation} \label{eq:infgen}
 Q_{ij} = 
 \begin{cases}
  \delta          &\begin{cases}
             j = i - 2^{m-1}; m = 1, 2, ..., N \\
             \text{and} \hspace{2mm} y_m(i) = 1
            \end{cases} \\
  \lambda \sum_{k=1}^N \A_{mk} y_k(i)     &\begin{cases}
                     j = i - 2^{m-1}; m = 1, 2, ..., N \\
                     \text{and} \hspace{2mm} y_m(i) = 1
                    \end{cases} \\
  -  \sum_{k=1}^{2^N} Q_{kj} & i=j\\
  0    & \text{otherwise},
 \end{cases}
\end{equation}
whose solution is given as
\begin{equation}
 s^T(t) = s^T(0) \exp{(Q t)},
\end{equation}
where $s_i(t) = \Prob({Y_1(t) = y_1, Y_2(t) = y_2, ..., Y_n(t) = y_n})$ is the joint probability that defines the micro-state $i$. From the joint probability, we are able to obtain the nodal probability of $y_j = \Prob(Y_j = 1) = \E{Y_j}$ as the sum over all the states of all nodes except the node $j$, which can be written as
\begin{equation}
 y_j(y) = \Prob(Y_j(t) = 1) = \E{Y_j(t)} = \sum_{i=0}^{2^N-1} s_i(t) S_i(j),
\end{equation}
where $S_i(j)$ if the binary state of node $j$ in state $i$ (integer representation), where $i = \sum_{k=1}^N y_k(i) 2^{k-1}$. Thus, we only sum the states where the node $j$ is infected. The order parameter can also be represented as
\begin{equation} \label{eq:sis_exact_y}
 \rho(t) = y(t) = \frac{1}{N} s^T(0) e^{Qt} \M u,
\end{equation}
where $u$ is the all one vector and $\M$ is a $2^N \times N$ matrix with the bit-reversed binary state representation. Formally,
\begin{equation}
 \M = \begin{bmatrix}
       0 & 0 & 0 & \cdots & 0\\
       1 & 0 & 0 & \cdots & 0\\
       0 & 1 & 0 & \cdots & 0\\
       1 & 1 & 0 & \cdots & 0\\
       0 & 0 & 1 & \cdots & 0\\
       \vdots & \vdots & \vdots & \vdots & \vdots\\
       1 & 1 & 1 & \cdots & 1\\
      \end{bmatrix}.
\end{equation}
In Figure \ref{fig:sis_exact} (b) we show an example of the numerical solution of Equation \ref{eq:sis_exact_y}. In this example we used a small Erd\"os and R\'enyi network with $N = 10$, $\E{k} = 5$, $\delta = 1$ and $\lambda = 2$.

Probably the most important results extracted from the exact Markov chain are concerned with its absorbing state. The first observation is that the SIS dynamics presents only one absorbing state, $i=0$, or in its binary representation $S_0 = [0, 0, ..., 0]$. Thus, in a finite population when time goes to infinity, it will always reach the absorbing state. In \cite{Mieghem2014} the author argues that, for a rate $\tau = \frac{\lambda}{\delta}$ above the epidemic threshold, the number of infected nodes over time presents two regions: (i) for small times the process tends exponentially fast to the meta-stable\footnote{In a metastable state, there are a set of possible states (micro-configurations) where the system remains trapped and there is a stochastic variation over time. In addition, note that there are many different configurations for which the fraction of infected nodes is the same. More formally, there is a set of states which have a finite probability larger than zero, configuring a meta-state.} state (see Figure \ref{fig:sis_exact}) and (ii) after that it also decays exponentially in $t$, but with a very small rate, towards the absorbing state. On the other hand, below the epidemic threshold, the number of infected nodes decays exponentially fast to the absorbing state, as shown above.

Furthermore, observe that Equation \ref{eq:dif_E} allows us to perform a $n$-th order approximation by deriving the $n$-th random variables product equation and approximating the $(n+1)$-th products. Physically, what we are doing with this procedure is assuming that the $n$-th products are independent. These kinds of approximations are very common in the literature and are explored in Section \ref{sec:mfa}, especially, at first order which is the so-called quenched mean field (QMF), while the second-order approximation is the pair quenched mean field (PQMF). In Section \ref{sec:model_accuracy} we also perform a comparison of those two approximations and Monte Carlo simulations in order to validate and quantify the precision of both of them.

\section{Mean field approaches} \label{sec:mfa}

In this section, we explore the continuous and discrete time mean-field approximations in order to overcome the complexity of the exact approach, $\bigO{2^N}$, which is prohibitive for large systems. Regarding the continuous time formulations, we follow a path of increasing complexity. In Section \ref{sec:mf}, we present the classical homogeneous mean-field approximation, which does not account for a network structured population. Next, in Section \ref{sec:hmf} we extend this model taking into account a heterogeneous population, however neglecting the actual network and considering an ensemble of networks, which are described by their degree distributions. In Section \ref{sec:qmf}, we discuss a model which accounts for a fixed network structure. This model proved to be very efficient in many situations, as we will discuss later, however it neglects dynamical correlations. Again, to overcome such a limitation we follow the pair-approximation, whose description is presented in Section \ref{sec:pqmf}. Complementary to the continuous time approach, we consider a discrete time framework, following  a cellular automaton process in order to describe epidemic and rumor spreading, which is presented in Section \ref{sec:dtmc}. We also address other aspects in sections \ref{sec:msg}, \ref{sec:dtmc-async}, \ref{sec:generalization_multi}.

In the following, we mainly explore the SIS model, focusing on the framework and the key concepts of each of the previously mentioned approximations. This model allows extending the concepts discussed to other dynamical processes, which we will not explore in detail. For instance in Section \ref{sec:model_accuracy}, we apply them and compare some of these approaches with Monte Carlo simulations, systematically evaluating their accuracy. Additionally, in \ref{sec:appendix} we present a table with all the derived equations for all the mean field approaches.

\subsection{Mean Field (MF)}\label{sec:mf}

In a homogeneous population, the probability of contact between any pair of nodes is approximately the same. This is true in the complete graph and can be a reasonable approximation for Erd\"os--R\'enyi networks. In a complete graph, each vertex has $(N-1)$ connections. Thus, each individual can contact any other with the same probability. Regarding regular graphs, that is also a homogeneous structure, the probability of contact depends on the mean degree of the graph, $\E{k}$, which is the number of contacts each individual has. The dynamical system that describes the SIS epidemic spreading model in such a homogeneous population is described by the equation \cite{AnderssonBritton2000, Barrat08:book, Satorras015}
\begin{equation} \label{eq:sis_homogeneus}
  \frac{d\E{Y}}{dt} = \lambda \E{k} \E{Y} (1 - \E{Y}) - \delta \E{Y}
\end{equation}
where $\E{Y}$ is the fraction of the population that is infected and $\E{k}$ is the average number of connections in a regular graph. Observe that the fraction of susceptible individuals is given by $\E{X} = (1 - \E{Y})$. In the steady-state and neglecting second order terms, i.e., $\E{Y}^2$, which corresponds to a linear stability analysis around the only absorbing state of the exact SIS model, $\E{Y} = 0$, we obtain \cite{Barrat08:book}
\begin{equation}
 R_0 = \E{k} \frac{\lambda}{\delta} > 1,
\end{equation}
where $R_0$ is also called the basic reproduction number. This relationship characterizes a phase transition, where $R_0$ is the critical point, also called epidemic threshold. For $R_0 < 1$, the disease tends to die out and for $R_0 > 1$ it could be sustained in the population, i.e. an endemic state. For $R_0 > 1$,  the disease persists since there are more individuals being infected than recovering. Note that the non-zero stable solution represents a meta-stable state of the exact dynamics. In such a case, the steady state is not an absorbing state in the Markov sense, as there is a set of possible states (micro-configurations) where the system remains trapped and there is a stochastic variation over time. In addition, note that there are many different configurations for which the fraction of infected nodes is the same. More formally, there is a set of states above the threshold, which has a finite probability larger than zero, configuring a meta-state. The only absorbing state of this set of equations is thus the disease-free state since when it is reached the (micro and macro) dynamics stops. This meta-state will also be captured by other mean field approaches and the same discussion applies to any of the other approximations described here. Moreover, this phenomenology is clearer and even more interesting in the quenched and pair-quenched mean field, where the heterogeneity is directly accounted for in the model and individual probabilities are evaluated.

As mentioned earlier, although the SIS model might represent the dynamics of several diseases and captures interesting physical properties, there are many real diseases in which individuals acquire permanent immunity after being recovered. Aiming to model such diseases we use the SIR model. Here, in order to exemplify the case where the Markov chain has an infinite number of absorbing states in the thermodynamic limit we present the mean field approximation for the SIR model. In this case, another equation must be included, yielding \cite{AnderssonBritton2000, Barrat08:book, Satorras015},
\begin{equation} 
\begin{cases}
   \frac{d \E{X}}{dt} =& -\lambda \E{k} \E{Y} (1 - \E{Y} - \E{Z}) \\
   \frac{d \E{Y}}{dt} =& \lambda \E{k} \E{X} (1 - \E{Y} - \E{Z}) - \delta \E{Y} \\
   \frac{d \E{Z}}{dt} =& \delta \E{Y}
\end{cases}.
\end{equation}
Note that $\E{X} + \E{Y} + \E{Z} = 1$ and the sum of the derivatives is zero, which indicates that the population size is fixed. Moreover, observe that there is an infinite number of absorbing states, each one with a non-zero fraction of removed individuals. Intuitively, once an individual has recovered she no longer takes place in the dynamics. Formally, for any configuration where $\E{Y} = 0$ we have a stable solution. Interestingly, observe that $\frac{d \E{Y}}{dt}$ is the same for the SIR and SIS epidemic models, allowing us to translate some properties of the SIS model to the SIR context. This can be done near the critical point, for an infinite population: as the fraction of recovered individuals is greater than zero, the number of recovered individuals should be infinite. In other words, since the population is infinite and the fraction of recovered individuals is not null, it follows that we have an infinite number of recovered individuals. 

For the case of rumor spreading, let us consider a homogeneous population and the Maki--Thompson model. The set of equations that describes the dynamics is given as
\begin{equation} \label{eq:MT_MF}
\begin{cases}
 \dfrac{d \E{X}}{dt} =& -\lambda \E{Y} \E{X}\\ 
 \dfrac{d \E{Y}}{dt} =& \lambda \E{Y} \E{X} - \alpha \left( \E{Y} \E{Y} + \E{Y} \E{Z} \right)\\
 \dfrac{d \E{Z}}{dt} =& \alpha \left( \E{Y} \E{Y} + \E{Y} \E{Z} \right)
\end{cases},
\end{equation}
where variables $\E{X} = x$, $\E{Y} = y$ and $\E{Z} = z$ can be interpreted as fractions of the population (or the probability of uniformly sampling an ignorant, spreader or stifler individual, respectively). Considering that we have a homogeneous population, the independence of the random variables, $\E{Y X} = \E{Y} \E{X}$ (implicitly taken in the previous equation), is not a strong assumption, since correlations tend to vanish in the thermodynamic limit. This approach was initially proposed in \cite{makithompson1973} and largely explored in the literature (e.g. \cite{Lebensztayn2011}). Firstly, note that, as the SIR dynamics, the MT also has infinitely many absorbing states and a strong dependence on the initial conditions, since they limit the steady state. Next, we exemplify two fundamental results regarding the convergence of this process in the limit of an infinite population considering $\lambda = \alpha = 1$. Formally,
\begin{equation}
 \sqrt{N} \left( \frac{X_{\infty}}{N} - x_{\infty} \right) \Rightarrow \mathcal{N}(0, \sigma^2) \, \text{ as } \, N \to \infty,
\end{equation}
where $x_{\infty}$ is the steady state average fraction of ignorants on the population and $ \Rightarrow $ denotes convergence in distribution and $ \mathcal{N}(0, \sigma^2) $ is the Gaussian distribution with mean zero and variance
\begin{equation} \label{eq:mt_sigma}
 \sigma^2 = \frac{x_{\infty}(1 - x_{\infty})}{1 - 2x_{\infty}} = 0.27273575
\end{equation}
and
\begin{equation} \label{eq:mt_x}
 x_{\infty} = \frac{-W_0(-2\exp{(-2)})}{2} = 0.20318787,
\end{equation}
where $W_0$ is the principal branch of the Lambert function. These results can be verified in Section \ref{sec:Monte_Carlo} and Figure \ref{Fig:complete}, where we compare these theoretical predictions with Monte Carlo simulations in a finite graph. Following a different approach, in \cite{moreno2004}, the authors found analytically the same fraction of ignorants for $\lambda = \alpha = 1$. Additionally, they also reported other values for different parameters' values. Note that the system of equations \ref{eq:MT_MF} can be analytically explored. Following a similar analysis as \cite{moreno2004,Barrat08:book} \footnote{Note that in \cite{moreno2004,Barrat08:book} the parameters are scaled by the average degree of the network.}, from Equation \ref{eq:MT_MF} and in the steady state, i.e., in the infinite-time limit, we have that $y_{\infty} \rightarrow 0$. Thus, 
\begin{equation}
 \int_0^\infty  y(t) \text{dt} = \lim_{t \rightarrow \infty} z(t) = z_{\infty},
\end{equation}
which allows to obtain the following transcendental equation 
\begin{equation} \label{eq:zinf}
 z_{\infty} = 1 - e^{-\left(\frac{\lambda}{\alpha} z_{\infty} \right)}.
\end{equation}
From Equation \ref{eq:zinf}, whose non-zero solution exists if and only if \cite{moreno2004}
\begin{equation}
 {\dfrac{d}{z_{\infty}} \left( 1 - e^{-\left(\frac{\lambda}{\alpha} z_{\infty} \right)} \right) }\bigg\rvert_{z_{\infty} = 0} > 1,
\end{equation}
which translate into the condition $\frac{\lambda}{\alpha} > 0$. This result has the fundamental implication that there is no rumor threshold, i.e., that there will always be a (finite) fraction of the population that learned the rumor. This is in sharp contrast to disease spreading.

\subsection{Heterogeneous Mean Field (HMF)} \label{sec:hmf}

The heterogeneous mean field (HMF), also called degree-based mean field (DBMF), was the first theoretical framework used to analyze epidemic spreading processes in complex networks \cite{Satorras01:PRL, Boguna03:PRL, Barrat08:book, Satorras015}. This approach divides the population into groups according to the vertex degree, assuming that individuals with the same degree are statistically equivalent. Thus, this approximation describes the dynamics based on the network's degree distribution. Similarly to \ref{eq:sis_homogeneus} (SIS), the HMF dynamical equation for the fraction of infected individuals with a given degree $k$, $\E{Y_k}$, is formally expressed by
\begin{equation} \label{eq:sis_dbmf}
 \frac{d \E{Y_k(t)}}{dt} = - \E{Y_k(t)} + \tau k (1 - \E{Y_k(t)}) \sum_{k'}\Prob(k'|k) \E{Y_{k'}(t)},
\end{equation}
where $\tau$ is the effective spreading rate and the contacts are accounted for based on the conditional probability $\Prob(k'|k)$, which is the probability that a node with degree $k'$ has a neighbor with degree $k$. Note that the fraction of susceptible individuals for a given degree $k$ is $\E{X_k} = (1 - \E{Y_k})$, as these quantities are complementary. This system of equations cannot be solved in a closed form for a general degree distribution \cite{Satorras015} due to the difficulty in estimating $\Prob(k'|k)$ in correlated networks. Interestingly, observe that in homogeneous populations, the set of Equations \ref{eq:sis_dbmf}, reduces to a single equation given by \ref{eq:sis_homogeneus}. A quantity of interest in this system is the critical point, which can be obtained by means of linear stability analyses \cite{Boguna2002}, similarly as presented in the previous section. Considering only the first order terms of $\E{Y_k(t)}$ in Equation \ref{eq:sis_dbmf}, we have
\begin{equation}
 \frac{d \E{Y_k(t)}}{dt} \simeq \sum_{k'}J_{kk'} \E{Y_{k'}(t)},
\end{equation}
where $J_{kk'} = -\delta_{kk'} + \tau k \Prob(k'|k)$ is the Jacobian matrix and $\delta_{kk'}$ is the Kronecker delta function. The endemic state will emerge when $\tau \Lambda_M > 1$, where $\Lambda_M$ is the largest eigenvalue of the connectivity matrix, defined as \cite{Boguna2002}
\begin{equation} \label{eq:connectivity_matrix}
 C_{kk'} = k \Prob(k'|k),
\end{equation}
which has real eigenvalues and its largest eigenvalue is real and positive as a consequence of the Perron--Frobenius theorem. Therefore, the endemic state exists when
\begin{equation}
 \tau > \tau_c^{HMF} = \frac{1}{\Lambda_{1}},
\end{equation}
where $\Lambda_{1}$ is the largest eigenvalue of $C_{kk'}$.

Regarding uncorrelated networks, we can derive closed expressions for the epidemic threshold. In this case, the conditional probability does not depend on degree $k$. Formally, in uncorrelated networks we have $\Prob(k'|k) = \frac{k' \Prob(k')}{\E{k}}$ \cite{Barrat08:book}, which implies
\begin{equation} \label{eq:sis_dbmf_unc}
 \frac{d \E{Y_k(t)}}{dt} = - \E{Y_k(t)} + \tau k [1 - \E{Y_k(t)}] \Theta,
\end{equation}
where $\Theta$ is the probability of finding an infected node following a randomly chosen edge, which is given by
\begin{equation} \label{eq:Theta}
 \Theta = \sum_{k'}\frac{k' \Prob(k')}{\E{k}} \E{Y_{k'}(t)}.
\end{equation}

From Equation \ref{eq:sis_dbmf_unc} and imposing the steady state condition, $\frac{d \E{Y_k(t)}}{dt} = 0$, we obtain
\begin{equation} \label{eq:steady_dbmf}
 \E{Y_k(t)} = \frac{\tau k \Theta(\tau)}{1 + \tau k \Theta(\tau)}.
\end{equation}
Note the dependency of $\Theta$ on the effective spreading rate $\tau$. In addition, observe that the higher the node degree, the higher the probability of being infected. This observation is also a motivation to consider inhomogeneous degree distributions, as it clarifies the divergence between this approach and the homogeneous case. Combining Equations \ref{eq:Theta} and \ref{eq:steady_dbmf}, we obtain
\begin{equation} \label{eq:Theta_self_steady}
 \Theta(\tau) = \frac{1}{\E{k}} \sum_{k} k \Prob(k) \frac{\tau k \Theta(\tau)}{1 + \tau k \Theta(\tau)},
\end{equation}
which is a self-consistent equation admitting a non-zero solution when the threshold is satisfied \cite{Satorras01:PRL, Barrat08:book}. The critical point can be obtained using a geometrical argument, as shown in \cite{Barrat08:book} or considering an early time analysis \cite{Barrat08:book}. Mathematically it is given as
\begin{equation} \label{eq:tau_dbmf_unc}
 \tau_c > \tau^{HMF}_c = \frac{\E{k}}{\E{ k^2 }}.
\end{equation}
Furthermore, observe that the threshold in Equation \ref{eq:tau_dbmf_unc} can also be obtained from matrix $C_{kk'}$, in Equation \ref{eq:connectivity_matrix}, by noticing that $C_{kk'} = \frac{k k' \Prob(k')}{\E{k}}$ has a unique eigenvalue $\Lambda_{1} = \frac{\E{k}}{\E{k^2}}$ \cite{Boguna2002, Barrat08:book}. 

The results in \ref{eq:tau_dbmf_unc} imply a null epidemic threshold in networks with $\E{k^2} \rightarrow \infty$ in the limit of infinite system size, which are the cases for power-law degree distributions with exponent $2 < \zeta \leq 3$. This observation is crucial to understand the spreading of diseases in real networks. On the one hand, scale-free networks can be interpreted as the ideal medium for disease spreading, as the epidemic threshold goes asymptotically to zero (as a function of the network size). On the other hand, the network heterogeneity allows us to develop a more efficient immunization procedure since we can better identify the most influential spreaders. These results are presented in more detail in \cite{Barrat08:book}. Regarding the definition of a critical point, it is important to highlight that the concept of phase transition only applies to the infinite size limit (the thermodynamic limit). However, in the literature of complex network dynamics, especially for epidemic spreading, it is usual to use the terms critical point and phase transition on finite systems, as we find a behavioral change on that point.

Regarding the prevalence, given by $\rho_{\infty} = \sum_{k} \Prob(k) \E{X_k}$, we can analyze its behavior in the vicinity of the epidemic threshold by solving Equation \ref{eq:Theta_self_steady} \cite{Satorras2001}. After some manipulations, we get that $\rho_{\infty} \sim (\tau - \tau_c)^\nu$, where 
\begin{equation}
 \nu^{HMF} = 
 \begin{cases}
  \frac{1}{3-\zeta} & \text{for} \hspace{1cm} \zeta < 3 \\
  \frac{1}{\zeta - 3} & \text{for} \hspace{1cm} 3 < \zeta \leq 4 \\
  1 & \text{for} \hspace{1cm} \zeta \geq 4
 \end{cases}.
\end{equation}
On the one hand, the epidemic threshold vanishes for $2 < \zeta \leq 3$. On the other hand, it presents a slow growth of the prevalence near the critical point. The critical exponent, $\nu$, is larger than one in this case, which leads to a very slow growth near the critical point. This observation makes the spreading process on the top of these networks less threatening, as observed in \cite{Satorras01:PRL, Satorras015}.

Finally, for the SIR model, we note that as the disease reaches an individual through one of its edges \cite{Barrat08:book}, we should re-write Equation \ref{eq:Theta} as
\begin{equation} \label{eq:Theta_sir}
 \Theta = \sum_{k'}\frac{k' -1}{k'} \Prob(k'|k) \E{Y_{k'}(t)},
\end{equation}
which implies that $\Theta = \sum_{k'}\frac{k' -1}{\E{k}} \Prob(k') \E{Y_{k'}(t)}$ for uncorrelated networks. In terms of the critical point, it leads to $\tau^{HMF}_c = \frac{\E{k}}{\E{ k^2 } - \E{k}}$ for the SIR \cite{Barrat08:book}.

\subsection{Quenched Mean Field (QMF)} \label{sec:qmf}

Quenched mean field (QMF), also called the Individual-based mean field (IBMF) or N-Intertwined mean field approximation (NIMFA) \cite{Mieghem09}, is a theoretical approach based on nodal dynamics for a fixed network structure. This approach describes the time evolution of each node based on the state of the node and its neighbors. This method considers a set of $N$ equations to describe the time evolution of the probabilities. Implicitly, it makes the assumption of independence between the state of the nodes. In addition, it is important to observe that QMF is always an upper bound to the exact equation \cite{Mieghem09} when considering a connected graph. This can be easily understood as a consequence of the approximation considered, $\E{X_i(t) Y_j(t)} \approx \E{X_i(t)} \E{Y_j(t)}$. Note that while considering that nodal probabilities are independent we are, implicitly, neglecting the fact that having an infected neighbor increases the probability of being infected. Furthermore, observe that this influence is always positive \cite{Mieghem09, Cator2014}\footnote{Formally, from \cite{Mieghem09}, denoting $Y_i(t)$ the state of the node $i$ and time $t$, $\Prob(Y_j(t)=1|Y_i(t)=1) \geq \Prob(Y_j(t)=1)$, while independence can be formally expressed as $\Prob(Y_j(t)=1,Y_i(t)=1) = \Prob(Y_j(t)=1) \Prob(Y_i(t)=1)$. Observing both equations, it is clear that the QMF is an upper bound. As a reference,  Cator and Van Mieghem proved rigorously that the states of any two nodes in the SIS model are non-negatively correlated for all finite graphs in \cite{Cator2014}.}. Additionally, it is worth mentioning that in \cite{Ferreira2012}, based on extensive numerical analyses, the authors state that the QMF is an improvement over HMF as it provides a better estimate of the epidemic threshold.

Besides the dynamical aspects, briefly described above, this framework also enables us to take into account a single fixed network, allowing to extend the spreading process in more complex structures, such as multilayer networks. It is noteworthy that dealing with multilayer networks using the HMF formalism would destroy the multilayer structure, turning it into a community structured network with heterogeneous spreading rates (inter and intra-layer rates). Thus, quenched formalisms are more suitable when modeling dynamical processes in multilayers. In this section, we will explore the basics of QMF in single-layer networks in Section \ref{sec:qmf-single} and extend it to multilayer networks in Section \ref{sec:qmf-mux}.

\subsubsection{Single-layer networks} \label{sec:qmf-single}

First of all, beginning with single-layer networks, similarly to the homogeneous case, Equation \ref{eq:sis_homogeneus}, and the HMF, Equation \ref{eq:sis_dbmf}, the dynamical equations that describe such a process are
\begin{equation} \label{eq:sis_ibmf}
  \frac{d \E{Y_i(t)}}{dt} = \lambda \sum_{j=0}^N \A_{ij} \E{Y_j(t)} - \E{Y_i(t)} \left(  \lambda \sum_{j=0}^N \A_{ij} \E{Y_j(t)} + \delta \right)
\end{equation}
or in matrix terms
\begin{equation} \label{eq:sis_ibmf_matrix}
 \frac{d V(t)}{dt} = \left( \lambda \text{ diag}(1 -\E{Y_i(t)}) \A - \delta \I \right) V(t),
\end{equation}
where $V(t) = [\E{Y_1(t)}, \E{Y_2(t)}, ..., \E{Y_n(t)}]$, i.e., $V(t)$ is a vector consisting of $y_i(t) = \E{Y_i(t)}$.
Note that for the complete graph, we have a set of $N$ identical equations, in which all are equal to Equation \ref{eq:sis_homogeneus}. Furthermore, we remark the relationship between QMF and HMF. The former takes into account a fixed network, considering a specified adjacency matrix. The latter considers an additional approximation, i.e., the annealed network, which assumes that every node with degree $k$ is statistically equivalent. From Equation \ref{eq:sis_ibmf}, we can recover the HMF approach considering the annealed adjacency matrix \cite{Satorras015}
\begin{equation}
 \bar{\A}_{ij} = \frac{k_j \Prob(k_i|k_j)}{N \Prob(k_i)}.
\end{equation}
Moreover, the uncorrelated matrix is given as $\bar{\A}_{ij} = \frac{k_i k_j}{N \E{k}}$. Additionally to these matrices, we have to consider the degree-based average $y_k = \sum_{i \in k} \frac{y_i}{N \Prob(k)}$.

We assume the existence of the steady state, and calculate the steady-state probabilities for each node. We impose the steady state we have that $\frac{d y_j(t)}{dt}\rvert_{t \rightarrow \infty} = 0$. Denoting by $y_{j\infty}$ the corresponding steady state solution and considering networks without self-loops ($A_{ii} = 0$) we got from Equation \ref{eq:sis_ibmf}, 
\begin{equation} \label{eq:ibmf_steady_node}
 y_{i\infty} = \frac{\lambda \sum_{j=0}^N \A_{ij} y_{j\infty}}{\lambda \sum_{j=0}^N \A_{ij} y_{j\infty} + \delta} = 1 - \frac{1}{1 + \tau \sum_{j=0}^N \A_{ij} y_{j\infty}}.
\end{equation}
The trivial solution implies that every node is healthy. However, if $\delta = 0$, then every node is infected. In fact, this situation represents the SI epidemic process, when there is no recovery. In addition, $y_{i\infty} = 1 - O(\tau^{-1})$ for large values of $\tau$. 

Interpreting Equation \ref{eq:ibmf_steady_node} as a recursive relationship we are able to obtain a useful bound \cite{Mieghem09} as
\begin{equation}
 y_{i\infty} = 1 - \frac{1}{1 + \tau \sum_{j=0}^N \A_{ij} y_{j\infty}} = 1 - \frac{1}{1 + \tau k_i - \tau \sum_{j=0}^N \A_{ij} ( 1 - y_{j\infty})} \leq 1 - \frac{1}{1 + \tau k_i}.
\end{equation}
This equation can be continuously iterated. Thus, for any effective spreading rate $\tau = \frac{\lambda}{\delta} > 0$ we have the following bound for the probability of a node being infected
\begin{equation}
 0 \leq  y_{i\infty} \leq 1 - \frac{1}{1 + \tau k_i}.
\end{equation}
Moreover, this bound can be refined when $\tau \geq \frac{1}{k_{min}}$ as shown in \cite{Mieghem09}, leading to
\begin{equation} \label{eq:ibmf_bounds}
 1 - \frac{1}{1 + \frac{k_i}{k_{min}} \left( \tau k_{min} - 1 \right) } \leq  y_{i\infty} \leq 1 - \frac{1}{1 + \tau k_i}. 
\end{equation}

Finally, in order to evaluate the critical behavior of the system, we can also consider the steady state of Equation \ref{eq:sis_ibmf_matrix}, reducing it to 
\begin{equation} 
 \A V_{\infty} -  \text{diag}(y_{i\infty}(t)) \left( \A V_{\infty} + \tau^{-1} u \right) V_{\infty} = 0,
\end{equation}
where $u$ is an all-one vector. Next, after some algebra \cite{Mieghem09}, we have
\begin{equation} \label{eq:ibmf_steady_node_k}
 \frac{1}{\tau} V_{\infty} + \frac{1}{\tau} \sum_{k=2}^\infty V_{\infty}^k = \A V_{\infty}.
\end{equation}
From such an equation and in the vicinity of the epidemic threshold, $\tau = \tau_c + \epsilon$, for an arbitrarily small constant $\epsilon > 0$ we have $V_{\infty} = \epsilon v$, where $v$ is the eigenvector belonging to the largest eigenvalue of the adjacency matrix $\A$. Considering Equation \ref{eq:ibmf_steady_node_k} divided by $\epsilon$ we have
\begin{equation*}
 \A v = \frac{1}{\tau} v + \frac{\epsilon}{\tau} v^2 + O(\epsilon^2).
\end{equation*}
As for the epidemic threshold, for a sufficiently small $\epsilon > 0$, where only the first order terms remain, it is obtained that
\begin{equation}
 \A v = \frac{1}{\tau} v,
\end{equation}
which implies that $v$ is an eigenvector of $\A$, belonging to the eigenvalue $\frac{1}{\tau}$. The Perron--Frobenius theorem states that the components of the eigenvector associated with the largest eigenvalue, $\Lambda_1$, are all positive. Moreover, there is only one eigenvector of $\A$ with nonnegative elements. Hence, $\tau_c = \Lambda_1^{-1}$ and for $\tau < \tau_c$ the only possible solution is $V_{\infty} = 0$. On the other hand, for $\tau > \tau_c$ there is a non-zero solution. 

For networks with a power-law degree distribution, i.e., $P(k) \sim k^{-\zeta}$ we can use previous results on the spectrum of graphs \cite{ChungPNAS2003} and derive the following expression for the epidemic threshold,
\begin{equation}
 \tau_c^{QMF} \simeq 
 \begin{cases}
  \frac{1}{\sqrt{k_{max}}} &\hspace{1cm} \zeta > \frac{5}{2} \\
  \frac{\E{k}}{\E{k^2}} &\hspace{1cm} 2 <\zeta < \frac{5}{2}
 \end{cases}
\end{equation}
where $k_{max}$ is the maximum degree of the network. Note that, at the thermodynamic limit, there is a vanishing threshold for every network in which the maximum degree is a growing function of the network size, which is often the case for all random, non-regular networks, as pointed out in \cite{Satorras015}.

The above formalism also allows to obtain an expression for the prevalence near the critical point (similarly to Section \ref{sec:hmf}). In \cite{Mieghem2012} the author shows that the steady state fraction of infected nodes, $y_{\infty} = \rho = \frac{1}{N} \sum_i y_{i\infty}(\tau)$, depends on $\Lambda_1$, which is the leading eigenvalue with the associated eigenvector $v_1$, obeying
\begin{equation} \label{eq:ibmf_critical_exp}
 \rho = \frac{1}{N} \frac{u^T v_1}{\Lambda_1 \sum_{j=1}^N (v_1)_j^3} \left( \tau_c^{-1} - \tau^{-1} \right) + O \left( \tau_c^{-1} - \tau^{-1} \right)^2,
\end{equation}
considering sufficiently small values of $\tau$ and approaching from above. In addition, we can derive some bounds for the term $\sum_{j=1}^N (v_1)_j^3$. Using the H\"{o}lder inequality and considering a connected graph, after some algebra we have \cite{Mieghem2012},
\begin{equation}
 \frac{1}{\sqrt{N}} \leq \sum_{j=1}^N (v_1)_j^3 \leq \max_{1 \leq j \leq N} (v_1)_j.
\end{equation}
Note that $(v_1)_j = \frac{1}{\sqrt{N}}$ in the regular graph \cite{Mieghem2012}, which implies that this bound is the best possible since both sides are reached in this case. Expression \ref{eq:ibmf_critical_exp} allows us to study the critical properties as a function of the leading eigenvalue and its respective eigenvector.

Finally, an important phenomenon can be seen in QMF, namely the localization of diseases, as pointed out in \cite{Goltsev2012}. Using the inverse participation ratio, we can infer if the disease is localized in a few sets of nodes, a localized state, or if the disease is active in the network as a whole. Formally, if the inverse participation ratio \footnote{The inverse participation ratio is defined as $\mathrm{IPR}(f(\Lambda_i)) = \sum_i^N f(\Lambda_i)^4$, where $f(\Lambda_i)$ is the normalized eigenvector associated with the eigenvalue $\Lambda_i$.} of the leading eigenvector of the adjacency matrix (a normalized eigenvector), $\mathrm{IPR}(f(\Lambda_1)) \sim \bigO{1}$, then the disease is localized in a few nodes, the hubs, and the epidemic grows slowly as a function of $\lambda$. On the other hand, if  $\mathrm{IPR}(f(\Lambda_1)) \sim \bigO{\frac{1}{\sqrt{N}}}$, the disease extends to the whole network, configuring a true active state \cite{Goltsev2012, Wang2017}.

\subsubsection{Multilayer networks: SIS model} \label{sec:qmf-mux}

Following the QMF approach, given a fixed network, we model the spreading process using the tensorial notation defined in Section \ref{sec:tensorial_notation}, the equations describing the system's dynamics read as \cite{Arruda2017}
\begin{equation} \label{eq:SIS_tensor}
\dfrac{d \E{Y_{\beta \tilde{\delta}}(t)}}{dt} = - \E{Y_{\beta \tilde{\delta}}(t)} +\left( 1 - \E{Y_{\beta \tilde{\delta}}(t)} \right) \tau \mathcal{R}_{\beta \tilde{\delta}}^{\alpha \tilde{\gamma}}(\lambda, \eta)  \E{Y_{\alpha \tilde{\gamma}}(t)},
\end{equation}
where the supra contact tensor is defined as
\begin{equation} \label{eq:adj_contact_tensor}
\mathcal{R}_{\beta \tilde{\delta}}^{\alpha \tilde{\gamma}}(\lambda, \eta) = M_{\beta \tilde{\sigma}}^{\alpha \tilde{\eta}} E^{\tilde{\sigma}}_{\tilde{\eta}}(\tilde{\gamma} \tilde{\delta})  \delta^{\tilde{\gamma}}_{\tilde{\delta}} +
\frac{\eta}{\lambda} M_{\beta \tilde{\sigma}}^{\alpha \tilde{\eta}} E^{\tilde{\sigma}}_{\tilde{\eta}}(\tilde{\gamma} \tilde{\delta}) (U^{\tilde{\gamma}}_{\tilde{\delta}} - \delta^{\tilde{\gamma}}_{\tilde{\delta}}),
\end{equation}
which encodes the contacts. It has a similar role to the matrix $R$ in \cite{Cozzo:2013} (see Section \ref{sec:dtmc-mux} for more details). Notice that we have implicitly assumed that the random variables $Y_{\beta \tilde{\delta}}$ are independent. 
Naturally, the order parameter, also called macro-state variable, is defined as the average of the individual probabilities, formally given by
\begin{equation} \label{eq:qmf-macro}
 \rho = \frac{1}{nm}\E{Y_{\beta \tilde{\delta}}} U^{\beta \tilde{\delta}}.
\end{equation}

Furthermore, regarding the epidemic threshold, analogously to the results for single-layer systems we have a critical point given as 
\begin{equation} \label{eq:threshold}
 \left(\frac{\delta}{\lambda} \right)_c = \Lambda_1,
\end{equation}
where $\Lambda_1$ is the largest eigenvalue of $\mathcal{R}$. The complete derivation is similar to the single-layer, presented in the previous subsection. Observe that the eigen-structure of the tensor $\mathcal{R}$ is the same as the matrix $R$ in \cite{Cozzo:2013}, since it can be understood as a flattened version of the tensor $\mathcal{R}_{\beta \tilde{\delta}}^{\alpha \tilde{\gamma}}(\lambda, \eta)$. As argued in \cite{DeDomenico2013}, the supra-adjacency matrix
corresponds to a unique unfolding of the fourth-order tensor $\mathcal{R}$ yielding square matrices. Moreover, if $\eta M_{\nu \tilde{\delta}}^{\xi \tilde{\gamma}} E_{\xi}^{\nu}(\beta \beta) \ll \lambda M_{\beta \tilde{\gamma}}^{\alpha \tilde{\xi}} E_{\tilde{\xi}}^{\tilde{\gamma}}(\tilde{\delta}\tilde{\delta})$, the critical point is dominated by the individual layer behavior and the epidemic threshold is approximated to that of a SIS model in single-layers, when considering the union of $m$ disjoint networks. Consequently, the epidemic threshold is determined by the largest eigenvalue, considering all layers. The same conclusion was reached in \cite{Cozzo:2013} using a perturbative approach. 

Finally, extending the bounds we found for single-layer networks, the nodal probability on the steady state can be bounded by \cite{Arruda2017}
\begin{equation} \label{eq:bounds}
 1 - \frac{1}{1 + \frac{d_{\beta \tilde{\delta}}}{d^{\text{min}}} \left[ \left( \frac{\lambda}{\delta} \right) d^{\text{min}} - 1 \right]} \leq \E{Y^{\infty}_{\beta \tilde{\delta}}} \leq 1 - \frac{1}{\left( \frac{\lambda}{\delta} \right) d_{\beta \tilde{\delta}} + 1},
\end{equation}
where $\E{Y^{\infty}_{\beta \tilde{\delta}}}$ denotes the probability that node $\beta$ in layer $\tilde{\delta}$ is in the steady state regime with $d_{\beta \tilde{\delta}} = \mathcal{R}_{\beta \tilde{\delta}}^{\alpha \tilde{\gamma}}(\lambda, \eta)  U_{\alpha \tilde{\gamma}}$ and $d^{\text{min}} = \text{min}\{d_{\beta \tilde{\delta}}\}$. The derivation and analysis of these bounds are similar to the ones presented in the previous subsection. Observe that the higher $d^{\text{min}}$, the closer the lower and upper bounds. In the extreme case $\left( \frac{\lambda}{\delta} \right) \rightarrow \infty$ the bounds approach each other and all nodes tend to be infected. Phenomenologically, the latter parameter models the limiting case of a SI-like scenario, where $\delta = 0$. In such a dynamical process, all individuals are infected in the steady state.

The model presented in this subsection, considering SIS spreading processes in multilayer networks using a continuous time approach was initially proposed by Arruda et al. \cite{Arruda2017}.  In that paper, the authors observed the phenomena of layer-wise localization and the barrier effect. Note that the localization phenomenon can also be explained by means of eigenvector localization along the same lines as \cite{Goltsev2012}, as mentioned in the previous section.

\subsubsection{Multilayer networks: SIR model} \label{sec:qmf-mux-sir}

For the sake of completeness, besides the SIS epidemic model, we can also consider the SIR model. In contrast to the SIS case, which has just one absorbing state (inactive), the SIR has many absorbing states. In fact, as noted before, for an infinite population we have an infinite number of absorbing states. We introduce the recovered and susceptible states, denoted here by $X_{\beta \tilde{\delta}}$ and $Z_{\beta \tilde{\delta}}$, respectively. Then, using a similar notation as in the latter section and associating Poisson processes to nodes and edges, we have the  dynamical set of equations
\begin{align} \label{eq:SIR_tensor}
  \begin{cases}
 \dfrac{d \E{Y_{\beta \tilde{\delta}}}}{dt} &= -\delta \E{Y_{\beta \tilde{\delta}}} + \E{Z_{\beta \tilde{\delta}}} \lambda \mathcal{R}_{\beta \tilde{\delta}}^{\alpha \tilde{\gamma}}(\lambda, \eta)  \E{Y_{\alpha \tilde{\gamma}}} \\ 
 \dfrac{d \E{X_{\beta \tilde{\delta}}}}{dt} &= \delta \E{Y_{\beta \tilde{\delta}}} \\
 \dfrac{d \E{Z_{\beta \tilde{\delta}}}}{dt} &= - \E{Z_{\beta \tilde{\delta}}} \lambda \mathcal{R}_{\beta \tilde{\delta}}^{\alpha \tilde{\gamma}}(\lambda, \eta)  \E{Y_{\alpha \tilde{\gamma}}}.
  \end{cases}
\end{align}
Note that, similarly to the previous cases, the Poisson processes on the nodes model the recovering, whereas on the edges, model the spreading. Additionally, observe that the critical point obtained for the SIS case also applies here. Note, however, that the physical meaning of such a transition changes. After the critical point, the number of recovered individuals scales with the number of nodes in the network. In other words, the fraction of recovered individuals is larger than zero. On the other hand, before the critical point, this fraction goes to zero in the thermodynamic limit. Interestingly, the number of recovered individuals will be at least the number of individuals initially infected due to the nature of the dynamics.

\subsection{Pair Quenched Mean Field (PQMF)} \label{sec:pqmf}

As shown in Section \ref{sec:exact}, we can obtain a $n$-th order approximation, iterating Equation \ref{eq:dif_E}. Here we extend the quenched mean field to the pair approximation. In other words, we evaluate the probabilities $\E{Y_i(t) Y_j(t)}$ for each edge on the network. The approximation in this case considers that the pairs $\E{Y_i(t) Y_j(t)}$ are independent, contrary to the node independence assumption on the QMF. This enables us to account for dynamical correlations, improving the accuracy of the QMF approach. This framework was previously studied in \cite{Cator2012b, Mata2013}. Regarding our notation, here the time dependence of our random variables is often suppressed in order to have shorter equations.

In the following section, we discuss the PQMF formulation analytically and develop a new set of equations, relating this approximation with the non-backtracking matrix (for more details, see Section \ref{sec:spectral_characterization}). Additionally, we show the relationship between the presented formalism with the message passing approach for recurrent state, proposed in \cite{Shrestha2015}. Next, in Section \ref{sec:pqmf-num}, we explore this model numerically, emphasizing its approximations. Finally, in Section \ref{sec:pqmf-critical} we discuss the predictions for the critical point.

Here we will only consider single-layer networks to avoid unnecessary complexity, but the model can be easily extended to multilayer networks in two different ways: (i) considering a supra-adjacency matrix instead of the adjacency matrix, but taking into account that the inter-layer elements are weighted by $\frac{\eta}{\lambda}$ or (ii) re-writing the following equations in tensorial form and considering the tensor $\mathcal{R}_{\beta \tilde{\delta}}^{\alpha \tilde{\gamma}}(\lambda, \eta)$ instead of $\A$. Both approaches are equivalent and were previously explored in this review. For more on this issue, the reader is refereed to Section \ref{sec:generalization_multi}.

\subsubsection{Analytical aspects} \label{sec:pqmf-analytical}

Assuming a fixed network, for each node, we express its probabilities over time following Equation \ref{eq:sis_exact} and describe the evolution of $\E{X_i Y_j}$ using relation \ref{eq:dif_E}. Hence, for each edge we have
\begin{equation} \label{eq:pqmf-xy}
 \frac{d\E{X_i Y_j}}{dt} = \E {-(\delta + \lambda ) X_i Y_j + \delta Y_i Y_j + \lambda \sum_{j\leftarrow k} \B_{(i\leftarrow j, j\leftarrow k)} X_i X_j Y_k - \lambda \sum_{i\leftarrow k} \Hm_{(i\leftarrow j, i\leftarrow k)} X_i Y_j Y_k },
\end{equation}
\begin{equation} \label{eq:pqmf-yy}
 \frac{d\E{Y_i Y_j}}{dt} = \E {-2\delta Y_i Y_j +  \lambda X_i Y_j + \lambda \sum_{j\leftarrow k} \B_{(i\leftarrow j, j\leftarrow k)} Y_i X_j Y_k + \lambda \sum_{i\leftarrow k} \Hm_{(i\leftarrow j, i\leftarrow k)} X_i Y_j Y_k },
\end{equation}
where $\B_{(i\leftarrow j, j\leftarrow k)}$ is the non back-tracking matrix (for more, see Section \ref{sec:spectral_characterization_B}) and $\Hm_{(i\leftarrow j, i\leftarrow k)}$ is a new matrix defined as
\begin{equation}
 \Hm_{(i\leftarrow j, k\leftarrow l)} = 
 \begin{cases}
  1 \qquad \text{if} \quad i = k \quad \text{and} \quad j \neq l \\
  0 \qquad \text{otherwise}.
 \end{cases}
\end{equation}
Note that we denote the indices of the matrix by arrows, e.g. $i\leftarrow j$, in order to easily understand the physical meaning of these terms. Interestingly, in the literature, in \cite{Cator2012b} all possible pairs were accounted for in the modeling, generating a system of $N^2 \times N^2$. On the other hand, in \cite{Mata2013}, the authors describe the system only for the edges, reducing the complexity to $(M+N) \times (M+N)$, where $M$ is the number of edges. Note that although for dense networks $M \in \bigO{N^2}$, in sparse networks, this consideration drastically reduces the computational cost of numerical evaluations of those equations. Although both models are the same, they differ in terms of their approximations of the product of three variables, i.e., $\E{Y_i Y_j Y_k}$. This is discussed in the next section.

\subsubsection{Numerical aspects} \label{sec:pqmf-num}

So far we have dealt with analytical aspects of the PQMF equations. It is also enlightening to numerically evaluate our models. First of all, we must discuss how to properly approximate the expectation of the product of three random variables. In \cite{Cator2012b} the approximation used was
\begin{equation} \label{eq:approx-pair-1}
\E{Y_i Y_j Y_k} = \E{Y_i Y_j} \E{Y_k},
\end{equation}
which is not the standard, as commented in \cite{Mata2013}, where the authors use the standard approximation, given as
\begin{equation}
 \E{Y_i Y_j Y_k} = \frac{\E{Y_i Y_j}\E{Y_j Y_k}}{\E{Y_j}}.
\end{equation}
Throughout this review, we will use the standard approximation scheme, since it is the most appropriate and accurate. In \cite{Cator2012b}, the authors also discuss the approximation~\ref{eq:approx-pair-1} by observing their limitations for small networks.

Thus, Equation \ref{eq:sis_exact} is kept the same, but the expectation operator is constrained to the random variables, as we are dealing with static networks. Mathematically,
\begin{equation}
 \frac{d\E{Y_i}}{dt} = -\delta \E{Y_i} + \lambda \sum_{k=1}^N \A_{ki} \E { X_i Y_k },
\end{equation}
where the terms $\E { X_i Y_k }$ are calculated using the standard approximation. After some algebra we have,
\begin{equation} \label{eq:pqmf-approx}
\scalefont{0.8}
 \frac{d\E{X_i Y_j}}{dt} \approx \delta (\E{Y_j} - \E{X_i Y_j}) - (\lambda + \delta) \E {X_i Y_j} + \lambda \frac{1 - \E{Y_i} - \E{X_i Y_j}}{1 - \E{Y_j}} \sum_{j\leftarrow k} \B_{(i\leftarrow j, j\leftarrow k)} \E{X_j Y_k} - \lambda \frac{\E{X_i Y_j}}{1 - \E{Y_i}} \sum_{i\leftarrow k} \Hm_{(i\leftarrow j, i\leftarrow k)} \E {X_i Y_k},
\end{equation}
which can be solved using a numerical approach, such as Runge-Kutta. Intuitively, note that the probability
\begin{equation}
 \Prob(Y_i = 1 | Y_j = 1) = \frac{\Prob(Y_i = 1 \cap Y_j = 1)}{\Prob(Y_j = 1)} = \frac{\E{Y_i Y_j}}{\E{Y_j}},
\end{equation}
which is the same that appears multiplying the summations. Thus, the first summation is the positive contribution of the neighbors of $j$, given that $i$ and $j$ are both susceptible. On the other hand, for the second summation it contributes negatively because it accounts for the probability that both ends of the edge have infected individuals.

\subsubsection{Critical point prediction} \label{sec:pqmf-critical}

PQMF is an improvement of the traditional QMF approach, providing a better upper bound to the average value. However, the critical point provided by this approach presents some important difficulties that should be discussed. In \cite{Mata2013} the authors present a non closed expression, which allows a precise estimation of the critical point. Recall that in this review we showed a matrix formulation for the PQMF, and thus some of the steps presented here regarding the calculation of the critical point might differ. In order to avoid any confusion we derive the critical point criteria here. First of all, performing a linear stability analysis on the pair approximation, as done in \cite{Mata2013}, we have
\begin{equation} \label{eq:pqmf-angelica-p2}
\frac{d\E{X_i Y_j}}{dt} \approx (\E{Y_j} - \E{X_i Y_j}) - (\tau + 1) \E {X_i Y_j} + \tau \sum_{j\leftarrow k} \B_{(i\leftarrow j, j\leftarrow k)} \E{X_j Y_k}.
\end{equation}
Next, considering Equation \ref{eq:sis_exact} at the steady-state, i.e., $t \rightarrow \infty$ and $\frac{d\E{Y_i}}{dt} \rightarrow 0$, we have
\begin{equation} \label{eq:pqmf-angelica-p1}
 \E {Y_j} =  \tau \sum_{k=1}^N \A_{jk} \E { X_j Y_k }.
\end{equation}
Furthermore, note that we can relate equations \ref{eq:pqmf-angelica-p2} and \ref{eq:pqmf-angelica-p1} using the following relationship \footnote{This step connects our approach with the evaluations presented in \cite{Mata2013}.}
\begin{equation} \label{eq:pqmf-angelica-p3}
 \tau \left( \sum_{j\leftarrow k} \B_{(i\leftarrow j, j\leftarrow k)} \E{X_j Y_k} \right) = \tau \left( \sum_{k=1}^N \A_{jk} \E { X_j Y_k } - \E { X_j Y_i } \right) =   \E {Y_j} - \tau \E { X_j Y_i }.
\end{equation}
In this way, from Equation \ref{eq:pqmf-angelica-p2} and the relation obtained in Equation \ref{eq:pqmf-angelica-p3} we have
\begin{equation} \label{eq:pqmf-angelica-p4}
 \E{X_i Y_j} \approx \frac{(2 + \tau) \E{Y_j} - \tau \E{Y_i}}{2(1 + \tau)},
\end{equation}
where we used the identity $\E{X_j Y_i} = \E{Y_i} - \E{Y_j} +\E{X_j Y_i}$. Finally, substituting Equation \ref{eq:pqmf-angelica-p4} in \ref{eq:pqmf-angelica-p1}, one find the Jacobian matrix
\begin{equation} \label{eq:pqmf-angelica}
 \J_{ij} = - \left( 1 + \frac{\tau^2 k_i}{2 \tau + 2}\right) \delta_{ij} + \frac{\tau (2 + \tau)}{2\tau + 2} \A_{ij}.
\end{equation}
Thus, as shown in \cite{Mata2013}, the critical point can be obtained numerically finding the null eigenvalues of $\J_{ij}$. In order to obtain a good approximation to the critical point, we employed a binary search on $\tau$, seeking $\lambda_{\max}(\J) = 0$ with a tolerance parameter of $\text{tol} = 10^{-6}$. In most of the networks evaluated our method converged in less than 20 iterations. These estimations are denoted here as $\tau_c^{PQMF}$. In Section \ref{sec:FSS}, we evaluate the accuracy of this expression and compare it with the other mean field approximations discussed throughout the review.

Although we do not present a closed expression for the critical point for the general case, in \cite{Mata2013} the authors evaluated some expressions regarding regular networks. Those cases are important since they might provide insights into real problems. The first case is the Random Regular Network (RRN), whose critical point is $\left( \tau^{PQMF}_{RRN} \right)_c \approx \frac{1}{k-1}$. The second is the star graph, whose critical point is given as \footnote{Note that the differences between the equations presented here and the original paper \cite{Mata2013} are due to the definition of $N$. Here $N$ is the number of nodes, while in \cite{Mata2013} it is the number of leaves.} $\left( \tau^{PQMF}_{Star} \right)_c = \frac{\sqrt{2N - 3}+1}{N-3}\approx \sqrt{\frac{2}{N-1}}$.

\subsubsection{An alternative critical point prediction} \label{sec:pqmf-critical-alt}

In the previous section, we presented the standard method to obtain the critical point. However, as previously mentioned, it is not in a closed form. In this section, we take advantage of the matrix formulation, in terms of the non-backtracking matrix presented in Section \ref{sec:pqmf-analytical} to obtain an approximation of the critical point in a closed form. We might also emphasize that, as a drawback of our approach, it is neither an upper nor a lower bound for the critical point. Interestingly, the method presented in the previous section is an improved upper bound, if compared to the QMF.

Firstly, we must recall some basic probability and expectation properties based on the relation $X_i + Y_i = 1$, resulting on the following identity,
\begin{equation} \label{eq:pqmf-id1}
 \E{X_i Y_j} = \E{Y_j} - \E{Y_i Y_j}.
\end{equation}
Then, deriving Equation \ref{eq:pqmf-id1} and substituting Equations \ref{eq:pqmf-xy} and \ref{eq:pqmf-yy}, we have
\begin{equation} \label{eq:pqmf-final}
 \frac{d\E{Y_j}}{dt} = \E {-\delta Y_j + \lambda \sum_{j\leftarrow k} \B_{(i\leftarrow j, j\leftarrow k)}  X_j Y_k },
\end{equation}
where $\E{Y_j}$ are the components of a vector with $M$ elements, each one associated with an edge. Physically, it is the probability that an edge from $j$ to $i$ has a susceptible source $j$. This equation is not intuitive to analyze, but it is convenient for a linear stability analysis near the absorbing state, $Y_i = 0, \forall i \in \mathcal{V}$. The fixed point is  $\E{Y_i} = \E{Y_i Y_j} = \E{Y_i X_j} = 0$. In order to simplify \ref{eq:pqmf-final}, we use the relation \ref{eq:pqmf-id1} implying that
\begin{equation} \label{eq:pqmf-final2}
 \frac{d\E{Y_j}}{dt} = \E {-\delta Y_j + \lambda \left( \sum_{j\leftarrow k} \B_{(i\leftarrow j, j\leftarrow k)} Y_k - \sum_{j\leftarrow k} \B_{(i\leftarrow j, j\leftarrow k)} Y_j Y_k \right) }.
\end{equation}
Finally, assuming that $\E{Y_j} \approx \E{X_i Y_j} \approx \epsilon$, we also have that $\E{Y_i Y_j} \approx 0$, due to relation \ref{eq:pqmf-id1}. Physically, we can imagine a scenario where just a small fraction of individuals in our population are infected. Thus, it is expected that, by randomly sampling an edge, it is more likely to sample an edge connecting a spreader and a susceptible ($\E{Y_i X_j}$), than two spreaders ($\E{Y_i Y_j}$)\footnote{Complementarily, observe that $\E{X_i X_j} = \E{X_j} - \E{X_i Y_j}$. Then assuming $\E{X_i X_j} \approx 1$ and $\E{X_j} \approx 1 - \epsilon$ (by construction, as it is the fraction of infected individuals in the population and we are performing our analysis around the absorbing state), we have $\E{X_i Y_j} \approx \epsilon$}. Considering this and proceeding with the linear stability analysis we have
\begin{equation} \label{eq:pqmf-final-fact}
 \frac{d \boldsymbol{\epsilon}}{dt} = \left( \lambda \B_{(i\leftarrow j, j\leftarrow k)} - \delta \I \right) \boldsymbol{\epsilon},
\end{equation}
where $\boldsymbol{\epsilon}$ is a vector. Next, observe that expression \ref{eq:pqmf-final2} presents repeated elements, weighting the individual states differently. In order to correct this expression and show that it also captures the critical behavior we can write the order parameter as
\begin{equation} \label{eq:bilinear-rho}
 \frac{d \rho^*}{dt} \approx \frac{1}{N} d^T \left( \lambda \B_{(i\leftarrow j, j\leftarrow k)} - \delta\I \right) \boldsymbol{\epsilon},
\end{equation}
where $\rho^* = \frac{1}{N}\sum_i^N d_i \boldsymbol{\epsilon}_i$ is the approximated order parameter near the critical point, $\boldsymbol{\epsilon}$ is a vector with expectation $\E{Y_i} \approx \boldsymbol{\epsilon}_i$, with the appropriate dimensions and $d^T = [\left(k_l^*\right)^{-1}]$, $\forall l = 1, 2, ..., M$ is a vector composed by $k_l^*$, which is the degree of node $j$, from which the edge $l$ points to. In other words, for the directed edge $l$, which connects $i$ to $j$,  $k_l^* = k_j$. Note that this vector has repeated components, that weigh the infection probability, correcting the contributions in Equation \ref{eq:pqmf-final2}. Next, in order to determine the critical point, observe that $d^T$ is composed by positive elements. Thus, the smallest value of $\left(\frac{\delta}{\lambda} \right)$ beyond which the product $\left( \B_{(i\leftarrow j, j\leftarrow k)} - \frac{\delta}{\lambda} \I \right) \boldsymbol{\epsilon}$ is positive is the leading eigenvalue of $\B$. Considering this, near the critical point we can re-write Equation \ref{eq:bilinear-rho} as
\begin{equation} \label{eq:bilinear-rho-eig}
 \frac{d \rho^*}{dt} \approx \frac{1}{N} \left( \lambda \Lambda_1 - \delta \right) d^T \boldsymbol{\epsilon} = \left( \lambda \Lambda_1 - \delta \right) \rho^*,
\end{equation}
where $\boldsymbol{\epsilon}$ is approximated by the leading eigenvector of $\B$ near the critical point, as also shown in Equation \ref{eq:pqmf-final-fact}. Therefore, this equation implies that this alternative critical point for the PQMF is given as
\begin{equation} \label{eq:tau_pqmf}
 \tau_c^{aPQMF} = \frac{1}{\Lambda_1},
\end{equation}
where $\Lambda_1$ is the leading eigenvalue of the non-backtracking matrix, $\B$. In order to distinguish between both predictions, we denote this alternative expression as $\tau_c^{aPQMF}$, where ``\emph{aPQMF}'' stands for alternative PQMF critical point estimation. Interestingly, as argued in \cite{Shrestha2015}, $\Lambda_1(\B) \leq \Lambda_1(\A)$ for single-layer networks. Besides, Equation \ref{eq:bilinear-rho-eig} explicitly shows the critical point. Observe that the Perron-Frobenius theorem assures that the leading eigenvector consists of positive elements. Furthermore, Equation \ref{eq:bilinear-rho-eig} emphasizes the contributions of the eigenvector to the critical behavior. Finally, the product $d^T \boldsymbol{\epsilon}$ is geometrically interpreted as $\cos (\theta) =  \frac{d^T \boldsymbol{\epsilon}}{\vertiii{d} \vertiii{\boldsymbol{\epsilon}}}$, where $\theta$ is the angle between $d$ and $\boldsymbol{\epsilon}$. 

Although  the solution of \ref{eq:pqmf-angelica} agrees with Equation \ref{eq:tau_pqmf} in the predictions for the random regular case, there is no general map between Equations \ref{eq:tau_pqmf} and \ref{eq:pqmf-angelica}. It is important to emphasize that the approximations performed to obtain the critical point in Equations \ref{eq:tau_pqmf} and \ref{eq:pqmf-angelica} are different.  In fact, they diverge on many cases, as numerically shown in Section \ref{sec:FSS}. The formulation of Section \ref{sec:pqmf-analytical} is interesting because it connects epidemic spreading processes with the non-backtracking matrix. This relationship was previously observed in \cite{Radicchi2016}, where the authors proposed a centrality measure based on the non-backtracking matrix as a tool for the identification of influential spreaders and also in \cite{Shrestha2015}, where the authors proposed a recurrent-state message-passing approach for the SIS epidemic process. On the other hand, regarding the critical point estimation presented in Equation \ref{eq:tau_pqmf}, it is not a bound for the critical point. In other words, it can be above or below the real critical point, contrasting with Equation \ref{eq:pqmf-angelica}, which is an improved upper bound (in comparison with the QMF predictions). It is also worth noting that this alternative critical point prediction also coincides with the recurrent-state message-passing approach, presented in the following section.

\subsection{Recurrent-state message-passing approach} \label{sec:msg}

Newman \cite{Newman2010} proposed a message passing formulation to study the SIR epidemic spreading process. In such formalism, the diseases are interpreted as messages in the network. Interestingly, this formalism is exact on trees. However, it does not translate directly to SIS epidemic spreading processes due to the recurrence of node states. For instance, in the SIR, once a node is recovered it passes from $Y \rightarrow Z$ and it will remain in that state. On the other hand, in the SIS, a node is always changing from $Y \rightarrow X$ and $X \rightarrow Y$. In this context, aiming to overcome these limitations, in \cite{Shrestha2015} the authors proposed a recurrent-state message-passing approach. Recently, in \cite{Wang2017}, the authors wrote a short literature review, summarizing some of the previous results and pointing to \cite{Wang2016}, where there is a classification of these methods into three categories:(i) Mean-Field like; (ii) Quenched Mean field and (iii) dynamical message passing. The above-mentioned references show the importance of message-passing approaches and in the rest of this section we will compare this approach with the pair approximation in order to see their similarities and differences.

\subsubsection{Analytical aspects}

In \cite{Shrestha2015}, the authors studied the SIRS, SEIS and the SIS models. Here we only discuss the SIS case, but all the conclusions can also be extended to those models. Reproducing the equations proposed in \cite{Shrestha2015} and using their notation (exception made to the rates, where we use our standard notation in order to avoid confusion), we have the following set of dynamical equations
\begin{equation} \label{eq:mp-1}
   \frac{d I_i}{dt} = - \delta I_i + \lambda S_i \sum_{j \in \partial i} I_{j\leftarrow i},
\end{equation}
\begin{equation} \label{eq:mp-2}
  \frac{dI_{i\leftarrow j}}{dt} = - \delta I_{i\leftarrow j} + \lambda S_j \sum_{k \in \partial j \setminus i} I_{k\leftarrow j}
\end{equation}
where the contacts are directed, $I_i(t)$ is the probability that node $i$ is infected on time $t$, complementarily, $S_i(t)$ is the probability that $i$ is susceptible and $I_{j\leftarrow i}$ is the probability that $j$ is infected by its neighbors at time $t$ in absence of node $i$ \cite{Wang2017}.

\subsubsection{Critical point prediction} \label{sec:msg-analytical}

From this framework,  using linear stability analysis, in \cite{Shrestha2015}, the authors derived a closed expression for the critical point. It is evident that from Equation \ref{eq:mp-2}, the Jacobian matrix of the linearized system is
\begin{equation}
 \J = \left( \B_{(i\leftarrow j, j\leftarrow k)} - \frac{\delta}{\lambda} \I \right),
\end{equation}
hence, the critical point can be expressed as
\begin{equation} \label{eq:tau_rmp}
 \tau_c^{rMP} = \frac{1}{\Lambda_1},
\end{equation}
where $\Lambda_1$ is the leading eigenvalue of the non-backtracking matrix. Note that this result also help us understand the connections behind the non-backtracking matrix and spreading processes. Additionally, it is important to observe that such an analysis agrees with the alternative PQMF derivation (see Section \ref{sec:pqmf-critical-alt}).

\subsubsection{Comparison between PQMF and the recurrent-state message-passing approach} \label{sec:pqmf-comp}

For comparison purposes, we will assume that equations \ref{eq:sis_exact} and \ref{eq:mp-1} represent similar quantities since both aim to predict the same dynamical behavior. Next, we compare how these quantities evolve in both contexts, stating their differences. This might highlight the differences between both models, regarding their premises and approximations. Thus, the comparison between equations \ref{eq:sis_exact} and \ref{eq:mp-1} and the application of Bayes theorem, leads to the following relation
\begin{equation} \label{eq:mp-pqmf}
 \E{X_i Y_k} = \Prob(Y_k = 1 | X_i = 1) \Prob(X_i = 1) = S_i I_{k\leftarrow i},
\end{equation}
where we included the term $S_i$ inside the summation on \ref{eq:mp-1}. Note that $\Prob(X_i = 1)$ represents $S_i$, which suggests that $I_{j\leftarrow i}$ can be interpreted as the conditional probability $\Prob(Y_k = 1 | X_i = 1)$ in the Markov chain context, for comparison purposes.  In order for both approaches to be the same we must have that $\frac{d S_i I_{i\leftarrow j}}{dt}$ corresponds to Equation \ref{eq:pqmf-xy}, up to some approximation. Note that in Equation \ref{eq:pqmf-approx} we followed the standard pair approximation, which might be different from the recurrent-state message-passing approximation. From the chain rule we have, 
\begin{equation}
 \frac{d S_i I_{i\leftarrow j}}{dt} = S_i \frac{d I_{i\leftarrow j}}{dt} + \frac{d S_i}{dt} I_{i\leftarrow j},
\end{equation}
where we can substitute equations \ref{eq:mp-1} and \ref{eq:mp-2}, yielding to
\begin{equation}
 \frac{d S_i I_{i\leftarrow j}}{dt} = S_i \left( - \delta I_{i\leftarrow j} + \lambda S_j \sum_{k \in \partial j \setminus i} I_{k\leftarrow j} \right) + I_{i\leftarrow j} \left( \delta I_i - \lambda S_i \sum_{k \in \partial i} I_{k\leftarrow i} \right),
\end{equation}
which can be reorganized as
\begin{equation} \label{eq:mp-deriv}
 \frac{d S_i I_{i\leftarrow j}}{dt} = - \delta S_i I_{i\leftarrow j} + \delta I_i I_{i\leftarrow j} + \lambda S_i S_j \sum_{k \in \partial j \setminus i} I_{k\leftarrow j} - \lambda S_i I_{i\leftarrow j} \sum_{k \in \partial i} I_{k\leftarrow i}.
\end{equation}
Now analyzing term by term and finding the correspondence between equations \ref{eq:pqmf-xy} and \ref{eq:mp-deriv}, we can infer that: (i) $S_i I_{i\leftarrow j}$ corresponds to $\E{X_i Y_j}$, which is our assumption on Equation \ref{eq:mp-pqmf} (ii) $\lambda S_i S_j \sum_{k \in \partial j \setminus i} I_{k\leftarrow j}$ corresponds to $\E{\lambda \sum_{j\leftarrow k} \B_{(i\leftarrow j, j\leftarrow k)} X_i X_j Y_k}$, where the PQMF follows a different approximation if compared to the message passage approach, (iii) $\lambda S_i I_{i\leftarrow j} \sum_{k \in \partial i} I_{k\leftarrow i}$ corresponds to  $\E{\lambda \sum_{i\leftarrow k} \Hm_{(i\leftarrow j, i\leftarrow k)} X_i Y_j Y_k}$, where, again, the the PQMF and message passing formalism follow different approximations and (iv) $\delta I_i I_{i\leftarrow j}$ should correspond to $\E{Y_i Y_j}$. It is obvious from the correspondent terms that the approximations themselves are actually different from the standard pair approximation used in Equation \ref{eq:pqmf-approx}. However, the main point of this comparison is to emphasize that the term $\lambda \E{X_i Y_j}$ was neglected. Observe that in \cite{Shrestha2015}, by construction, the authors avoid the contribution of the interaction of edge $i\leftarrow j$ on $I_{i\leftarrow j}$, while it appears naturally on the summations of $\E{X_i Y_k}$. As a drawback we might mention that this formalism cannot be understood as a lower nor as an upper bound. In \cite{Shrestha2015}, it was shown that the recurrent-state message-passing performs well in many simulations, which suggests to us that neglecting this term does not drastically affect the overall result. Additionally, the authors also showed that this method performed better, compared with Monte Carlo simulations, in their experiments. However, it is important to emphasize that in their experiments just small networks were used, the Zachary's karate club, with $N=34$ and an Erd\"os and R\'enyi network with $N=100$ and average degree $\E{k} = 3$. Thus, we believe that further investigations regarding the use of this approach are needed. Concerning the precision of the critical point, in Section \ref{sec:FSS} we extensively and systematically tested and compared the analytical predictions with Monte Carlo estimations.

\subsection{Synchronous cellular automaton} \label{sec:dtmc}

Besides continuous time approaches, we can also model spreading processes as a discrete time Markov chains (DTMC). There is more than one possible epidemic/rumor spreading model that consider discrete time. Here, we discuss cellular automaton. Briefly, synchronous cellular automaton (CA) treats time as discrete and nodal states are updated synchronously. In other words,  the state of every individual in the population is updated together in a synchronous manner. In contrast, an asynchronous cellular automaton is able to update individuals independently, so that the new state of a cell affects the calculation of states in neighboring cells. Our main interest resides in the synchronous case, which will be presented for single and multilayer networks in Sections \ref{sec:dtmc-single} and \ref{sec:dtmc-mux}, respectively. Furthermore, we briefly discuss the practical aspects of the asynchronous case in Section \ref{sec:dtmc-async}, as it was widely used in the literature as a simulation method.

Importantly, in continuous time approaches $\lambda$ and $\delta$ are rates and, in this context, the probabilities are obtained after the multiplication by the time interval, i.e. $\lambda dt$ and $\delta dt$. Here, in discrete time formalisms, they assume a different role and are interpreted directly as probabilities. Thus, throughout this section and whenever we use CA formalisms, $\lambda$ and $\delta$ represent probabilities.

\subsubsection{Single-layer networks} \label{sec:dtmc-single}

The cellular automaton (CA) approach was initially proposed in \cite{Gomez2010}, where the authors consider a discrete time approach for the SIS dynamics. This method has some similarities with the formalism presented in Section \ref{sec:qmf-single}, as both consider a fixed network. However, there are some fundamental differences, such as reinfection. In other words, an individual can be cured and reinfected in the same time step. Another important point of this formalism is the synchronous behavior. This means that at every time step, each infected node intends to spread the disease, which contrasts to the continuous time, where the events occur based on exponential time. In fact, this is the most important difference between these methods. 

The cellular automaton approach also allows us to introduce a parameter that controls the activity of the individuals. This parameter drives the dynamics from a contact process (CP)\footnote{This is not a standard CP as it considers the reinfection.}, where just one contact is performed (one contact per individual, but all the infected individuals try to spread the disease), to a reactive process (RP), where every contact is performed at the same time (each infected individual tries to spread the disease to all of their neighbors). For the SIS model, we have the following discrete-time system equations
\begin{equation} \label{eq:dtmc}
 y_i(t+1) = (1-y_i(t))(1 -s_i(t)) + y_i(t)(1-\delta) + y_i(t)\delta(1- s_i(t)),
\end{equation}
where $y_i(t)$ is the probability that node $i$ is infected at time $t$, while $s_i(t)$ is the probability of node $i$ not being infected at time $t$, formally given by
\begin{equation} \label{eq:dtmc_contact}
 s_i(t) = \prod_{j=1}^N \left(1-\lambda \R_{ji} y_j(t)\right).
\end{equation}
The term $(1-y(t))$ is the probability that the node is susceptible at time $t$ and $(1 -s_i(t))$ indicates the probability of this node becoming infected by at least one of its neighbors. Analyzing our model term by term, we have that $y(t)(1-\delta)$ takes into account the recovering of an infected node. Finally, $y(t)\delta(1- s_i(t))$ indicates the probability of a node being cured and also reinfected in the same time interval. Note that one might prefer a model without reinfections. This can easily be done by just removing the term $y_i(t)\delta(1- s_i(t))$ in Equation \ref{eq:dtmc}. The authors argue that their formulation generalizes the previous approaches. Furthermore, the activity can be modeled in terms of random walkers leaving the node $i$ at each time step, hence,
\begin{equation} \label{eq:Sis_mat_R}
 \R_{ij} = 1 - \left( 1 - \frac{\W_{ij}}{\sum_j \W_{ij}} \right)^\kappa,
\end{equation}
where $\kappa$ represents the number of trials to spread the disease. Note that the extreme cases are $\kappa = 1$, which implies $\R_{ij} = \Pd_{ij}$, the probability transition matrix (see Section \ref{sec:spectral_characterization}), which represents the contact process (CP) and $\lim_{\kappa \rightarrow \infty} \R_{ij} = \A_{ij}$, that represents the reactive process (RP).

The steady state in a discrete time system implies $y_i(t+1) = y(t)$. Considering this condition in Equation \ref{eq:dtmc} we have
\begin{equation} \label{eq:dtmc_steady}
 y_i = (1 - s_i) - (1 - \delta) y_i s_i,
\end{equation}
where the time dependency is suppressed. Moreover, the prevalence can be obtained in a similar way to the QMF considering the average value as
\begin{equation}
 \rho = \frac{1}{N} \sum_{i=1}^N y_i.
\end{equation}
We assume the existence of the critical point $\tau_c$ and that the parameters $\lambda$, $\delta$ and $\kappa$ are the same for every node. The epidemic threshold can be obtained by means of a linear stability analyses neglecting second-order terms. Equation \ref{eq:dtmc_contact} is thus approximated as
\begin{equation}
 s_i(t) \approx 1 - \lambda \sum_{j=1}^N \R_{ji} y_j(t).
\end{equation}
Inserting $s_i(t)$ in \ref{eq:dtmc_steady} and considering an arbitrary small value of $0 < p_i \ll 1$, we have the system
\begin{equation}
 (\R - \tau^{-1} \I)\ \boldsymbol{p} = 0,
\end{equation}
where $\boldsymbol{p}$ is a vector. This system has non-trivial solutions only if $\tau^{-1}$ is an eigenvalue of $\R$. Hence, the lowest value of $\tau$ is
\begin{equation}
 \tau_c^{CA} = \frac{1}{\Lambda_1},
\end{equation}
where $\Lambda_1$ is the largest eigenvalue of $\R$. Note that the RP case coincides with the QMF predictions (see Section \ref{sec:qmf-single}). Moreover, in \cite{Gomez2010} the authors also evaluated this formalism over the HMF approximations in order to derive the epidemic threshold. They showed that the influence of the reinfection terms for the CP and RP cases is given by non-linear terms of spreading and recovering rates.

\subsubsection{Multilayer networks} \label{sec:dtmc-mux}

Once again, here we opt for the tensorial notation for the multilayer network modeling, firstly presented in its matrix form \cite{Cozzo:2013}, next extended to multilayers in \cite{deArruda2016}. However it is important to emphasize that this choice is merely a matter of taste. We consider the contact tensor, $\mathcal{R}^{\beta \tilde{\delta}}_{\alpha \tilde{\gamma}}$, defined on Section \ref{sec:qmf-mux}. Observe that we focus on the RP case here, but the CP can be obtained, simply changing the contact tensor. Denoting the probability of node $\beta$, in layer $\tilde{\delta}$, becoming infected at time $t$ as $y_{\beta \tilde{\delta}}(t)$, the discrete time evolution equation for this probability is described as
\begin{eqnarray} \label{eq:sis_disc}
 y_{\beta \tilde{\delta}}(t+1)&=&(1 - y_{\beta \tilde{\delta}}(t))(1 - q_{\beta \tilde{\delta}}(t))+(1 - \delta )y_{\beta \tilde{\delta}}(t)\nonumber\\
&+&\delta(1 - q_{\beta \tilde{\delta}}(t))y_{\beta \tilde{\delta}}(t),
\end{eqnarray}
where the probability that a node will not be infected by any of its neighbors at time $t$ is given as
\begin{equation}
 q_{\beta \tilde{\delta}}(t) = \prod_{\alpha} \prod_{\tilde{\gamma}} \left( 1 - \lambda \mathcal{R}^{\beta \tilde{\delta}}_{\alpha \tilde{\gamma}}(\lambda, \gamma) y_{\alpha \tilde{\gamma}} \right).
\end{equation}
Observe that in Equation \ref{eq:sis_disc}, the indices $\beta \tilde{\delta}$ are not dummy and there is no summation on it. A more formal notation would be obtained substituting $y_{\beta \tilde{\delta}}$ by $y_{\eta \tilde{\sigma}} E^{\eta \tilde{\sigma}}(\beta \tilde{\delta})$. The implicit summation has only one term different from zero, which is $y_{\beta \tilde{\delta}}$.

Finally, the macro-state variable is given by the average of the individual probabilities, similar to Equation \ref{eq:qmf-macro}
\begin{equation} \label{eq:dtmc-macro}
 \rho = \frac{1}{nm}\E{Y_{\beta \tilde{\delta}}} U^{\beta \tilde{\delta}},
\end{equation}
where $U^{\beta \tilde{\delta}}$ is the all one tensor.

It is interesting to compare the spreading of a disease on the aggregate network with the full multiplex structure. This can be done by changing the contact tensor, $\mathcal{R}^{\beta \tilde{\delta}}_{\alpha \tilde{\gamma}}$, by the tensor $P_\beta^\alpha$, defined in Equation \ref{eq:agregated}\footnote{Note that this example concerns the RP case. The CP case can be obtained after some manipulations on the tensor $P_\beta^\alpha$.}. It worth noticing that when the process is on the aggregated network, each individual chooses a layer with uniform probability, then spreads the disease to all neighbors in that layer. On the other hand, in when we account for the full multilayer, Equation \ref{eq:sis_disc}, each node can infect its neighbors in any layer. Furthermore, the critical point also changes, yielding to
\begin{equation}
 \left(\tau_c^{CA}\ \right)_{P} = \frac{1}{\Lambda_1},
\end{equation}
where $\Lambda_1$ is the largest eigenvalue of the aggregated adjacency matrix. Interestingly, applying the interlacing results (see Section \ref{sec:Spectra_interlacing}) we have
\begin{equation}
 \left(\tau_c^{CA}\ \right)_{P} \geq \tau_c^{CA}.
\end{equation}
This result implies that the spreading process in the multilayer structure is more efficient, or in the worst case as efficient as the process in the aggregate network \cite{Garcia2014}. Note that we do not conclude anything about the whole phase diagram. We just use the term efficient in terms of the critical point. This comparison was formally presented in \cite{Garcia2014, Cozzo2015}. The examples shown here exemplify the importance of considering the multilayer structure and the role of the aggregated network. A similar conclusion can be obtained using the network of layers and its interlacing properties. Finally, note that the tensor $P_\beta^\alpha$ of the aggregated network have rank 2 and might be interpreted as a weighted matrix and the framework presented in this section reduces to the single-layer case, presented in the previous section.

\subsection{Asynchronous cellular automaton} \label{sec:dtmc-async}

As mentioned before, we will briefly describe and discuss some concepts of this approach, but without going into detail. The main aim of this section is to present this scheme, which was widely used in the literature, especially in the first studies on epidemic \cite{Satorras01:PRL, Satorras2001, Moreno2003} and rumor spreading \cite{moreno2004, Borge12}. Recently, more sophisticated simulation methods were proposed, as we will discuss and compare in Section \ref{sec:Monte_Carlo}. The simulations of the cellular automata described in Sections \ref{sec:dtmc-single} and \ref{sec:dtmc-mux} will be discussed in Section \ref{sec:mc_ca}.

Let us begin with the SIS model. The simulation can be easily described in two steps. For each infected node, which can be selected sequentially or in a random order: (i) spread the disease to all its susceptible neighbors with probability $\lambda$. After that (ii) the infected node recovers and becomes susceptible again with probability $\delta$. A SIR model only differs in the second step, in which, with a probability $\delta$, the infected individual turns into recovered.

This framework has also been applied to study rumor spreading. However, it presents a couple of subtleties. Regarding the Maki--Thompson model, each spreader, which can be selected sequentially or in a random order, contacts her neighbors and, if: (i) her neighbor is ignorant then the rumor is spread with probability $\lambda$, but, if (ii) the neighbor is another spreader or a stifler she becomes a stifler with probability $\alpha$. If we stop contacting the neighbors as soon as the spreader becomes a stifler, then we have a truncated process (TP). However, if we contact just one of the neighbors, we call this a contact process (CP) \cite{Borge12}. In addition to truncated and contact processes, the only change we need to perform in order to implement the Daley Kendall model concerns the contact between two spreaders. In the MT model, just the initial spreader can become a stifler, while in the DK model both informed nodes can change their states.

We must emphasize that this simulation method captures many interesting features, for instance, it was used to study  influential spreaders in epidemic and rumor spreading in \cite{Borge12, deArruda2014}. However, it also presents a downside, it does not predict the time correctly and is mainly applicable for steady state analysis. As mentioned before, more sophisticated methods are presented on Section \ref{sec:Monte_Carlo}.

\subsection{Comments on the generalization to multilayer networks} \label{sec:generalization_multi}

It is important to observe that the use of MF and HMF approaches in the analysis of dynamical processes in multilayer networks is debatable. To begin with, let us use the HMF as an example. Observe that the HMF considers that a collection of nodes is statistically equivalent and interacts (with a  given rate) with other groups in the same way. Thus, it is impossible to distinguish between a multilayer network of a single-layer network with a community structure. In other words, the equations would be the same if we have a rate for the spreading inside and outside the communities. The same argument also applies to the MF. Going further, note that in the MF the coarse-graining would account for even less information. Obviously, these approaches might be useful in certain circumstances and are also mathematically correct, but we focus on the QMF and the PQMF, which preserve the multilayer structure.

First of all, considering the QMF and the SIS or SIR, one can obtain a multilayer model just by changing the adjacency matrix by
\begin{equation} \label{eq:A_mux_dyn}
 \A^{Supra}_{ij} (\lambda, \eta) =
 \begin{cases}
  1 \hspace{1cm} \text{if $i$ is connected to $j$ by an intra-layer edge} \\
  \frac{\eta}{\lambda} \hspace{1cm} \text{if $i$ is connected to $j$ by an inter-layer edge} \\
  0 \hspace{1cm} \text{otherwise}
 \end{cases},
\end{equation}
where $\lambda$ and $\eta$ are the intra and inter-layer spreading rates, respectively. Regarding the MT process, one might use the matrix defined in \ref{eq:A_mux_dyn} for the spreading contributions (in the terms that appear on the multiplications by $\lambda$) and another matrix, but similarly defined, for the stifling contributions (in the terms that appear in the multiplications by $\alpha$). Mathematically, we consider $\A^{Supra}_{ij} (\alpha, \nu)$, where $\nu$ is the inter-layer stifling rate. In the experiments of Section \ref{sec:model_accuracy}, we consider $\eta = \nu$.

Next, for the PQMF approach, we follow a similar path, redefining the non backtracking matrix as
\begin{equation}
 \B_{(i\rightarrow j, k\rightarrow l)}^{Supra}(\lambda, \eta) = \begin{cases}
            1 \hspace{1cm} \text{if $j = k$, $i \neq l$ and $k\rightarrow l$ is an inter-layer edge}\\
            \frac{\eta}{\lambda} \hspace{1cm} \text{if $j = k$, $i \neq l$ and $k\rightarrow l$ is an intra-layer edge}\\
            0 \hspace{1cm} \text{otherwise},
           \end{cases}
\end{equation}
where $\lambda$ and $\eta$ are the intra and inter-layer spreading rates. Besides the non backtracking matrix, we also must consider 
\begin{equation}
 \Hm_{(i\leftarrow j, k\leftarrow l)}^{Supra} = 
 \begin{cases}
  1 \hspace{1cm} \text{if $i = k$ and $j \neq l$ and $k\rightarrow l$ is an inter-layer edge} \\
  \frac{\eta}{\lambda} \hspace{1cm} \text{if $i = k$ and $j \neq l$ and $k\rightarrow l$ is an intra-layer edge} \\
  0 \hspace{1cm} \text{otherwise}.
 \end{cases}.
\end{equation}
Note that a similar comment as the one above also applies to the MT dynamics, where one might use different matrices for spreader and stifling processes.

\section{Comparison of continuous-time and cellular automaton models} \label{sec:comp}

So far we have been using a constructive and intuitive way to show the models and analyze them. In this section, we compare continuous-time and synchronous cellular automaton models. One might think that both formalisms model the same process (in terms of simulations), differing just by a proper translation of the rates into probabilities. However, this is not always true. Observe that in continuous time processes only one event takes place at a time, in fact, the probability of two events happening at the same time goes to zero quadratically as the time interval goes to zero, $\bigO{\Delta t^2}$. On the other hand, the opposite behavior is expected on the cellular automaton case, where every node performs its contacts during the time interval $\Delta t$.

Let us, denote the infinitesimal generator of the continuous Markov chain given by the QMF as \cite{Mieghem09} 
\begin{equation}
 Q_i(t) = P_i(t) - \I = \begin{bmatrix}
      -\E{q_1^i(t)} & \E{q_1^i(t)} \\
      \delta & -\delta \\
          \end{bmatrix},
\end{equation}
where
\begin{equation} \label{eq:eq1}
 \E{q_1^i(t)} = \lambda \sum_i \A_{ij} \E{Y_j}.
\end{equation}
Next, a possible discretization of the SIS process is the sampled time Markov chain \cite{Mieghem2014} of our QMF formalism, Equation \ref{eq:sis_ibmf}. The obtained equation is given as
\begin{equation} \label{eq:qmf-sampled}
 \E{Y_i(t+ \Delta t)} - \E{Y_i(t)} = -\delta \Delta t \E{Y_i} + \lambda \Delta t \sum_{k=1}^N \A_{ki} \E{X_i(t)} \E{Y_k(t)},
\end{equation}
where $\Delta t$ is the time interval and $\delta \Delta t$ is the recovering probability, while $\lambda \Delta t$ is the spreading probability. Note that this discretization coincides with the Euler method for numerical solutions of differential equations. In \cite{Mieghem2014} it was shown that for a proper value of $\Delta t < \frac{1}{\max_i q_i}$, where $q_i = \sum_{j=1, j \neq i}^{N} Q_i{ij} = - Q_i{ii} \geq 0$, the sampled time Markov chain is exact (not approximated) to the QMF model \cite{Mieghem2014}. Following this rule, the sampling rate will always be greater than the fastest possible transition rate, $\max_i q_i$.

The transition probabilities of the sampled time Markov chain can be written as 
\begin{equation} \label{eq:p-disc}
 P_i(t) = \begin{bmatrix}
      1-\E{q_1^i(t)}\Delta t & \E{q_1^i(t)}\Delta t \\
      \delta \Delta t  & 1-\delta \Delta t \\
          \end{bmatrix},
\end{equation}
where $\E{q_1^i(t)}$ is given by Equation \ref{eq:eq1}. On the other hand, the transition probability matrix of the cellular automaton can be written as
\begin{equation} \label{eq:p-ca}
  P_i^{CA}(t) = \begin{bmatrix}
      s_i(t) & (1- s_i) \\
      \delta s_i(t) & (1 - \delta) + \delta (1 - s_i(t))\\
          \end{bmatrix},
\end{equation}
where $s_i(t)$ is the probability of node $i$ not being infected at time $t$, defined in \ref{eq:dtmc_contact}. Thus, comparing Equations \ref{eq:p-disc} and \ref{eq:p-ca}, it seems that both Markov chains are different, representing different processes. Despite their differences, both were designed to represent the SIS process and follow the same set of local rules. In order to further explore their similarities, it is enlightening to analyze an approximate version of the RP case without reinfections\footnote{The CA without reinfections is described by $y_i(t+1) = (1-y_i(t))(1 - s_i(t)) + y_i(t)(1-\delta)$.}. Here it is important to consider only the first order terms in $s_i$, denoted as $\tilde{s}_i(t) \approx 1 - \lambda \sum_i \R_{ij} y_j$. Observe that this approximation was on the linear stability analysis in order to find the critical point of our model in \ref{sec:dtmc-single}. Following this approximation, the transition probability can be written as:
\begin{eqnarray} \label{eq:p-ca2}
  P_i^{CA'}(t) &=& \begin{bmatrix}
      \tilde{s}_i(t) & (1- \tilde{s}_i(t)) \\
      \delta  & (1 - \delta)\\
          \end{bmatrix} = \nonumber \\
          &=& \begin{bmatrix}
      1 - \lambda \sum_i \A_{ij} y_j & \lambda \sum_i \A_{ij} y_j \\
      \delta  & (1 - \delta)\\
          \end{bmatrix},
\end{eqnarray}
which is the same transition probability matrix in Equation \ref{eq:p-disc}. This approximated model explains why the QMF and the CA(RP) models have the same critical point. Beyond the comparison between continuous time approaches and cellular automata, we remark that the MF, the QMF and the PQMF are all course-grained versions of the same Markov chain, hence they should be understood as approximations of the same process.

In summary, following the infinitesimal generator approach of the QMF formalism, one can write the sampled time Markov chain and its transition probability matrix. By comparing this matrix with the respective matrix for the CA framework, we have shown evidence that both methods are not the same process. Despite their differences, the first-order approximation of the CA without reinfections has the same transition matrix, which explains why, even being different processes, both models predict the same critical point. Notice that our comparison does not suggest which  model is more suitable for a given application. We are just showing where they are similar and where they differ.

\section{Monte Carlo simulations} \label{sec:Monte_Carlo}

In this section, we discuss the main computational techniques used to model the processes described in Section \ref{sec:epidemic}. We will focus on multilayer networks since the single-layer scenarios are a special cases of them, i.e., a multilayer with just one layer. In what follows, the performance of the algorithms will not be our main concern, instead we present the algorithms in a didactic form.

Firstly, we describe continuous time simulations in Section \ref{sec:mc-cont}, followed by the discrete time approach, focusing on the synchronous cellular automaton case, in Section \ref{sec:mc_ca}. Next, we discuss formal and practical aspects of the quasi-stationary algorithm (QS) as a simulation method for the study of absorbing state Markov chains. To round up, we  extensively compare, in terms of accuracy, the numerical solutions of the QMF and the PQMF equations and Monte Carlo simulations, covering SIS, SIR and MT processes on top of single and multilayer structures, see Section \ref{sec:model_accuracy}.

\subsection{Continuous-time simulations: Poisson processes} \label{sec:mc-cont}

First of all, we must restate our assumptions in the continuous time formulations, but here extending to multilayer networks. The SIS dynamics is modeled associating a Poisson process to each of the elementary dynamical transitions: intra and interlayer spreading and the recovery from the infected state. The first two processes are associated with the edges of the graph and are characterized by the parameters $\lambda$ and $\eta$, respectively. The latter transition is modeled in the nodes, also via a Poisson process with parameter $\delta$. In the following, we present two statistically exact Gillespie-like algorithms \cite{Gillespie1977}.

\subsection{The standard algorithm} \label{sec:std_mc}

The method used in \cite{Ferreira2012, Mata2015} is adapted here to the case of multilayer networks. For more references on this algorithm, we refer the reader to \cite{Ferreira2016, Cota2017}. At each time step, time is incremented by $\Delta t = \frac{1}{(\delta N_i + \lambda N_k + \eta N_m)}$, where $N_i$ is the number of infected nodes, and $N_k$ and $N_m$ are the number of intra-layer and inter-layer edges emanating from them, respectively. With probability $\frac{\delta N_i}{(\delta N_i + \lambda N_k + \eta N_m)}$, one randomly chosen infected individual becomes susceptible. On the other hand, with probability $\frac{\lambda N_k}{(\delta N_i + \lambda N_k + \eta N_m)}$, one infected individual, chosen with a probability proportional to its intra-layer degree, spreads the disease to an edge chosen uniformly random. Finally, with probability $\frac{\eta N_m}{(\delta N_i + \lambda N_k + \eta N_m)}$ one infected individual, chosen with a probability proportional to its inter-layer degree, propagates the disease to an edge chosen uniformly. If an edge between two infected individuals is selected during the spreading, nothing happens, only time is incremented. The process is iterated following this set of rules, simulating the continuous process described by the SIS scenario.

It is noteworthy that a generalization from this framework to other dynamical processes is trivial. Beginning with the SIR, the procedure is basically the same, however, instead of curing the node and classifying it as susceptible again, we must label it as recovered. Once there are no more spreaders the dynamics stops. On the other hand, in order to simulate the MT process, we must adapt our framework. This process is modeled associating a Poisson process to each of the elementary dynamical transitions: intra and interlayer spreading and stifling. The first two processes are associated with the edges of the graph and are characterized by the parameters $\lambda$ and $\eta$, respectively. The latter two transitions are also modeled on the edge via a Poisson process with parameters $\alpha$ and $\nu$. Contrasting with the SIS and SIR dynamics, the annihilation mechanisms are based on the contact between individuals, requiring us to also associate these processes to the edges. We also remark that the DK dynamics would also present a similar modeling, however when the annihilation processes take place, both ends of an edge must change, as the process is undirected, at variance with the MT approach. Here in this review we only follow the MT models.

From the computational viewpoint, the MT dynamics is modeled as follows. Firstly, given an initial condition, at each time step, the time is incremented by $\Delta t = \frac{1}{\left((\lambda + \alpha)N_k + (\eta + \nu) N_m  \right)}$, where $N_k$ and $N_m$ are the number of intra-layer and inter-layer edges emanating from spreaders, respectively. With probability $\frac{\lambda N_k}{\left((\lambda + \alpha)N_k + (\eta + \nu) N_m  \right)}$ a spreader, chosen with a probability proportional to its intra-layer degree, spreads the rumor/information to an edge chosen uniformly random. On the other hand, with probability $\frac{\alpha N_k}{\left((\lambda + \alpha)N_k + (\eta + \nu) N_m  \right)}$ a spreader, chosen with a probability proportional to its intra-layer degree, can be stifled by an edge chosen uniformly at random if the contacted neighbor is another spreader or a stifler. Similar actions are taken regarding inter-layer edges, however with probabilities $\frac{\eta N_m}{\left((\lambda + \alpha)N_k + (\eta + \nu) N_m  \right)}$, for spreading and probability $\frac{\nu N_m}{\left((\lambda + \alpha)N_k + (\eta + \nu) N_m  \right)}$ for stifling processes. The next time is incremented by $\Delta t$  and this set of rules is applied until we reach the absorbing state. In other words, until there are no spreaders on the population.

\subsection{A different implementation}

\begin{figure}[t]\centering
\includegraphics[width=0.75\columnwidth]{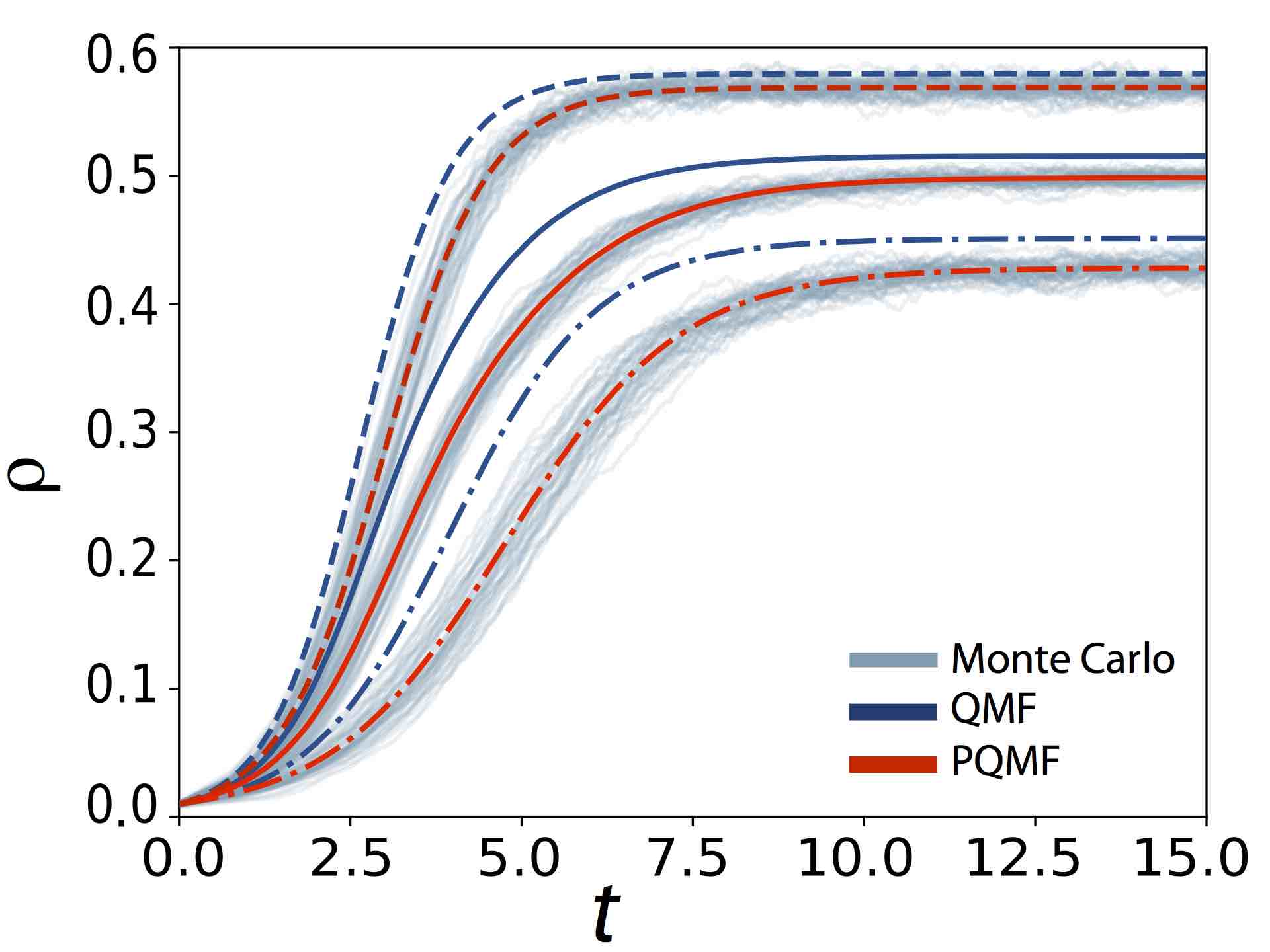}
\caption{Multiplex network composed by two Erd\"os and R\'enyi layers, $m=2$, and $n=5 \times 10^3$ nodes on each layer, where the first has $\E{k} = 16$, while the second has, $\E{k} = 12$. The effective spreading rate is $\tau = 0.15$. The continuous lines represent the first (QMF) and second (PQMF) order approximations on the whole population for the order parameter $\rho$, while the dot-dashed and dashed lines represent the average of infected individuals on the first and second layers, respectively. Each gray line represents one of the 50 independent Monte Carlo simulations.} \label{Fig:MC_ex}
\end{figure}

\begin{figure}[t]\centering
\includegraphics[width=0.65\columnwidth]{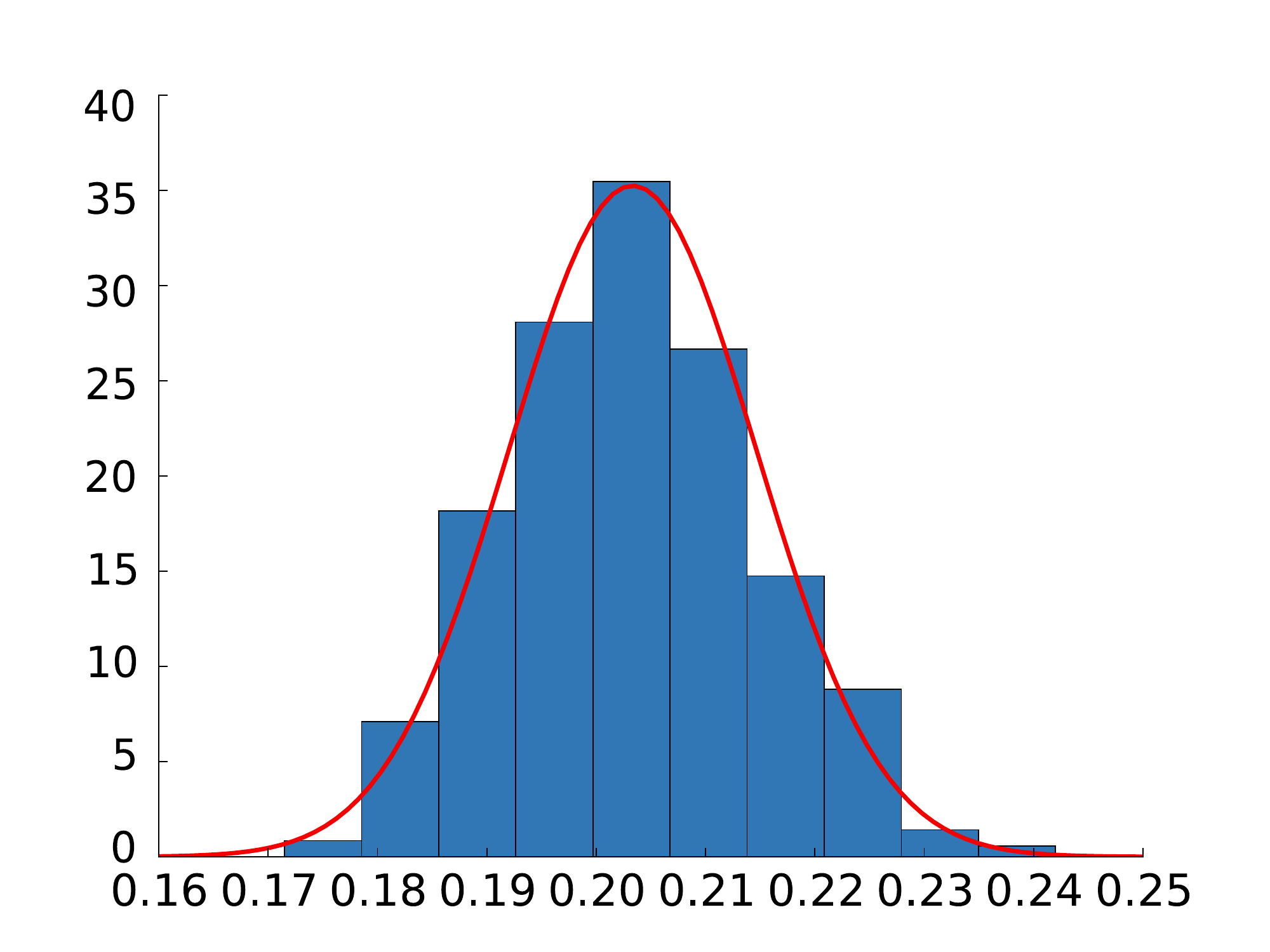}
\caption{Distribution of 500 simulations over a complete graph with $N = 2 \times 10^3$ nodes in order to validate the simulation. Comparing with the results shown in Section \ref{sec:mf} (Equations \ref{eq:mt_sigma} and \ref{eq:mt_x}) we obtained an error of $0.00014$ for the average and $0.0166$ for the variance of the simulations. Additionally, we performed the Kolmogorov--Smirnov test obtaining $D = 0.0408$, suggesting that our data is normally distributed.} \label{Fig:complete}
\end{figure}

Alternatively, instead of choosing uniform probabilities in order to decide which Poisson process we should execute, we can also do this simulation by an exponential time sampling for each process. These times refer to the next event (note that the times must be accumulative), thus choosing the smallest among all times. Considering the SIS as an example, a spreader has random exponential times associated with its edges, whose parameter is $\lambda$, $T + \Delta t$ (where $T$ is the actual time of the simulation and $\Delta t$ is the random exponential time), and an exponential random number, with parameter $\delta$, modeling the cure, similarly $T + \Delta t$. The same is reproduced for all infected individuals. Assuming that the next event regards a spreading, as the edge is considered to be directed, it is guaranteed that it emanates from a spreader, then her respective neighbour becomes infected at that time. If she was already infected, nothing happens. Next, the time is updated, $T' = T + \Delta t$. Moreover, after spreading the disease, a new time is associated, allowing the individual to spread another time on that same edge, $T' + \Delta t'$ (where $\Delta t'$ is the new exponentially sampled random number, the time for the next event). On the other hand, if the next event is associated with a recovery process, the node makes its state susceptible and all its Poisson processes are deleted and, obviously, time must be incremented, $T' = T + \Delta t$. Note that choosing a collection of exponentially distributed values, as described in the previous section, leads to the same outcome. Thus, the main advantage of this approach is to capture the time behavior of the process slightly more precisely being more intuitive, as we implement each Poisson process independently. Additionally, it makes the algorithm easier to study non-Markovian and heterogeneous processes, since we only need to change the probability distribution of the time between events -- note that Markovian processes present an exponential distribution of intervals between events.

In Figure \ref{Fig:MC_ex}, we show an example of this algorithm compared to the first, QMF, and second-order approximations, PQMF, over time. In this case we considered a multiplex network composed by two Erd\"os and R\'enyi layers, $m=2$, and $n=5 \times 10^3$ nodes on each layer, where the first has $\E{k} = 16$, while the second has, $\E{k} = 12$. We simulated a SIS with an effective spreading rate $\tau = 0.15$.

A similar approach also applies to the other dynamical processes, e.g., SIR and MT. We have to consider the rates associated with the processes and the rules that must be applied. For instance, in Figure \ref{Fig:complete}, we show the simulations over a complete graph with $N = 2 \times 10^3$ nodes in order to validate our simulations. Here we compare the results obtained over 500 simulations and Equations \ref{eq:mt_sigma} and \ref{eq:mt_x} for the theoretical expected fraction of ignorants and its variance. The errors obtained were $0.00014$ for the average and $0.0166$ for the variance of the simulations. Finally, we performed a Kolmogorov--Smirnov test obtaining $D = 0.0408$, suggesting that our data is also normally distributed, as expected by the predictions shown in Section \ref{sec:mf}. As for the computational cost, we must emphasize that this algorithm is much more costly than the standard one.

\subsection{Quasi-stationary algorithm (QS)} \label{sec:qs_def}

In this section, we first formally define the Quasi-stationary algorithm (QS), followed by some numerical examples on stars and random regular networks, which allow us to discuss important features related to the critical behavior of epidemic spreading in networks.

\subsubsection{Definition}

In this section, we explore the quasi-stationary algorithm (QS) in two steps. To begin with, we discuss some of its basic aspects, showing the main mechanism behind this algorithm. This analysis follows the approach shown in \cite{Dickman2002}. Next, from the computational point of view, we show how it is implemented and how we can use it to precisely determine the critical point on the SIS dynamics \cite{Ferreira2012, Mata2015}. Near the critical point, fluctuations are often high, driving the system to the absorbing state on finite system. This complication made the determination of such a point, not a trivial task. In order to avoid it we might use the quasi-stationary distribution, where the system is restricted to active states, but still represents the original process.

Firstly, consider a general stochastic process with an absorbing state, denoted by $\left( Y^*_t \right)_{t \geq 0}$, where $Y^*_t = 0$ is the absorbing state. The probability that the process does not fall into the absorbing state, also called survival probability, is given as \cite{Dickman2002}
\begin{equation} \label{eq:Ps}
 P(t) = \sum_{n^* \geq 1} p_{n^*}(t),
\end{equation}
where $P(t)$ is the survival probability, while $p_{n^*}(t)$ is the probability of having ${n^*}$ infected individuals (or occupied sites in a more general context) in the population. Thus, we can define the QS probability, $\bar{p}_{n^*}$ from the following condition
\begin{equation}
 p_{n^*}(t) = P(t) \bar{p}_{n^*} \hspace{1cm} {n^*} \geq 1,
\end{equation}
where $\bar{p}_{n^*}$ is time-independent, since we are taking the limit of infinity time. Note that, by construction, $\bar{p}_0 = 0$. The normalization
\begin{equation}
 \sum_{{n^*} \geq 1} \bar{p}_{n^*} = 1,
\end{equation}
also applies.

On a complete graph, our process was defined in a single variable denoting the number of infected individuals, ${n^*}$. Thus, denoting the transition rates as $W_{{m^*},{n^*}}$, where the dynamics transits from $Y^* = {n^*}$ to $Y^* = {m^*}$, for the epidemic spreading we have
\begin{equation}
\begin{cases}
  W_{{n^*}-1,{n^*}} = n^* \\
  W_{{n^*}+1,n{n^*}} = \tau n^*(N - n^*).
\end{cases}
\end{equation}
Therefore, we have the following master equation
\begin{equation} \label{eq:QS_master}
  \frac{d p_{n^*}}{dt} = W_{{n^*},{n^*}+1} p_{{n^*}+1} + W_{{n^*},{n^*}-1} p_{{n^*}-1} - W_{{n^*}-1,{n^*}} p_n - W_{{n^*}+1,{n^*}} p_{n^*},
\end{equation}
which describes $p_{n^*}$. Derivating Equation \ref{eq:Ps} and applying the relation contained in the master Equation \ref{eq:QS_master}, we get $\frac{d P}{dt} = - p_1(t)$, which under the QS can be expressed as
\begin{equation}
\frac{1}{P} \frac{d P}{dt} = - p_1(t),
\end{equation}
which leads to \cite{Dickman2002}
\begin{equation}
 \tau = \frac{1}{\bar{p}_1},
\end{equation}
allowing us to precisely obtain the critical point by means of the QS distribution. Moreover, observe that the order parameter may also be obtained through the QS distribution as $\rho^{QS} = \sum_{n \geq 1} \bar{p}_{n^*}$. Interestingly, in the thermodynamic limit $\rho^{QS}$ converges to $\rho$. Before the critical point, the distribution $\bar{p}_n$ is concentrated on a finite number of individuals, which means that just one, or a finite fraction, of individuals is infected. Thus, in the thermodynamic limit the order parameter goes to zero, $\rho^{QS} = \frac{m^*}{N} \rightarrow 0 = \rho$, where $m^*$ is a finite number.

Next, from a computational point of view, the quasi-stationary method \cite{Ferreira2012, Mata2015} restricts the dynamics to non-absorbing states and can be implemented as follows. Every time the process tries to visit an absorbing state, it is substituted by an active configuration previously visited and stored in a list with $M$ configurations, constantly updated. With a probability $p_r$ a random configuration in such a list is replaced by the actual configuration. In order to extract meaningful statistics from the quasi-stationary distribution, denoted by $\bar{p}_{n^*}$, where $n^*$ is the number of infected individuals, the system must be in the stationary state and a large number of samples must be extracted. Thus, we let the simulations run during a relaxation time $t_r$ and extract the distribution $\bar{p}_{n^*}$ during a sampling time $t_a$. The threshold can be estimated using the modified susceptibility \cite{Ferreira2012}, given by
\begin{equation}
 \chi = \frac{ \E{ (n^*)^2 } - \E{ n^* }^2 }{\E{ {n^*} }} = nm \left( \frac{ \E{ (\rho^{QS})^2 } - \E{ \rho^{QS} }^2 }{\E{ \rho^{QS} }} \right),
\end{equation}
where $\rho^{QS}$ is the quasi-stationary distribution $\bar{p}_{n^*}$. As argued in \cite{Ferreira2012, Mata2015} the susceptibility presents a peak at the phase transition in finite systems. This measure is the coefficient of variation of the temporal distribution of states over time in the steady state. Note that the magnitude of the susceptibility $\chi$ is not of primary interest to us, but rather the position of its maximum value with respect to $\lambda/\delta$, as it will give the value of critical threshold for sufficiently large systems.

\subsubsection{QS evaluation: a complete graph analysis and comments}

\begin{figure*}
\includegraphics[width=0.96\textwidth]{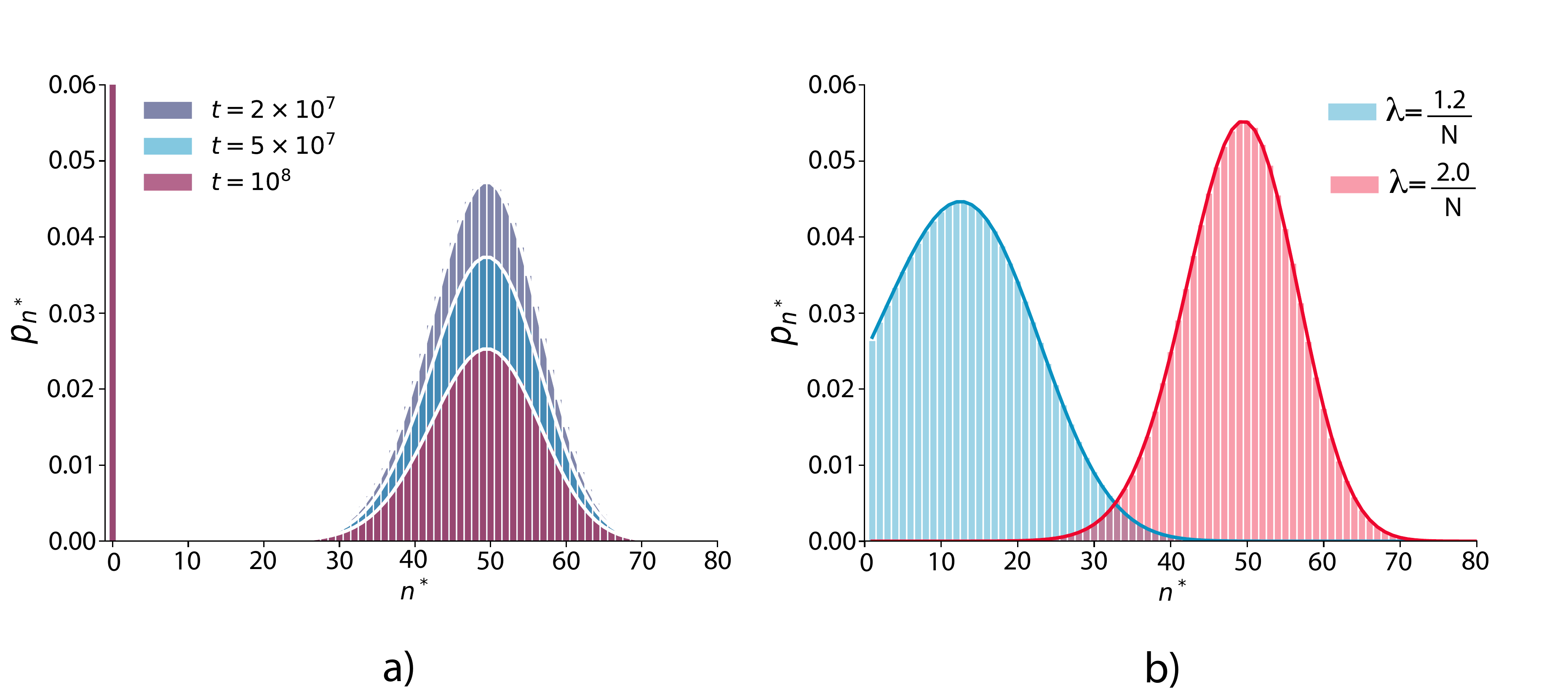}
\caption{We show in (a) the solution of the set of equations \ref{eq:QS_master} considering different times and not restricting the system to active states (in other words, we allow the system to achieve the absorbing state). In this case we used $\lambda = \frac{2.0}{N}$. In (b) we restrict the set of equations \ref{eq:QS_master} to non-absorbing states and show the steady state solution (at $t = 10^4$), represented by the continuous lines, and compare with the QS distribution, represented by the bars of the histogram. The parameters used in the QS algorithm are $t_r = 10^4$ and $t_a = 10^5$ for $\lambda = \frac{1.2}{N}$ and $t_a = 10^3$ for $\lambda = \frac{2.0}{N}$. In both figures the underlying network is a complete graph with $N = 100$ and the recovery rate is $\delta = 1$.} \label{fig:qs_complete}
\end{figure*}

In this section we present the analysis for the complete graph, which is very convenient since we can obtain an exact numerical solution allowing us to compare with simulations. As previously mentioned, the main reason to use the QS algorithm is to avoid the absorbing state, which is often visited when the spreading rate is close to the critical point. Although less often, it is also possible to get to the absorbing state above the critical point. In this case, three variables play an important role: (i) the ratio between the spreading and recovery rates, $\tau$, (ii) time and (iii) the system size. Thus, the larger $\tau$ less likely to visit the absorbing state. More importantly,  the larger the system, the less likely is to get trapped in the absorbing state. In fact, the notion of a phase transition and, consequently the definition of a critical point, only applies in the thermodynamic limit, $N \rightarrow \infty$. In practical terms, all systems are finite and, specially if we are studying the critical behavior, we must consider the finite size effects. In this context, the QS algorithm acts restricting the system to only active states, thus providing reliable information about the dynamics on top of a given structure. In this spirit, we use a rather small system, with $N =100$ nodes, in order to emphasize those finite size effects.

In order to exemplify the time effects on the probability distribution of infected nodes, in Figure \ref{fig:qs_complete} (a) we show the solution of the set of equations \ref{eq:QS_master} for different times. Note that we do not restrict the system to active states and allow the system to visit the absorbing state. It is evident that as we increase time, the probability to reach the absorbing state, $n^* = 0$, also increases. Consequently, the bulk of the distribution also changes, but the average of the distribution does not. This example is instructive because it also allows to discuss a fundamental concept behind epidemic spreading in finite systems, the meta-state. There are many different configurations for which the fraction of infected nodes is the same and the system fluctuates around this average value. In the Markov sense the average value is not an absorbing state, but the dynamics is ``traped'' around this average value, as seen in Figure \ref{fig:qs_complete} (a).

Next, in order to avoid finite-size effects we use the QS algorithm, restricting ourselves to non-absorbing states, $n^* > 0$. Analytically we can solve Equation \ref{eq:QS_master} for a long time (in order to achieve the steady state), then we define $p_{n^*} = 0$ and re-normalize the probabilities in order to obtain $\sum_{n^*} p_{n^*} = 1$. Since we are dealing with the complete graph the obtained results are exact and can be compared directly with the QS probabilities. This comparison is shown in \ref{fig:qs_complete} (b). Note that the agreement is remarkable, but also expected. Interestingly, it is also worth comment that the closer to the critical point, the larger the variance of the curves is, as can be seen comparing the distributions for $\lambda = \frac{1.2}{N}$ and $\lambda = \frac{2.0}{N}$. We have chosen this set of parameters in order to allow the reader to compare our curves with the original ones (Figure 1 on in \cite{Dickman2002}). Finally, we emphasize that in \cite{Dickman2002} the authors consider a contact process\footnote{The contact process (CP) is similar to the SIS. The CP is often considered as a process where each node has a different spreading rate, given as $\lambda_i = \frac{\lambda}{k_i}$, but every node has the same recovery rate $\delta_i = \delta$. This process shares some similarities with the SIS model, but should present different dynamical behavior, specially in heterogeneous structures. Note that this is not the same concept used in the Cellular automata context.} instead of the traditional SIS. However, on the complete graph both processes are the same, up to a normalization factor of the spreading rate.

\subsubsection{Example: a numerical evaluation}

\begin{figure*}[!t]
\includegraphics[width=\textwidth]{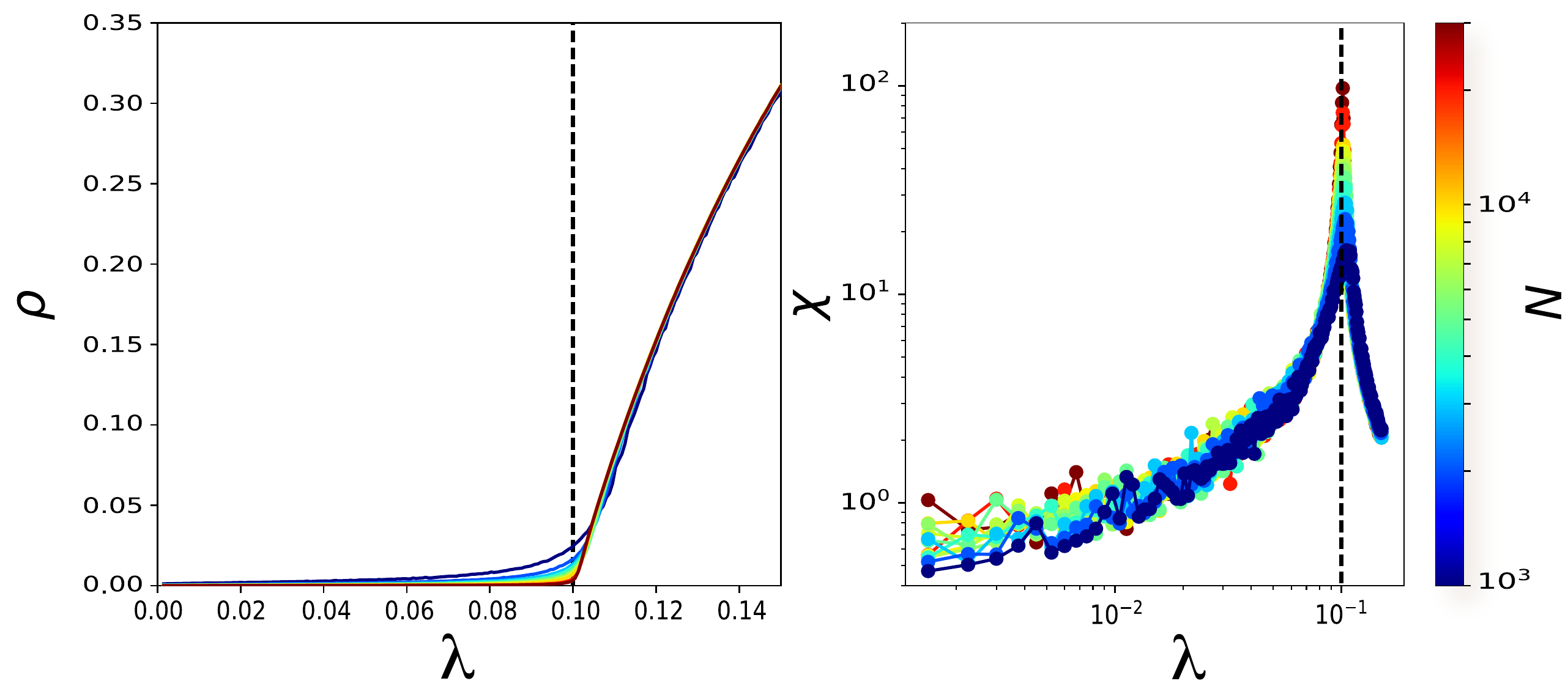}
\caption{Finite size analysis for a Erd\"os--R\'enyi network with $\E{k} = 10$ from $N = 10^3$ to $N = 3 \times 10^4$ considering the fraction of infected individuals (left panel) and susceptibility (right panel) as a function of the spreading rate $\lambda$ and recovering rate $\delta = 1$. The dashed lines represent the QMF predictions for the critical point.} \label{fig:RRN_QS}        
\end{figure*}

To complement the previous section, here we show a practical application of the QS algorithm in two paradigmatic examples: (i) Erd\"os--R\'enyi (for more, see Section \ref{sec:er}) and (ii) a star graph. The first is a completely homogeneous network, where every node has, on average, the same connectivity, $\E{k_i} = k$, for $k_i = 1, 2, ..., N$, while the star is a heterogeneous case, where the central node has $N$ neighbors, while the leaves have only one connection. Despite its simplicity, this model was previously used in \cite{deArruda2014} as an approximation for hubs in scale-free networks. In \cite{Mata2013} the authors used these networks to discuss the accuracy of the PQMF predictions of the critical point.

\begin{figure*}[!t]
\includegraphics[width=\textwidth]{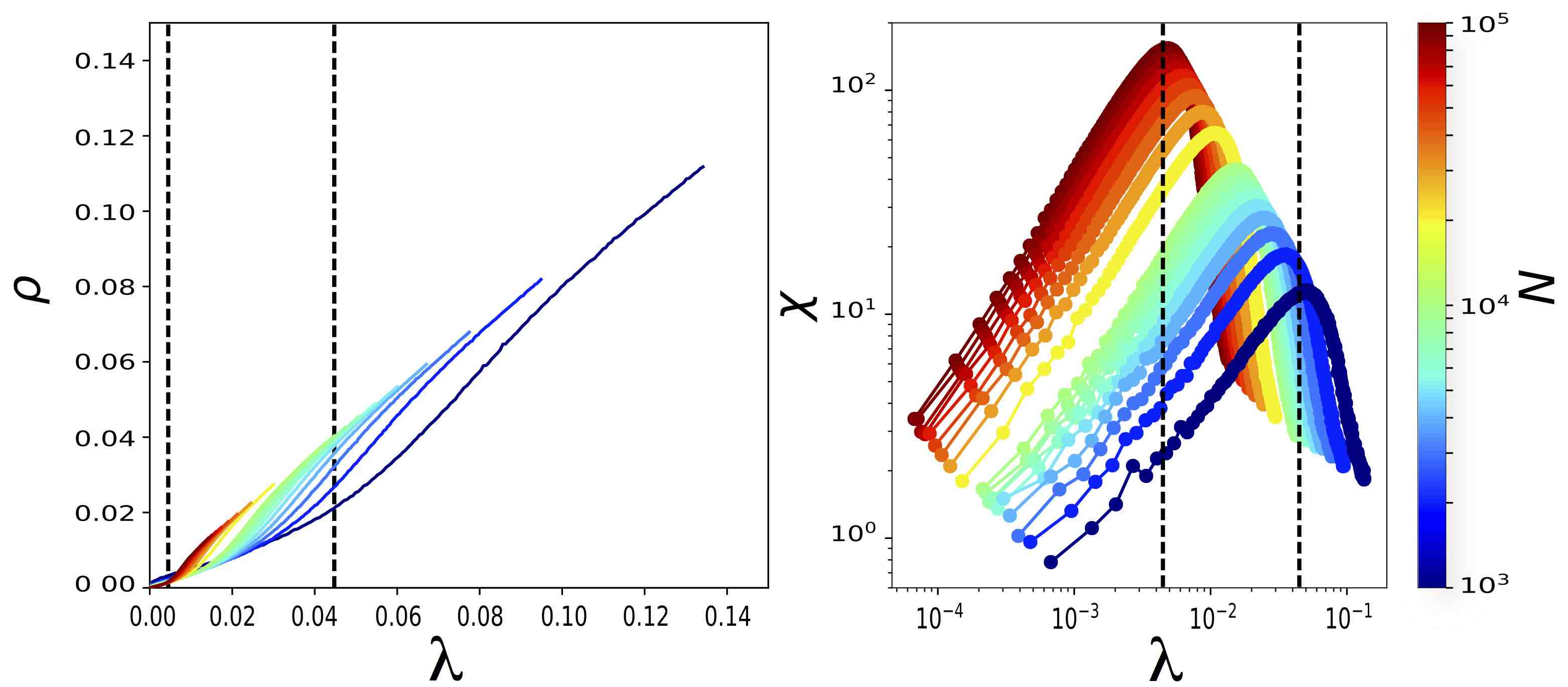}
\caption{Finite size analysis for a star from $N = 10^3$ to $N = 10^5$ considering the fraction of infected individuals (left panel) and susceptibility (right panel) as a function of the spreading rate $\lambda$ and recovering rate $\delta = 1$. The dashed lines represent the PQMF predictions for the critical point.} \label{fig:Star_QS}        
\end{figure*}

Figure \ref{fig:RRN_QS} shows the fraction of infected individuals and their respective susceptibility as a function of the spreading rate and size for ERs. As expected by the mean field approaches, the critical point remains fixed as we change the system size. Additionally, we also observe that the peak of susceptibility also scales with the system size, suggesting that it diverges on the thermodynamic limit. Consequently, these results suggest a second order phase transition. Similar results are also expected for random regular networks and complete graphs. We also show the predictions of the QMF for the critical point, which seem to be reasonable in this experiment. Note that in reference \cite{Mata2013}, the authors compared the QMF and PQMF predictions in RRNs, showing that the PQMF is more accurate than the QMF in this class of networks. This is indeed true and verified in our experiments (results not shown). However, they conducted their experiments in sparse networks, $\E{k} = 6$ \cite{Mata2013}, while we used a denser network, $\E{k} = 10$. Finally, it should be mentioned that the denser the networks, the more precise QMF is. For the sake of our discussion, observe that $\left( \tau^{QMF}_{RRN} \right)_c \approx \frac{1}{k}$, while $\left( \tau^{PQMF}_{RRN} \right)_c \approx \frac{1}{k-1}$.

Figure \ref{fig:Star_QS} shows the fractions of infected individuals and their respective susceptibility as a function of the spreading rate and size for the stars. In contrast to the results for the homogeneous case, here the critical point tends to zero as the system size goes to infinity. Regarding the critical point predictions, in the star graph the QMF approach does not present an accurate prediction, confirming the results of \cite{Mata2013}. In Figure \ref{fig:Star_QS}, the dashed lines present the PQMF prediction, $\left( \tau^{PQMF}_{Star} \right)_c \approx \sqrt{\frac{2}{N-1}}$. For the sake of comparison, the QMF prediction is given as $\left( \tau^{QMF}_{Star} \right)_c \approx \frac{1}{\sqrt{N-1}}$. It is interesting to observe that both formulas present a different behavior as a function of $N$ \cite{Mata2013}.

Finally, it is worth mentioning that more in-depth and proper evaluations of the accuracy of the critical point predictions in power-law networks were performed in \cite{Ferreira2012}, where the authors compared numerical and theoretical results for the HMF and QMF approaches.

\subsection{Discrete-time simulations: Cellular automaton} \label{sec:mc_ca}

In addition to the continuous time algorithms, we next discuss the particularities of the cellular automaton formalism. As shown in Section \ref{sec:comp}, the most traditional approaches in epidemic spreading modeling, the continuous time and the cellular automaton, represent different Markov chains and therefore different processes. Although both follow the same set of local rules, their formal process differs. Thus, we must also provide a Monte Carlo simulation that corresponds to the cellular automaton formalism. Here we focus on the synchronous case. 

Let us first discuss the SIS dynamics. Denoting by $Y_i(t)$ the state of node $i$ at time $t$, we have that $Y_i(t) = 1$ if node $i$ is infected and $0$ otherwise. Thus, starting from an initial condition $S_i(0)$ for all $i=1, \ldots, N$, the algorithm is described in four steps. At each time step, $t$: (i) the state of every node is copied, $Y_i(t+1) = Y(t)$, next (ii) every infected individual at time $t$, $\{ i | Y_i(t) = 1 \}$, is recovered with probability $\delta$, yielding $Y_i(t+1) = 0$, then (iii) every infected individual at time $t$, $\{ i | Y_i(t) = 1 \}$, will contact all its neighbors, denoted by $j$ (considering the neighbor status at time $t+1$, which allows  reinfection\footnote{Note that we are not changing the state of the node in the same time window. Thus, we are allowing reinfections. If we do not allow reinfection we should recover after the spreading processes, following a similar sequential algorithm. It is also interesting to mention that, without reinfection, just the state vector is necessary and the actions can be made in-place.}), in the RP or to a single random neighbor if in the CP case, spreading the disease to each susceptible neighbor contacted with probability $\lambda$, i.e., then $Y_j(t+1) = 1$; and (iv) time is incremented $t' = t+1$. The main difference between the SIS and a SIR process is that in the former case the infected individuals will become susceptible again, while in the latter they will be recovered and removed from the dynamics. These processes must be iterated for a fixed number of steps, $T_{max}$, or if there are no more infected individuals (an absorbing state).

\begin{figure*}
\includegraphics[width=0.96\textwidth]{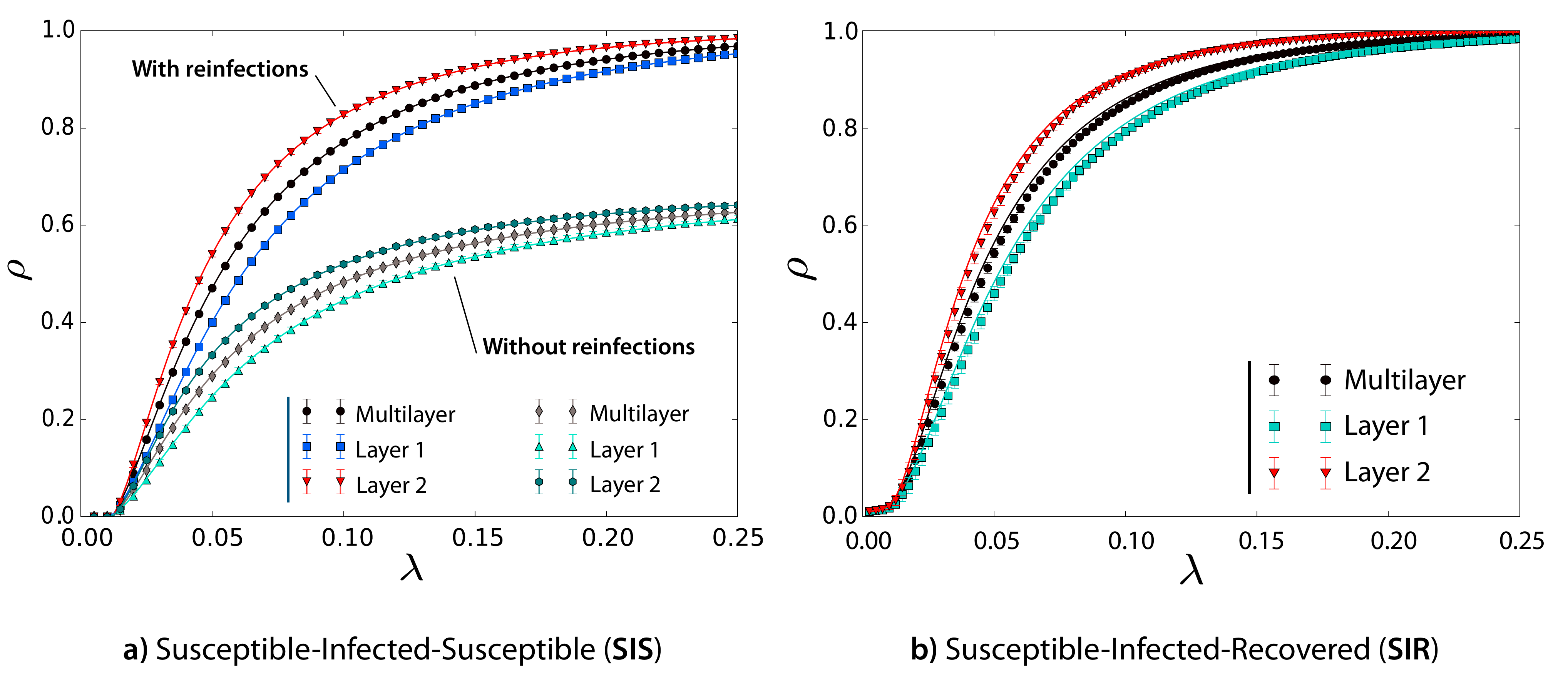}
\caption{Example of discrete Monte Carlo simulations. Phase diagrams for epidemic spreading considering the CA(RP) formalism. In (a) the SIS model, considering two cases, with or without reinfections at the same time step. In (b) the SIR model. The continuous lines are the analytical solutions, while the symbols are an average of $10^2$ the Monte Carlo simulation, $\eta=0.5$ fixed in all simulations. The multiplex considered is composed by two scale-free networks, $P(k) \sim k^{-\zeta}$, the first with $\zeta \approx 2.8$, while the second with $\zeta \approx 3.0$. The variance is the size of the symbols.} \label{fig:mc_disc}
\end{figure*}

Figure \ref{fig:mc_disc} shows three examples of phase diagrams for our algorithm, SIS with and without reinfection and SIR. We also show the comparison between the simulations results and the mathematical formalism presented in Section \ref{sec:dtmc}. Further experiments concerning the accuracy of these models are presented in Section \ref{sec:model_accuracy}.

Regarding the comparison between the synchronous and asynchronous cases, there is no proof that they would converge to the same results in specific cases. In fact, there are a couple of examples where both show different outcomes, as shown in \cite{Cornforth2005}. However, we observe that the asynchronous case is similar for the SIR case, as the removed nodes do not play a role in the dynamics. We remark that the time evolution of both approaches is different, but they lead to the same final fraction of infected individuals. Note also that asynchronous processes often perform first the spreading and then the recovery (see the description in Section \ref{sec:dtmc-async}), while here we opted for the opposite since the update is always synchronous and, in principle, the order would not change the outcome. Note that $Y_i(t+1)$ never affects $Y_i(t)$. However, curing first allows us to easily model the reinfections with no further steps in our algorithm.

\subsection{A systematic accuracy evaluation} \label{sec:model_accuracy}

Similarly to \cite{Gleeson2012}, where the author evaluates the accuracy of many mean field models, here we also perform an accuracy evaluation. Specifically, we focus on the QMF and the PQMF approaches, since it was already shown that the QMF is qualitatively better than the HMF \cite{Ferreira2012}, providing a better estimate of the critical point. Along with this section, we perform an extensive analysis regarding three models: SIS, SIR and MT. Moreover, we evaluate different structures, including single and multilayer networks, with and without correlations and triangles. Firstly, in Section \ref{sec:acc_single}, we study the accuracy of these models in single layers. Secondly, we use the previously analyzed networks to build 2-Layer multiplex networks. Then, in Section \ref{sec:acc_mux}, based on those structures we evaluate their accuracy while changing the coupling parameter.

\subsubsection{Single-layer analyses} \label{sec:acc_single}

\begin{figure*}[t]
\begin{center}
\includegraphics[width=\textwidth]{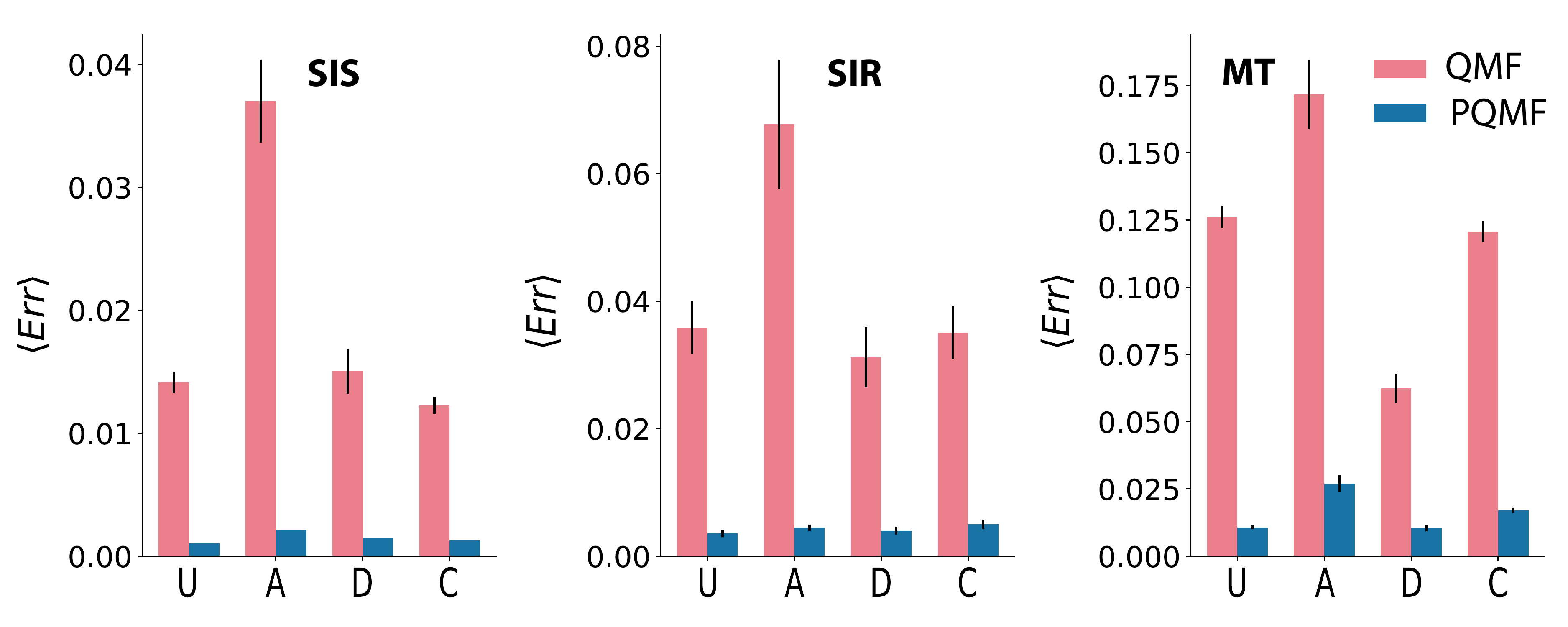}
\caption{Estimated errors in the bars and their standard error in the deviations. From left to right, we present the error of the SIS, SIR and MT processes with respect to the QMF and PQMF mean field approaches in four different structures, uncorrelated, assortative, disassortative and a clustered network. The networks have $N=10^4$, see text for more details on their structural properties.} \label{Fig:MC_Num_Err_single}
\end{center}
\end{figure*}

In order to compare the accuracy of different mean field approximations, we extract the phase diagram, $(\rho, \lambda)$ (with $\delta = 1$, hence $\tau = \lambda$), from the mean field approach and compare it with Monte Carlo simulations. The error is quantified/estimated using the average of the absolute error among the Monte Carlo runs, 
\begin{equation}
 Err_i = \frac{\sum_i \left| \rho^{MC}_i - \rho^{MF}\right|}{n_{\text{exp}}},
\end{equation}
where $n_{\text{exp}}$ is the number simulations performed, $\rho^{MC}_i$ is the result of one Monte Carlo simulation and $\rho^{MF}$ is the prediction of the corresponding mean field approach. The standard error\footnote{Note that we are estimating the error based on independent samples of it. The formula of $SE$ is derived from our knowledge about the variance of a sum of independent random variables. Now, suppose that $X_1, X_2 , \ldots, X_{n_p}$ are $n_p$ independent observations from a population that has a mean $\mu$ and standard deviation $\sigma$, then the variance of the total $T = (X_1 + X_2 + \cdots + X_{n_p})$ is $n_p\sigma^2$. Hence, the variance of $T/{n_p}$ is given as $\frac{1}{{n_p}^2}n_p\sigma^2=\frac{\sigma^2}{n_p}$, thus its standard deviation is $\sigma/{\sqrt{n_p}}$.} is calculated as
\begin{equation}
 SE = \frac{s}{\sqrt{n_{\text{points}}}},
\end{equation}
where $n_{\text{points}}$ is the number of points in the phase diagram $(\rho, \lambda)$ and $s$ is the sample standard deviation of $Err_i$. In these experiments we performed $50$ independent Monte Carlo runs for each $\lambda$ from $\lambda = 0.01$ to $\lambda = 0.49$ in steps of $\lambda = 0.01$. The simulations and numerical experiments were executed until $t_{max} = 10^2$. Additionally, aiming to reduce the noise on the curve of the SIS, we consider a temporal average on the meta-state. Formally,
\begin{equation}
 \rho^{MC} = \frac{1}{t_s} \int_{t_s}^{t_{max}} \rho(t) \text{d}t = \frac{1}{t_s} \sum_{t \geq t_s}^{t \leq t_{max}} \rho(t) \Delta t,
\end{equation}
where $t_s$ is the sampling time, which was assumed to be $t_s = 50$ and on the discrete equation $\rho(t) \Delta t$ is interpreted as the macro-state times the time it was present in the meta-state. In words, $\rho^{MC}$ is the temporal average of the states visited in the meta-state, observing that they are properly weighted by the time the system spent on each state. This can be done because we know that there is a stochastic fluctuation over the average value, which is the value we are aiming to capture by the mean field methods. Note that it does not apply to the SIR and MT dynamics due to the nature of their absorbing states.

From the structural point of view, in this subsection we focused on four scale-free networks: (i) scale free uncorrelated network with $P(k) \sim k^{-2.7}$, $N = 10^4$ and $\E{k} \approx 7.85$, assortativity and clustering coefficient close to zero, denoted as (U), (ii) correlated scale free network with $P(k) \sim k^{-2.7}$, $N = 10^4$ and $\E{k} \approx 7.85$, disassortative with $\rho^P = -0.39$, and clustering coefficient close to zero, denoted by (D), (iii) correlated scale free network with $P(k) \sim k^{-2.7}$, $N = 10^4$ and $\E{k} \approx 7.85$, assortative with $\rho^P = 0.30$, and clustering coefficient close to zero, denoted by (A) and (iv) scale free network with clustering coefficient \cite{Holme2002}, $N = 10^4$ and $\E{k} \approx 8$, assortativity coefficient $\rho^P = -0.07$ and clustering coefficient $\E{ cc } = 0.215$, denoted by (C).

In Figure \ref{Fig:MC_Num_Err_single}, we summarize our experiments in terms of the errors. The observed errors are consistently small for the second-order approximation, PQMF, as expected. It is arguable that the first order approximation, QMF, is a good approximation in most of the cases. However, in the MT dynamics, it predicts the process' dynamics poorly, exhibiting a huge error. Regarding the structural correlations, we also observed assortative structures seem to pose a bigger challenge for the approximations, even when compared with disassortative and networks with a high clustering coefficient.

It is also important to comment on both approaches' computational cost. Considering a graph with $N$ nodes and $M$ edges, the QMF can be solved using $\bigO{N}$ equations, while the PQMF, $\bigO{N + M}$\footnote{Note that the number of equations are $N$ (QMF) and $N + M$ (PQMF) for the SIS and $2N$ (QMF) and $2N + 2M$ (PQMF) for SIR and MT.}. If this graph is dense, it also implies that $M \in \bigO{N^2}$. Thus, in dense graphs, the PQMF has a huge computational cost. However, luckily, in dense networks, the QMF is expected to perform better, since the dynamical correlations tend to vanish in a complete graph on the thermodynamic limit. In other words, for a large and dense enough graph the dynamical correlations are expected to be smaller and the approximations $\E{XY} \approx \E{X} \E{Y}$ become more accurate as we approach the thermodynamic limit \cite{Mieghem09}.

\subsubsection{Multilayer analyses} \label{sec:acc_mux}

\begin{figure*}[t]
\includegraphics[width=0.97\textwidth]{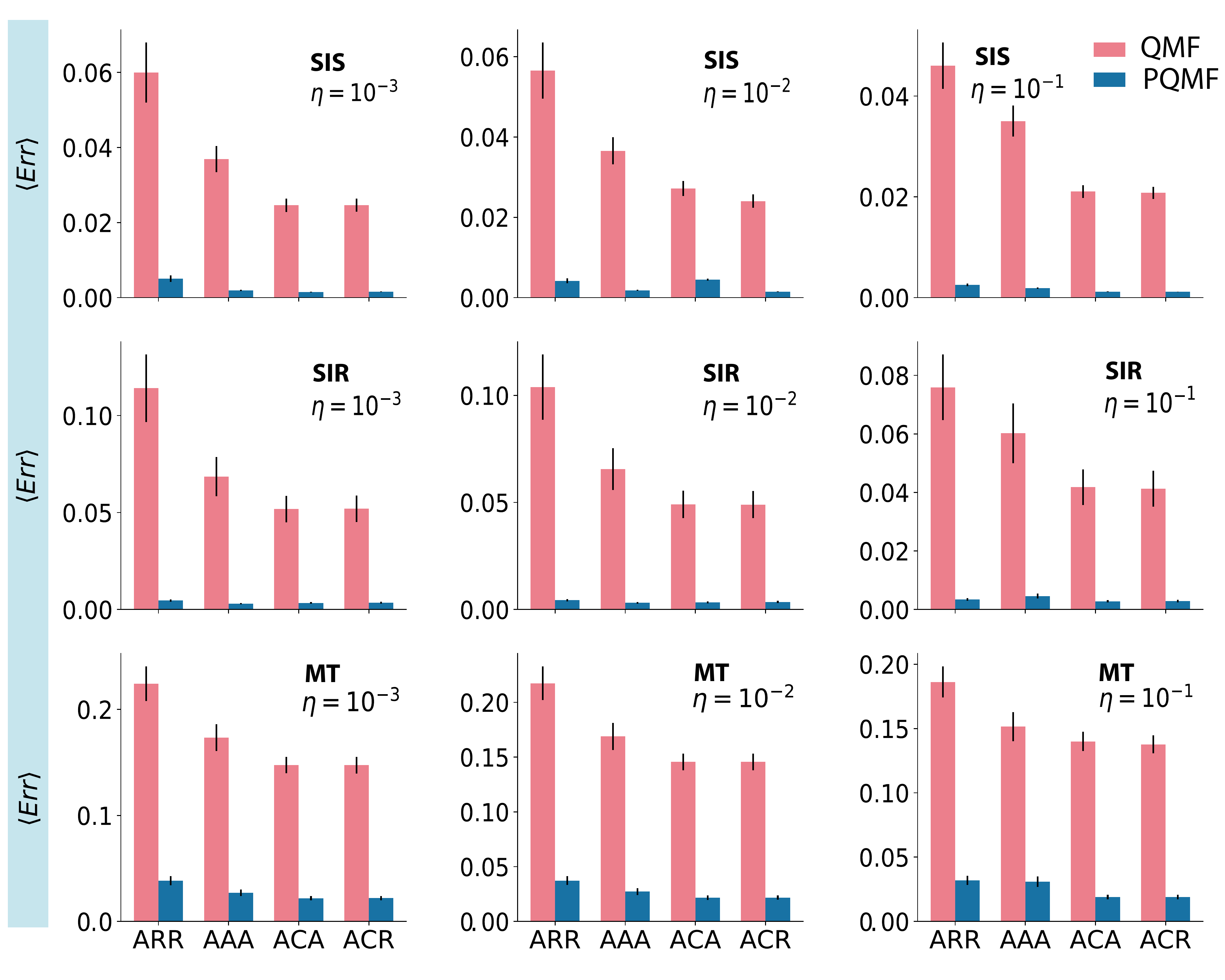}
\caption{Estimated errors in the bars and their standard errors in the deviations. From top to bottom, we present the error of the SIS, SIR and MT processes with respect to the QMF and PQMF mean field approaches in four different structures: ARR (Assortative+RRN -- randomly coupled), AAA (Assortative+Assortative -- Assortativelly coupled), ACA (Assortative+Clustered -- Assortatively coupled) and ACR (Assortative+Clustered -- randomly coupled). From left to right, we change the coupling parameter $\eta = 10^{-3}$, $\eta = 10^{-2}$, $\eta = 10^{-1}$. The networks have $n=10^4$ and $m=2$, see text for more details on their structural properties.} \label{Fig:MC_Num_Err_mux}
\end{figure*}

\begin{figure*}[t]
\includegraphics[width=0.97\textwidth]{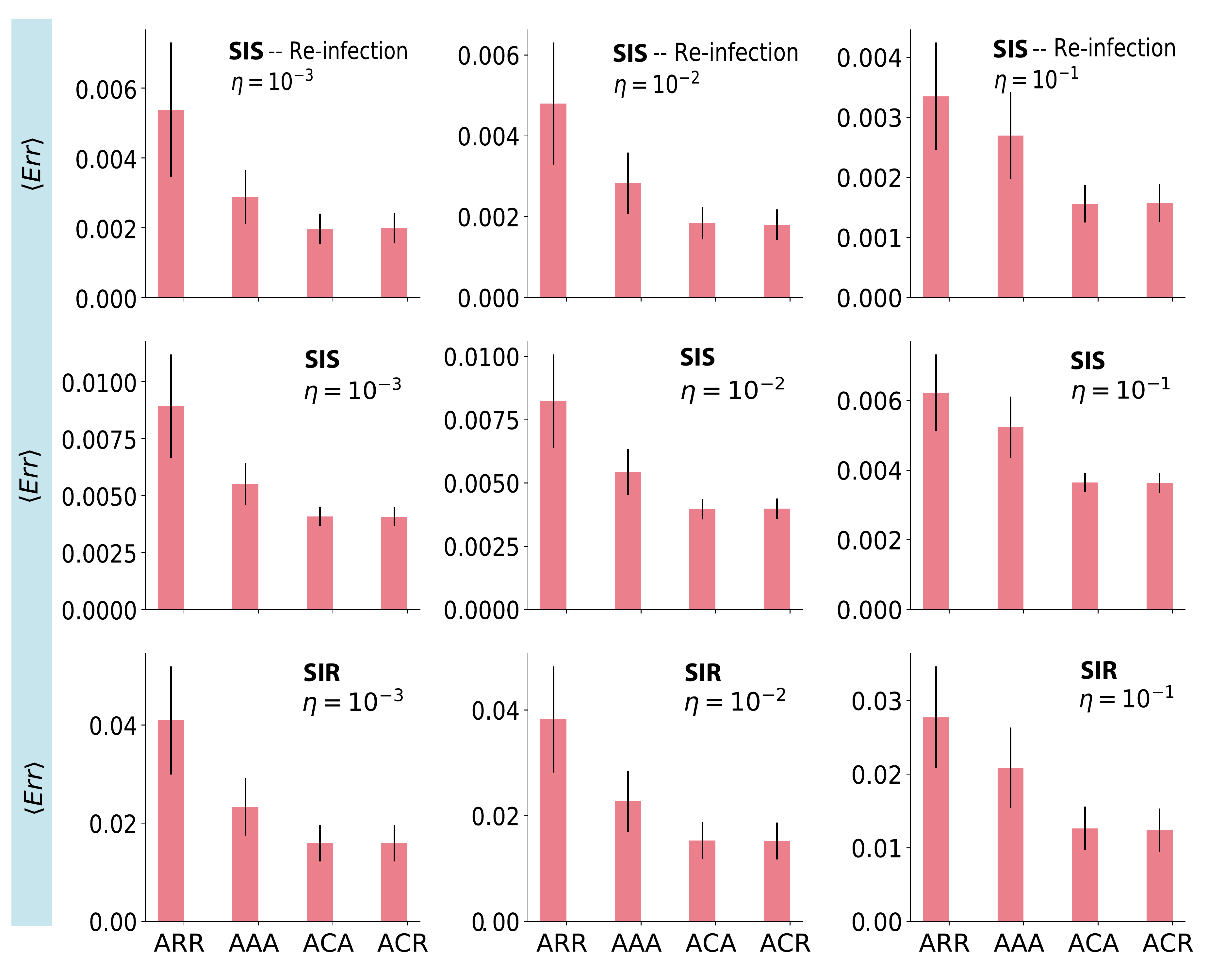}
\caption{Estimated errors in the bars and their standard errors in the deviations. From above to bottom, we present the error of the SIS with reinfections, SIS and SIR processes with respect to the CA approach in four different structures: ARR (Assortative+RRN -- randomly coupled), AAA (Assortative+Assortative -- Assortativelly coupled), ACA (Assortative+Clustered -- Assortatively coupled) and ACR (Assortative+Clustered -- randomly coupled). From left to right, we change the coupling parameter $\eta = 10^{-3}$, $\eta = 10^{-2}$, $\eta = 10^{-1}$. The networks have $n=10^4$ and $m=2$, see text for more details on their structural properties.} \label{Fig:MC_Num_Err_CA}
\end{figure*}

Extending the results of the previous section to the multilayer framework, we are able to quantify the accuracy of mean field approaches in this structure. In addition to the networks previously evaluated, we also considered the RRN (for more, see Section \ref{sec:models}), since it presents a well defined critical point. Besides, since every node in such a network has the same degree, this network may also present some dynamical correlations too, being a good proxy for our experiments. Thus, we tested our methods in four different multiplex networks: (i) ARR (Assortative+RRN -- randomly coupled), (ii) AAA (Assortative+Assortative -- Assortatively coupled), (iii) ACA (Assortative+Clustered -- Assortatively coupled) and (iv) ACR (Assortative+Clustered -- randomly coupled).

Figure \ref{Fig:MC_Num_Err_mux} shows the results for the comparison of Monte Carlo simulations and the QMF and PQMF mean field approaches in terms of the errors. Again, the errors observed are consistently small for the second-order approximation, PQMF, as expected. We also note that this observation suggests that one of the main causes of this error is the existence of dynamical correlations that are neglected by the first order approximation. Also similarly to the single layer case, here it is arguable that the QMF can be considered a good approximation in most of the cases. Obviously, it always depends on the application, but in general, the QMF can qualitatively predict the behavior of our dynamics. Moreover, it has the advantage of being an upper bound for the average value. However, in the MT dynamics, it predicts the dynamics poorly, showing a huge error as for the layer case. In this case, the most appropriate choice is the PQMF. Interestingly, we do not observe a big difference while changing the coupling parameter $\eta$.

Regarding the structural correlations, we also observed that random regular and assortative structures seem to pose the biggest challenge for the approximations. It is interesting to mention that regular networks (lattices) are still a big challenge for predictions based mean on field approaches due to their long shortest paths. Intuitively, the state of nodes in lattices presents plenty of dynamical correlations, especially in sparse cases.

\begin{figure*}[!t]
\includegraphics[width=0.96\textwidth]{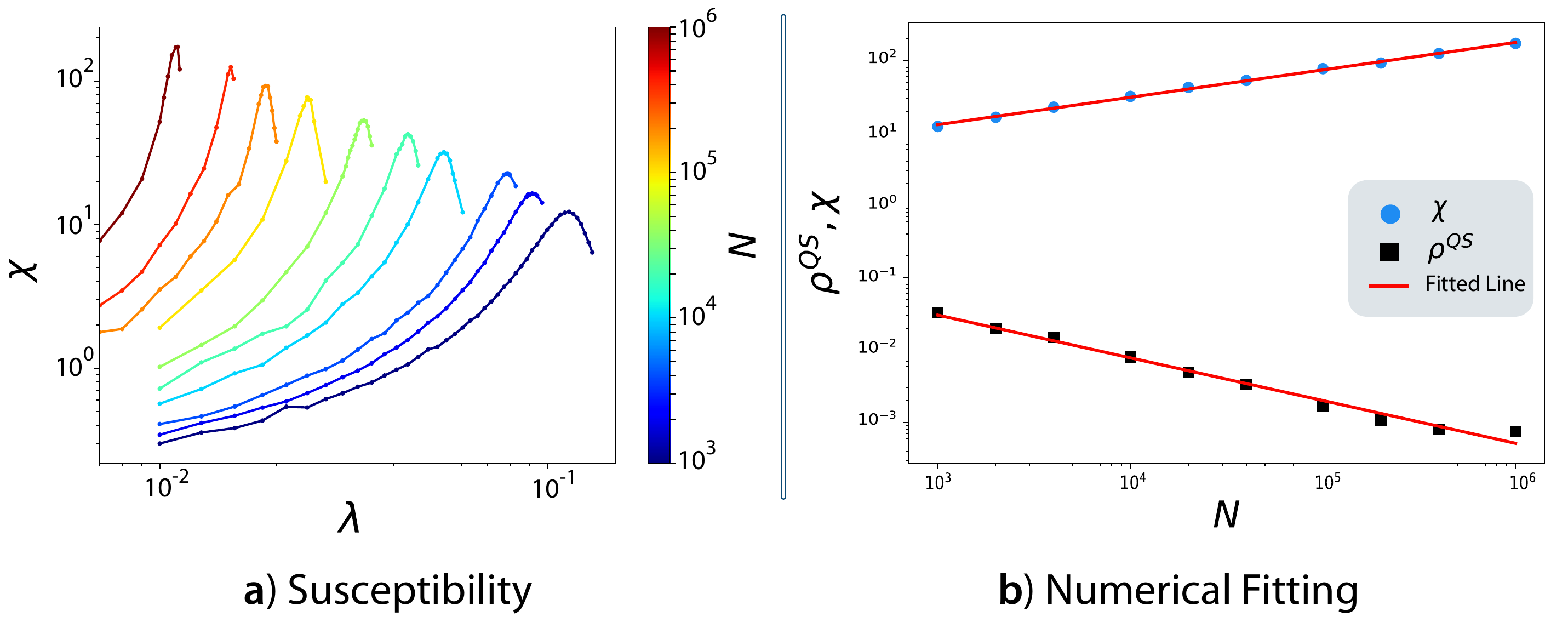}
\caption{Finite size scaling for a power-law network with $\Prob(k) \sim k^{-2.25}$. In (a) the susceptibility as a function of the spreading rate $\lambda$ with $\delta =1$, while in (b) the scaling of the susceptibility peak and its respective fraction of infected individuals, both obtained using the QS algorithm.} \label{fig:FSS_225}
\end{figure*}

\begin{figure*}[!t]
\includegraphics[width=0.96\textwidth]{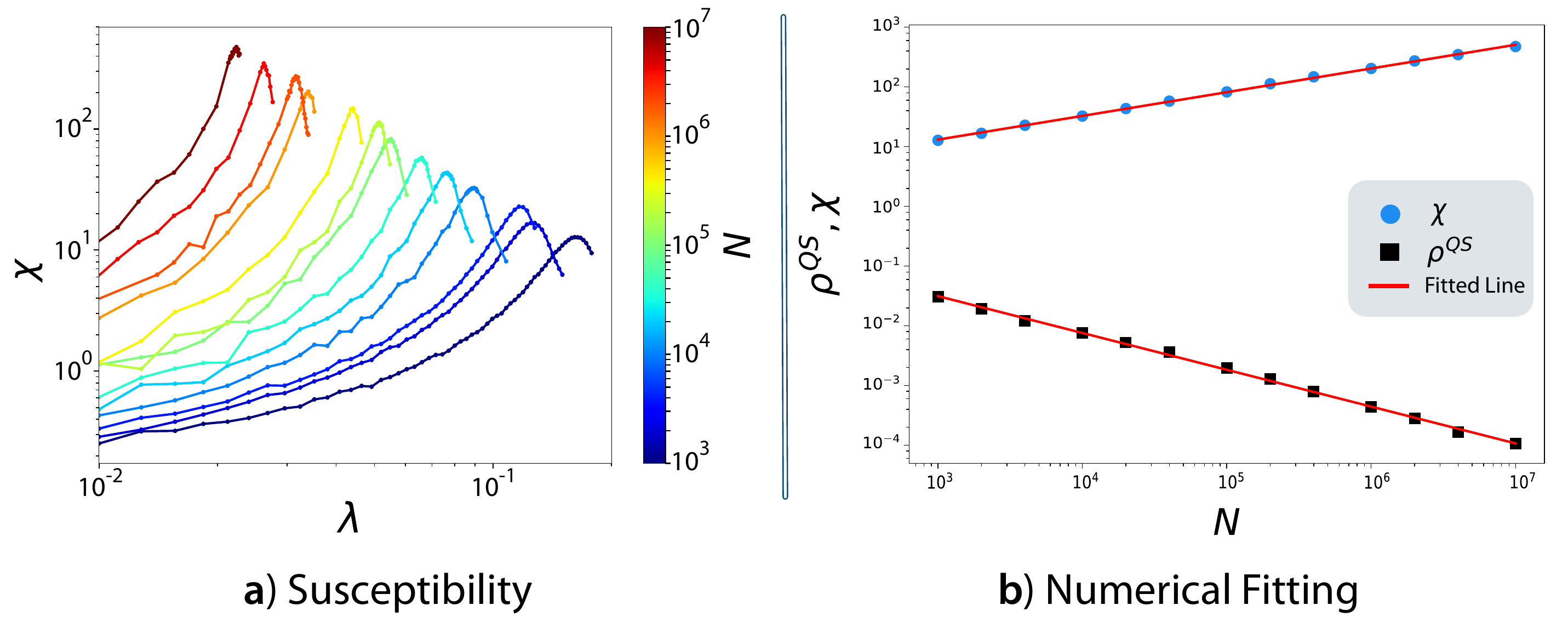}
\caption{Finite size scaling for a power-law network with $\Prob(k) \sim k^{-2.75}$. In (a) the susceptibility as a function of the spreading rate $\lambda$ with $\delta =1$, while in (b) the scaling of the susceptibility peak and its respective fraction of infected individuals, both obtained using the QS algorithm.} \label{fig:FSS_275}
\end{figure*}

\begin{figure*}[!t]
\includegraphics[width=0.96\textwidth]{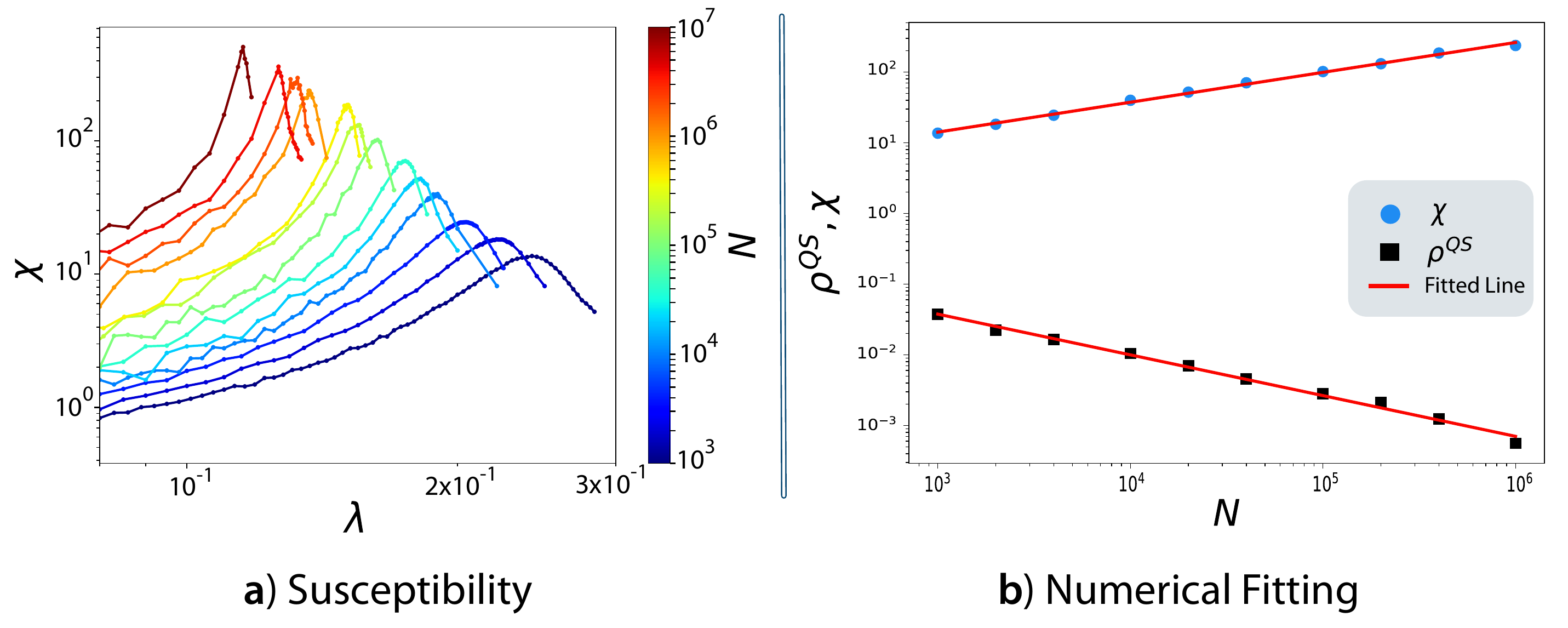}
\caption{Finite size scaling for a power-law network with $\Prob(k) \sim k^{-3.5}$. In (a) the susceptibility as a function of the spreading rate $\lambda$ with $\delta =1$, while in (b) the scaling of the susceptibility peak and its respective fraction of infected individuals, both obtained using the QS algorithm.} \label{fig:FSS_35}
\end{figure*}

Next, we present results for the CA(RP) methods using the simulation scheme described in Section \ref{sec:Monte_Carlo}. In Figure \ref{Fig:MC_Num_Err_CA}, we compare results obtained for of Monte Carlo simulations and CA(RP) mean field approaches in terms of the errors. As it can be seen, CA approaches are really precise. The observed errors are remarkably small. In this case, the SIR models present the largest error among all the CA approaches studied. Note that the CA follows a first order approximation and the error is comparable to the second order models on continuous time approaches. Here we focused on the RP case, where the error is very small. However, it is also worth mentioning that the CP cases show a poor result. These results are not presented and the reader is refereed to \cite{Arruda2017b} for more details.

\section{Finite size analysis} \label{sec:FSS}

In this section, we present an example of finite size analysis on SIS dynamics on top of uncorrelated power-law single layer networks. In order to estimate the susceptibility and the order parameter, we use the QS algorithm. For the simulations in these sections we used the QS list with $M = 25$ elements and $t_r$ values from $10^5$ to $10^4$, while $t_a$ varies between  $10^7$ and $10^5$. An important consequence of the critical behavior of a dynamical system is its scaling as the system size grows. As discussed in \cite{Ferreira2012}, for a system with a finite critical point in the thermodynamic limit, it is expected that $\tau_p(N) - \tau_c(\infty) \sim N^{\frac{-1}{\bar{\nu}}}$, where $\tau_p(N)$ is the spreading ratio, which for size $N$ and $\tau_c(\infty)$ is the same quantity in the thermodynamic limit. Note that $\tau_p(N)$ is subject to finite size effects. In this way, at the critical point, we have that
\begin{equation*}
 \rho^{QS} \sim N^{\frac{\beta'}{\bar{\nu}}}, \hspace{1cm} \chi \sim N^{\frac{\beta'+\gamma'}{\bar{\nu}}}.
\end{equation*}
From the previous equations, we can estimate the exponents $\frac{\beta'}{\bar{\nu}}$ and $\frac{\beta'+\gamma'}{\bar{\nu}}$, simply estimating the susceptibility peaks for different system sizes and fitting a line in the log-log plots of the desired quantity versus $N$. Such experiments are presented in the following sections for power-law networks. 

The finite size scaling experiments were performed in the same way as in \cite{Ferreira2012}, where the authors compare the theoretical critical point predictions with their respective numerical simulations. Similarly to \cite{Ferreira2012}, here we also study three different power-law degree distributions, $\Prob(k) \sim k^{-\zeta}$, for $\zeta = 2.25$, $\zeta = 2.75$ and $\zeta = 3.5$. Note that these exponents were chosen intentionally due to the spectral properties of their adjacency matrices and their consequences on the critical point predictions, as shown in sections \ref{sec:A} and \ref{sec:qmf}. We recall that for $2 < \zeta < \frac{5}{2}$ the QMF and HMF predictions coincide, i.e. $\tau_c^{QMF} = \tau_c^{HMF} = \frac{\E{k}}{\E{k^2}}$, while for $\frac{5}{2} < \zeta < 3$ the QMF predictions scale with $\tau_c^{QMF} = \frac{1}{\sqrt{k_{max}}}$ (which implies that $\tau_c^{QMF} \sim N^{-\frac{1}{4}}$ for the uncorrelated case, for more, see Section \ref{sec:conf}) and $\tau_c^{HMF} = \frac{\E{k}}{\E{k^2}}$. Finally, for the case $\zeta > 3$, the second moment of the degree distribution is finite, yielding a finite critical point prediction in the HMF theory. However, the QMF theory still predicts a vanishing critical point in this case. Regarding the scenario $\zeta > 3$, we emphasize that multiple susceptibility peaks were initially observed in \cite{Ferreira2012} and studied in detail in \cite{Mata2015}, where the authors related such phenomena with localization properties of those networks, specifically in \cite{Mata2015}, the authors associated the role of the outliers on the degree distribution with the multiple susceptibility peaks. They also proposed a hard upper cutoff (also called rigid cutoff) for the degree distribution as $k_{\max} = k_{\min} N^{\frac{0.75}{\zeta-1}}$, which suppresses the emergence of outliers \cite{Mata2015}. Here we follow such a model for our scaling analysis in order to avoid the complications of $\zeta > 3$, but we refer the reader to \cite{Mata2015} for a deeper discussion on this type of structures.

Our results on power-law single layer networks are presented as follows:
\begin{itemize}
\item \textbf{\emph{Power-law $2 < \zeta < \frac{5}{2}$}}: We consider uncorrelated networks with a degree distribution of the form $\Prob(k) \sim k^{-2.25}$. Figure \ref{fig:FSS_225} shows (a) the susceptibility curves, and (b) the scaling as a function of the system size $N$. As noted before, by means of a simple linear fitting using the least squared method we obtain the following exponents
\begin{equation*}
 \frac{\beta'}{\bar{\nu}} \approx -0.619, \hspace{1cm} \frac{\beta'+\gamma'}{\bar{\nu}} \approx 0.378.
\end{equation*}
It is noteworthy that in \cite{Ferreira2012} the authors estimated $\frac{\beta'}{\bar{\nu}} \approx -0.65$ and $\frac{\beta'+\gamma'}{\bar{\nu}} \approx 0.37$, which are close to the obtained values, validating our analyses. Besides, in 
\cite{Ferreira2012} the authors evaluated systems up to $N = 10^7$, which may be one of the reasons of the small differences found.

\item \textbf{\emph{Power-law $\frac{5}{2} < \zeta < 3$}}: Next, we study uncorrelated networks with a degree distribution of the form $\Prob(k) \sim k^{-2.75}$. Figure \ref{fig:FSS_275} shows on (a) the susceptibility curves, while on (b) the scaling as a function of the system size $N$. As in the previous case, we obtain the following exponents
\begin{equation*}
 \frac{\beta'}{\bar{\nu}} \approx -0.617, \hspace{1cm} \frac{\beta'+\gamma'}{\bar{\nu}} \approx 0.397.
\end{equation*}
Note that in \cite{Sander2016} the authors estimated the exponent associated with $\rho^{QS}$ for $\zeta = 2.7$ and obtained $\frac{\beta'}{\bar{\nu}} \approx 0.612$, which again validate our analyses.

\item \textbf{\emph{Power-law $\zeta > 3$}}: Finally, we consider uncorrelated networks with a degree distribution of the form $\Prob(k) \sim k^{-3.5}$, however, adopting a hard cutoff $k_{max} = k_{min} N^{\frac{0.75}{\zeta-1}}$. Figure \ref{fig:FSS_35} shows in panel (a) the susceptibility curves, while in panel (b) the scaling as a function of the system size $N$ is depicted. For this case we get the following exponents
\begin{equation*}
 \frac{\beta'}{\bar{\nu}} \approx -0.591, \hspace{1cm} \frac{\beta'+\gamma'}{\bar{\nu}} \approx 0.395.
\end{equation*}
It is important to emphasize that systems with $\zeta > 3$ present a very rich dynamical behavior and here we showed a really simple case, where the outliers of the degree distribution play a smaller role than the uncorrelated case, i.e., $k_{\max} = N^{\frac{1}{2}}$. As previously mentioned, such a case was deeply studied in \cite{Mata2015}.
\end{itemize}

\begin{figure*}[!t]
\includegraphics[width=1\textwidth]{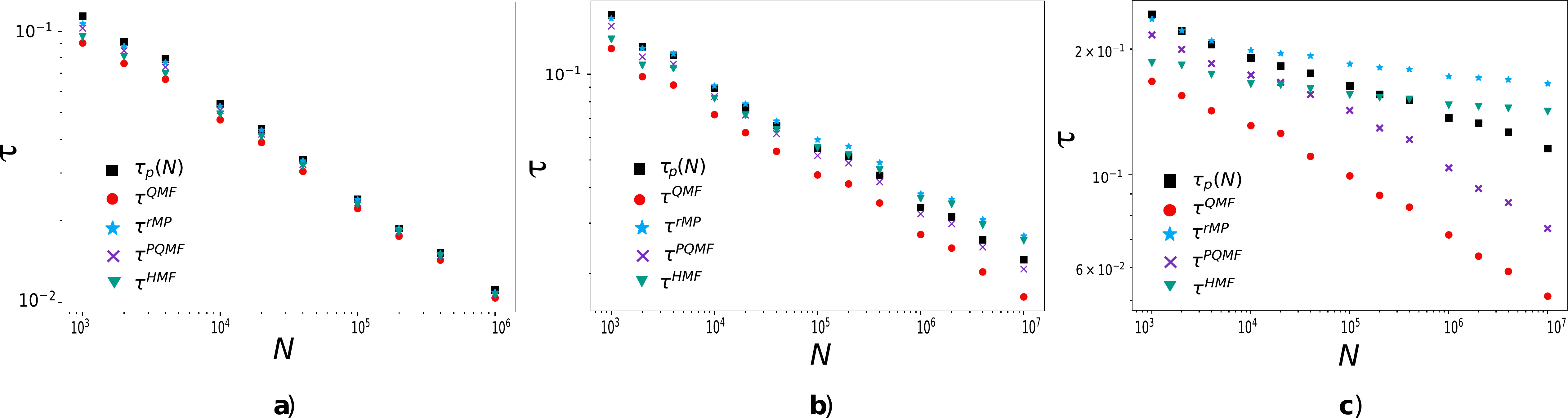}
\caption{Evaluation of the HMF, QMF and PQMF mean field predictions for the critical point considering different system sizes. The critical points were estimated using the QS algorithm on top of three different uncorrelated power-law networks with: $\zeta = 2.25$ in (a), $\zeta = 2.75$ in (b) and $\zeta = 3.5$ (rigid cutoff) in (c).} \label{fig:FSS_TEO}
\end{figure*}

Next, we also compare the previous scenarios with the mean field predictions, similarly to \cite{Ferreira2012, Mata2013}, but, aside from the QMF and HMF methods already considered in \cite{Ferreira2012}, we also evaluate the PQMF predictions, in a similar way as \cite{Mata2013} and also with the message passing predictions. The results are shown in Figure~\ref{fig:FSS_TEO}, where we considered three different uncorrelated power-law networks with: $\zeta = 2.25$ ($k_{\max} = N^{\frac{1}{2}}$) in (a), $\zeta = 2.75$ ($k_{\max} = N^{\frac{1}{2}}$) in (b) and $\zeta = 3.5$ (rigid cutoff -- $k_{\max} = k_{\min} N^{\frac{0.75}{\zeta-1}}$, similarly to \cite{Mata2015}) in (c) and their susceptibility plots are shown in figures \ref{fig:FSS_225}, \ref{fig:FSS_275} and \ref{fig:FSS_35} respectively. As for $\zeta = 2.25$, our results agree with similar experiments performed in \cite{Ferreira2012, Mata2013} and all the predictions are apparently exact in the thermodynamic limit. Besides, note that the QMF and HMF have the same scale since both coincide. This was already pointed out in \cite{Ferreira2012} for the QMF and HMF, but here we extend such an observation to the PQMF and recurrent message passing approaches. Additionally, the most accurate predictions are the inverse of the leading eigenvalue of the non-backtracking matrix, followed by the PQMF, HMF, and QMF. Next, for $\zeta = 2.75$ a similar result was presented in \cite{Mata2013}, where the authors validated  the PQMF predictions. In this case, the QMF and the PQMF predictions seem to scale correctly and the PQMF is a very good approximation for the critical point, in agreement with \cite{Mata2013}. Interestingly, the non-backtracking predictions seem to scale similarly to the HMF. Although it performs better in some cases, both seem to lead to wrong predictions in the large system limit. Finally, the last case analyzed is $\zeta = 3.5$  with a rigid cutoff $k_{\max} = k_{\min} N^{\frac{0.75}{\zeta-1}}$, similarly to the analysis performed in \cite{Mata2015}, where the authors were suppressing the effects of the outliers of the degree distribution. Our results are similar to the ones obtained in \cite{Mata2015}. These results are not shown here since it is out of the scope of this section. In this case, we observed that none of the predictions follows the correct trend, but the HMF and the non-backtracking approaches show a better results.

\section{A brief comment on the nature of the critical point}

To round up this review, we present a short discussion on the nature of the critical point and its relationship with state localization, hub activation and reactivation. We thus summarize some of the comments already made in this review but focusing on the physical/phenomenological point of view. This section is mainly centered in reference \cite{Boguna2013}, where the authors presented a very pertinent discussion on the nature of the critical point.

In \cite{Goltsev2012, Boguna2013} the authors analyzed the effects of eigenvector localization on the SIS steady state. They observed that on power-law networks with $\zeta < \frac{5}{2}$ the leading eigenvector is delocalized, implying that for $\tau > \tau_c^{QMF}$ the disease is spread over the whole network. In other words, it implies that there is a finite density of infected individuals. On the other hand, for $\zeta > \frac{5}{2}$ this eigenvector is localized, implying that the system is active in a few sets of nodes, comprising the hubs and their neighbors. This would lead to a number of infected nodes that scale sublinearly with the system size, thus it does not constitute a true active state \cite{Goltsev2012, Boguna2013}.

Next, following a similar view, in \cite{Lee2013}, the authors (partially) considered dynamical correlation effects, restricting themselves to hubs and their neighbors. In this way, the activity of a node is characterized by its lifetime, which is a function of the spreading rate and its degree, i.e., $\hat{\tau}(k, \lambda)$. When hubs are directly connected, the lifetime is sufficiently large to sustain the disease. In this scenario, above the critical point, $\tau_c^{QMF}$, there is an endemic state configured by hubs mutual reinfection. On the other hand, if the hubs are not directly connected this reinfection mechanism is not possible. Therefore, in \cite{Lee2013}, the authors suggest that above $\tau_c^{QMF}$ we have a Griffiths phase \cite{Vojta2006}. In this phase, the density of infected nodes decays more slower than exponentially, while the true phase transition is located at a higher value of $\tau$. Importantly, such an analysis points to a $\tau_c > 0$ in power-law networks with $\zeta >3$ and a vanishing critical point only for $\zeta < 3$.

Furthermore, in \cite{Boguna2013}, the authors presented numerical and analytical evidences of a vanishing critical point for any small world network with a degree distribution decaying slower than exponentially, which includes power-law networks with any $\zeta$. This result contrasts with the interpretations given in \cite{Goltsev2012, Lee2013}, where a finite threshold is predicted for $\zeta > 3$. Similarly to \cite{Lee2013}, in \cite{Boguna2013}, the authors also considered dynamical correlations, but they do not restrict themselves to hubs immediate neighbors. In fact, as pointed in \cite{chatterjee2009, Boguna2013} the direct connection is not a necessary condition to the hub reinfection mechanism. Moreover, the authors also presented arguments suggesting that perturbations can propagate up to distances of order $\ln(N)$. In summary, the main result of \cite{Boguna2013} relies on a null epidemic threshold in the thermodynamic limit in all random small world networks with degree distributions decaying slower than exponentially.

Finally, in \cite{Ferreira2016b}, the authors extend the ideas presented in \cite{Boguna2013} proposing a criterion to evaluate the nature of the critical point. Their analysis is based on the comparison between hub reinfection and recovery time scales. Formally, if the lifespan is larger than the infection time, the epidemic process is triggered by means of hub activation (hubs are reinfecting each other), leading to a vanishing threshold in the thermodynamic limit. Conversely, considering a scenario where the recovery timescale is smaller than the infection time scale the process is triggered by a collective process. In this way, it yields to a standard phase transition. Additionally, in \cite{Ferreira2016b}, the authors also argue that in the process triggered by hub activation and reactivation the QMF theory are qualitative correct because it accounts for the whole structure. On the other hand, when the transition is triggered by a collective process, the HMF should be qualitatively correct since it works on an annealed version of the network.

\section{Conclusions}

Along this review, we have presented the basic concepts methods and techniques to study spreading processes in (both single and multilayer) complex networks. We have made a special effort to properly classify and distinguish among the various approaches found in the literature. We classified the spreading processes into two classes according to the time nature: (i) continuous-time and (ii) cellular automata approaches. Moreover, the CA approaches can be classified further into two classes according to their updating scheme, namely, (a) synchronous; and (ii) asynchronous. Regarding the analytical aspects, for each process in this review, we started defining the local rules and, then formally defined their correspondent Markov chain. We used the exact SIS model to discuss some key points, such as finite size effects and the critical behavior. Obviously, the exact Markov chain is not practical due to its computational cost. In order to overcome this difficulty, we discussed the use of mean field approaches. In increasing order of complexity, we followed the simplest MF, the HMF, QMF and also the PQMF, which we also related to the non-backtracking matrix and the message passing approach. To the best of our knowledge, such relationship is not found in the previous literature. For a complete table of all the derived equations for all the mean field approaches and dynamical processes discussed here, we refer the reader to \ref{sec:appendix}.

On the other hand, we also discussed the peculiarities of doing proper simulations for each case (continuous-time or cellular automata). In this part of the review, we focused on the correct association between a given mathematical framework and its simulation algorithm. Going further, we also performed an accuracy analysis of mean field approaches (QMF and PQMF) in relation to Monte Carlo simulations. To round up, we also discussed the QS algorithm, which is a powerful tool in the analysis of stochastic processes with a single absorbing state, which is precisely the case of SIS epidemics.

Finally remark that the main aim of this review is not to perform an extensive literature review, but rather to present the different formalisms, clarifying the main ideas behind them and allowing the reader to deal with different specific situations.

\section{Acknowledgements}

GFA acknowledges Fapesp for the support provided (Grant 2012/25219-2). GFA also acknowledges Ang\'elica Sousa da Mata, Jos\'e Fernando Fontanari and Silvio da Costa Ferreira Junior for their fruitful comments. FAR acknowledges the Leverhulme Trust, CNPq (Grant No. 305940/2010-4) and FAPESP (Grants No. 2016/25682-5 and Grants 2013/07375-0) for the financial support given to this research. Y. M. acknowledges support from the Government of Arag\'on , Spain through a grant (E-19) to the group FENOL and by MINECO and FEDER funds (Grant FIS2017-87519-P). This research was developed using the computational resources of Centro de Ci\^encias Matem\'aticas Aplicadas \`a Ind\'ustria (CeMEAI) supported by FAPESP.

\bibliographystyle{elsarticle-num}
\bibliography{review}

\appendix
\section{} \label{sec:appendix}

\textbf{Summary of the mean field approximations:} In tables \ref{tab:summary1} and \ref{tab:summary2} we present a summary of the mean field approximations discussed here. Here we focus on the approximations and its numerical approximations. We also focused on uncorrelated networks, assuming that the dynamical correlations are neglectable. Obviously, each mean field approach follows a different approximation. We might emphasize that, in Table \ref{tab:summary1} for the HMF equations one should consider $\Prob(k'|k) = \frac{k' \Prob(k')}{\E{k}}$ for uncorrelated networks. In order to solve them one might use a numerical approach, such as Runge-Kutta. For the sake of an example, see Figure \ref{Fig:MC_ex} on the main text, where we compare the QMF and PQMF approaches with Monte Carlo approximations.

\begin{table*}[!th]
\scalefont{0.8}
\caption{Part I of the summary of mean field approximations for the SIS, SIR and MT processes.}
\label{tab:summary1}
\begin{adjustbox}{angle=90}
    \begin{tabular}{c|c|l|c}
    Approximation & Process & Formulation & Critical Point\\
    \hline
    MF & SIS & $\frac{d\E{Y}}{dt} = \lambda \E{k} \E{Y} (1 - \E{Y}) - \delta \E{Y}$ & $\tau^{MF} > 1$ \\
    \hline
    MF & SIR & $\begin{cases}
		\frac{d \E{X}}{dt} = -\lambda \E{k} \E{Y} (1 - \E{Y} - \E{Z}) \\
		\frac{d \E{Y}}{dt} = \lambda \E{k} \E{X} (1 - \E{Y} - \E{Z}) - \delta \E{Y} \\
		\frac{d \E{Z}}{dt} = \delta \E{Y}
		\end{cases}$ & $\tau^{MF} > 1$ \\
    \hline
    MF & MT &   $\begin{cases}
		\dfrac{d \E{X}}{dt} =& -\lambda \E{Y} \E{X}\\ 
		\dfrac{d \E{Y}}{dt} =& \lambda \E{Y} \E{X} - \alpha \left( \E{Y} \E{Y} + \E{Y} \E{Z} \right)\\
		\dfrac{d \E{Z}}{dt} =& \alpha \left( \E{Y} \E{Y} + \E{Y} \E{Z} \right)
		\end{cases}$ & \textbf{--} \\

    \hline
    \hline
    HMF & SIS & $\frac{d \E{Y_k(t)}}{dt} = - \delta \E{Y_k(t)} + \lambda k (1 - \E{Y_k(t)}) \sum_{k'}\Prob(k'|k) \E{Y_{k'}(t)}(t)$ & $\tau^{HMF}_c = \frac{\E{k}}{\E{ k^2 }}$ \\
    \hline
    HMF & SIR & $\begin{cases}
                  \frac{d \E{X_k(t)}}{dt} = - \lambda k \E{X_k(t)} \sum_{k'} \frac{k'-1}{k'} \Prob(k'|k) \E{Y_{k'}(t)} \\
                  \frac{d \E{Y_k(t)}}{dt} = - \delta \E{Y_k(t)} + \lambda k \E{X_k(t)} \sum_{k'} \frac{k'-1}{k'} \Prob(k'|k) \E{Y_{k'}(t)} \\
                  \frac{d \E{Z_k(t)}}{dt} = \delta \E{Y_k(t)}
                 \end{cases}$  & $\tau^{HMF}_c = \frac{\E{k}}{\E{ k^2 } - \E{k}}$ \\
    \hline
    HMF & MT & $\begin{cases}
                  \frac{d \E{X_k(t)}}{dt} = - \lambda k \E{X_k(t)} \sum_{k'} \frac{k'-1}{k'} \Prob(k'|k) \E{Y_{k'}(t)} \\
                  \frac{d \E{Y_k(t)}}{dt} = \lambda k \E{X_k(t)} \sum_{k'} \frac{k'-1}{k'} \Prob(k'|k) \E{Y_{k'}(t)} - \alpha k \E{Y_k(t)} \sum_{k'}\Prob(k'|k) (\E{Y_{k'}(t)}) + \E{Z_k(t)}) \\
                  \frac{d \E{Z_k(t)}}{dt} = \alpha k \E{Y_k(t)} \sum_{k'}\Prob(k'|k) (\E{Y_{k'}(t)}) + \E{Z_k(t)})
                 \end{cases}$ & \textbf{--} \\
    \hline
    \hline
    QMF & SIS & $\frac{d \E{Y_i(t)}}{dt} = \lambda \sum_{j=0}^N \A_{ij} \E{Y_j(t)} - \E{Y_i(t)} \left(  \lambda \sum_{j=0}^N \A_{ij} \E{Y_j(t)} + \delta \right)$ & $\tau_c^{QMF} = \frac{1}{\Lambda_1(\A)}$ \\
    \hline
    QMF & SIR & $\begin{cases}
                  \frac{d \E{X_i(t)}}{dt} = - \lambda \E{X_i(t)} \sum_{j=0}^N \A_{ij} \E{Y_j(t)} \\
                  \frac{d \E{Y_i(t)}}{dt} = - \delta \E{Y_i(t)} + \lambda \E{X_i(t)} \sum_{j=0}^N \A_{ij} \E{Y_j(t)} \\
                  \frac{d \E{Z_i(t)}}{dt} = \delta \E{Y_i(t)}
                 \end{cases}$ & $\tau_c^{QMF} = \frac{1}{\Lambda_1(\A)}$ \\
    \hline
    QMF & MT & $\begin{cases}
                  \frac{d \E{X_i(t)}}{dt} = - \lambda \E{X_i(t)} \sum_{j=0}^N \A_{ij} \E{Y_j(t)} \\
                  \frac{d \E{Y_i(t)}}{dt} = \lambda \E{X_i(t)} \sum_{j=0}^N \A_{ij} \E{Y_j(t)} - \alpha \E{Y_i(t)} \sum_{j=0}^N \A_{ij} (\E{Y_j(t)} + \E{Z_j(t)}) \\
                  \frac{d \E{Z_i(t)}}{dt} = \alpha \E{Y_i(t)} \sum_{j=0}^N \A_{ij} (\E{Y_j(t)} + \E{Z_j(t)})
                 \end{cases}$ & \textbf{--} \\
    \hline
    \end{tabular}
\end{adjustbox}
\end{table*}

\begin{table*}[!th]
\scalefont{0.75}
\caption{Part II of the summary of mean field approximations for the SIS, SIR and MT processes.}
\begin{adjustbox}{angle=90}
    \begin{tabular}{c|c|l|c}
    Approximation & Process & Formulation & Critical Point\\
    \hline
    PQMF & SIS & $\begin{cases}
		   \frac{d \E{Y_i(t)}}{dt} = - \delta Y_i(t) + \lambda \sum_{j=0}^N \A_{ij} \E{X_i Y_j(t) } \\
		   \frac{d\E{X_i Y_j}}{dt} \approx \delta \E{Y_j} - \lambda \E {X_i Y_j} + \lambda \frac{1 - \E{Y_i} - \E{X_i Y_j}}{1 - \E{Y_j}} \sum_{j\leftarrow k} \B_{(i\leftarrow j, j\leftarrow k)} \E{X_j Y_k} - \lambda \frac{\E{X_i Y_j}}{1 - \E{Y_i}} \sum_{i\leftarrow k} \Hm_{(i\leftarrow j, i\leftarrow k)} \E {X_i Y_k}
                  \end{cases}$ & $\tau_c^{PQMF} = \frac{1}{\Lambda_1(\B)}$\\
    \hline
    PQMF & SIR & $\begin{cases}
                   \frac{d \E{X_i(t)}}{dt} = - \lambda \sum_{j=0}^N \A_{ij} \E{X_i Y_j(t)} \\
                   \frac{d \E{Y_i(t)}}{dt} = - \delta \E{Y_i(t)} + \lambda \sum_{j=0}^N \A_{ij} \E{X_i Y_j(t)} \\
                   \frac{d \E{Z_i(t)}}{dt} = \delta \E{Y_i(t)} \\
                   \frac{d \E{X_i Y_j(t)}}{dt} =  \lambda \frac{\E{X_i(t)} - \E{X_i Y_j(t)} - \E{Y_i Z_j (t)}}{\E{X_j(t)}} \sum_{j\leftarrow k} \B_{(i\leftarrow j, j\leftarrow k)} \E{X_j Y_k(t) } - \delta \E{X_i Y_j(t)} - \lambda \sum_{i\leftarrow k} \Hm_{(i\leftarrow j, i\leftarrow k)} \frac{\E{X_i Y_j (t)} \E{X_i Y_k(t)}}{\E{X_i(t)}} \\
                   \frac{d \E{X_i Z_j(t)}}{dt} =  \delta \E{X_i Z_j(t)} + \lambda \sum_{i\leftarrow k} \Hm_{(i\leftarrow j, i\leftarrow k)} \frac{\E{X_i Z_j (t)} \E{X_i Y_k(t)}}{\E{X_i(t)}}
                  \end{cases} $ & $\tau_c^{PQMF} = \frac{1}{\Lambda_1(\B)}$ \\
    \hline
    PQMF & MT & $\begin{cases}
                  \frac{d \E{X_i(t)}}{dt} = - \lambda \sum_{j=0}^N \A_{ij} \E{X_i Y_j(t)} \\
                  \frac{d \E{Y_i(t)}}{dt} = \lambda \sum_{j=0}^N \A_{ij} \E{X_i Y_j(t)} - \alpha \sum_{j=0}^N \A_{ij} (\E{ Y_i Y_j(t)} + \E{Y_i Z_j(t)}) \\
                  \frac{d \E{Z_i(t)}}{dt} = \alpha \sum_{j=0}^N \A_{ij} (\E{Y_i Y_j(t)} + \E{Y_i Z_j(t)}) \\
                  \frac{d \E{X_i Y_j(t)}}{dt} =  \lambda \frac{\E{X_i(t)} - \E{X_i Y_j(t)} - \E{Y_i Z_j (t)}}{\E{X_j(t)}} \sum_{j\leftarrow k} \B_{(i\leftarrow j, j\leftarrow k)} \E{X_j Y_k(t) } - \lambda \sum_{i\leftarrow k} \Hm_{(i\leftarrow j, i\leftarrow k)} \frac{\E{X_i Y_j (t)} \E{X_i Y_k(t)}}{\E{X_i(t)}} \\
                  \hspace{1.4cm} - \alpha \sum_{j\leftarrow k} \frac{\E{X_i Y_j (t)}}{\E{Y_j(t)}} \B_{(i\leftarrow j, j\leftarrow k)} \left( \E{Y_j(t)} - \E{Y_j Z_k (t)} \right) \\
                  \frac{d \E{X_i Z_j(t)}}{dt} = \lambda \sum_{i\leftarrow k} \Hm_{(i\leftarrow j, i\leftarrow k)} \frac{\E{X_i Z_j (t)} \E{X_i Y_k(t)}}{\E{X_i(t)}} \\
		  \hspace{1.4cm} + \sum_{j\leftarrow k} \frac{\E{X_i Y_j (t)}}{\E{Y_j(t)}} \B_{(i\leftarrow j, j\leftarrow k)} \left( \E{Y_j(t)} - \E{Y_j Z_k (t)} \right)
                 \end{cases} $ & \textbf{--} \\
    \hline
    \hline
    CA & SIS & $\begin{cases}
                   y_i(t+1) = (1-y_i(t))(1 -s_i(t)) + y_i(t)(1-\delta) + y_i(t)\delta(1- s_i(t)) \\
                   s_i(t) = \prod_{j=1}^N \left(1-\lambda \R_{ji} y_j(t)\right)
                  \end{cases}$  & $\tau_c^{CA} = \frac{1}{\Lambda_1} (\A)$ \\
    \hline
    \end{tabular}
    \label{tab:summary2}
\end{adjustbox}
\end{table*}

\end{document}